\documentclass[aps,prd,twocolumn,floatfix,superscriptaddress,groupedaddress,nofootinbib,preprintnumbers]{revtex4-2}

\pdfoutput=1
\usepackage{graphicx}
\usepackage{latexsym}
\usepackage{amsfonts}
\usepackage{amssymb}
\usepackage{amsmath}
\usepackage{slashed}
\usepackage{feynmp}
\usepackage{hyperref}
\usepackage{url}
\usepackage{color}
\usepackage{cancel}
\usepackage{multirow,array}
\usepackage{bbold}
\usepackage{float}
\usepackage{ulem}
\usepackage{mathtools}
\usepackage{subfigure}
\graphicspath{ {./figs/} }
\newcommand{\unit}[1]{\ensuremath{\, \mathrm{#1}}}
\DeclareMathOperator{\sech}{sech}

\begin{document}
\date{\today}

\title{Limits on Dark Matter Annihilation from the Shape of Radio Emission in M31}

\author{Mitchell J.~Weikert}
\affiliation{Department of Physics and Astronomy, Rutgers University, Piscataway, NJ 08854, USA}

\author{Matthew R.~Buckley}
\affiliation{Department of Physics and Astronomy, Rutgers University, Piscataway, NJ 08854, USA}

\begin{abstract}
Well-motivated models of dark matter often result in a population of electrons and positrons within galaxies produced through dark matter annihilation -- usually in association with gamma rays. As they diffuse through galactic magnetic fields, these $e^\pm$ produce synchrotron radio emission. The intensity and morphology of this signal depends on the properties of the interstellar medium through which the $e^\pm$ propagate. Using observations of the Andromeda Galaxy (M31) to construct a model of the gas, magnetic fields, and starlight, we set constraints on dark matter annihilation to $b\bar{b}$ using the morphology of $3.6 \unit{cm}$ radio emission. As the emission signal at the center of M31 is very sensitive to the diffusion coefficient and dark matter profile, we base our limits on the differential flux in the region between $0.9-6.9 \unit{kpc}$ from the center. We exclude annihilation cross sections $\gtrsim 3 \times 10^{-25}$ cm$^3$/s in the mass range $10-500$ GeV, with a maximum sensitivity of $7\times 10^{-26}$ cm$^3$/s at $20-40$ GeV. Though these limits are weaker than those found in previous studies of M31, they are robust to variations of the diffusion coefficient.
\end{abstract}
\maketitle 

\section{Introduction} \label{sec:IB}

To date, all evidence for dark matter comes from its gravitational influence on the visible matter in the Universe. However, the majority of successful models for the production of dark matter require some level of non-gravitational interactions between the visible and dark sectors. Perhaps the best-known such scenario is that of thermally-produced dark matter, where a small interaction between dark matter and the Standard Model results in a relic population of non-relativistic particles due to thermal freeze-out during the early Universe. Weakly-Interacting Massive Particles (WIMPs) are the most well-known implementation of this class of dark matter models. In such models, the observed density of dark matter is obtained if the velocity-averaged annihilation cross section is $\langle \sigma v\rangle \sim 3\times 10^{-26}\,\unit{cm^3/s}$, for dark matter in the mass range $\sim 1 - 10^3\unit{GeV}$ \cite{2022PhRvD.106j3526B}.\footnote{Though model-specific details can easily change these numbers by ${\cal O}(1)$ factors or more.} 

Thermal relics, as well as other models with dark matter-Standard Model interactions of similar magnitude, result in a number of possible experimental signatures. Of particular interest to this work is indirect detection, where present-day residual annihilation or decay of dark matter into Standard Model particles gives visible signatures that can be detected here on Earth. Annihilation to Standard Model particles will generically result in cascade decays terminating in stable $e^\pm$, photons, neutrinos, and $p/\bar{p}$, evidence of which can reach Earth-based detectors from their astronomical point of origin. As the strength of these signals increases with dark matter density squared and decreases with the distance to target squared, the indirect detection targets with the greatest signal rate are the largest and closest conglomerations of dark matter. The highest intensity signals are therefore expected to be seen from our own Milky Way Galactic Center, but other nearby galaxies -- such as Andromeda (M31) \cite{2017ApJ...836..208A, 2021PhRvD.103b3027K}, the Large Magellanic Cloud (LMC) \cite{2015PhRvD..91j2001B}, the Small Magellanic Cloud (SMC) \cite{2016PhRvD..93f2004C}, and local dwarf galaxies \cite{2011PhRvL.107x1303G, 2011PhRvL.107x1302A, 2014PhRvD..89d2001A,2015PhRvD..91h3535G, 2015ApJ...809L...4D, 2015PhRvL.115h1101G, 2015PhRvL.115w1301A, 2021PhRvD.104h3026G} -- can have significant signals too. As the backgrounds and systematics for these systems differ from the Milky Way, they can be compelling targets despite the lower signal rate.

High-energy prompt photons, which can either come directly from the annihilation of dark matter or the cascade decays of annihilation products, travel largely unimpeded from where they were produced to Earth. Such photons are therefore the most straightforward indirect detection signal, with a morphology that is set only by the dark matter distribution. Interestingly, many groups have identified an excess of gamma rays in the energy range $1-3 \, \unit{GeV}$ from data collected by the Fermi Large Area Telescope (LAT) \cite{2009ApJ...697.1071A} in the Milky Way \cite{2011PhLB..697..412H, 2016PDU....12....1D, 2015JCAP...03..038C, 2014PhRvD..90b3526A, 2013PhRvD..88h3521G}, with morphology compatible with the dark matter expectations. Possible signals consistent with this gamma ray excess have been reported in M31 \cite{2017ApJ...836..208A}, and the LMC \cite{2015PhRvD..91j2001B}, though with less significance. These excesses can be well-fit by dark matter models with $m_\chi \sim\mathcal{O}(10 - 100 \, \unit{GeV})$ annihilating to either $b\bar{b}$ or $\tau^+\tau^-$ followed by cascade decays which result in the observed photons, with a thermally averaged cross section of $\langle \sigma v\rangle\sim 2 \times 10^{-26} \unit{cm^3/s}$ \cite{2016PDU....12....1D, 2015JCAP...03..038C, 2014PhRvD..90b3526A, 2013PhRvD..88h3521G, 2015PhRvD..91j2001B, 2018PhRvD..97j3021M}. 

However, the ultimate origin of this gamma ray excess remains unclear. An unresolved population of millisecond pulsars (MSPs) in the center of the Milky Way has been suggested as an alternate source of this signal \cite{2011JCAP...03..010A, 2013MNRAS.436.2461M, 2014JHEAp...3....1Y, 2019JCAP...09..042M, 2018NatAs...2..819B, 2018NatAs...2..387M}. Ref.~\cite{2015JCAP...05..056L} has argued that the distribution of gamma rays in the Galactic Center excess (GCE) in the Fermi-LAT data contains non-Poissonian statistics, suggestive of a MSP origin. At this time, debate appears to be far from settled, with questions about the spectrum of MSPs \cite{2013PhRvD..88h3009H}, morphology and background emission modelling \cite{2016PhRvL.117k1101C, 2017ApJ...840...43A, 2021PhRvD.103f3029D}, and the non-Poissonian statistics interpretation \cite{2019arXiv190408430L, 2020PhRvD.101b3014C} all remaining open. A recent analysis \cite{2021arXiv211006931M} suggests that the observed excess is best fit by a combination of point sources and a diffuse source, but uncertainties are large enough that either origin could dominate.

In this context, searches for indirect detection signals beyond prompt photons are especially interesting. Dark matter annihilation into Standard Model final states which decay into gamma rays will necessarily also have significant branching ratios into electrons and positrons. These $e^\pm$ will interact with galactic magnetic fields, ambient photons (from starlight, dust and the Cosmic Microwave Background (CMB)) and interstellar gas, losing energy and emitting a range of secondary photons ranging in energy from radio up to X-rays. These signals depend on the properties of the target beyond the dark matter distribution, introducing uncertainties that do not exist in prompt photon searches; however the systematics and backgrounds are largely distinct as well.

In this work, we set constraints on dark matter annihilation by analyzing a 3.6\,cm radio map of the Andromeda galaxy (M31) by the Effelsberg telescope \cite{2020A&A...633A...5B}. M31 has been a common target for dark matter indirect detection searches using radio emission from the center of the galaxy \cite{2013PhRvD..88b3504E, 2016PhRvD..94b3507C, 2021MNRAS.501.5692C, 2022PhRvD.106b3023E}; though the resulting constraints are sensitive to assumptions made about the astrophysical characteristics of the galaxy and the dark matter distribution. 

The $e^\pm$ injection rate (and the associated radio signal) at the center of M31 is dependent on the slope of the dark matter density distribution, which has considerable uncertainties \cite{2022arXiv221202999B}. The galactic center radio signal is also dependent on assumptions of the diffusion coefficient.
For example, Refs.~\cite{2013PhRvD..88b3504E, 2016PhRvD..94b3507C, 2021MNRAS.501.5692C} assume electrons and positrons lose all of their energy before diffusing a measurable distance, predicting larger signals in this region than analyses which assume greater diffusion (as considered in as Ref.~\cite{2018PhRvD..97j3021M}).

In order to set robust limits which are less sensitive to reasonable variations of the astrophysical parameters, we consider the morphology and intensity of the radio emission from the region of M31 between 0.9--6.9~kpc from the galactic center. We compute the expected synchrotron emission from electrons and positrons in M31 produced through dark matter in the mass range $6-500 \unit{GeV}$ annihilating to $b\bar{b}$ while varying the diffusion coefficient over the range of experimentally allowed values \cite{2014A&A...571A..61G, 2011ApJ...739...20C, 2012A&A...546A...4T, 2020A&A...633A...5B}. 
Commonly used tools for modeling the transport of $e^\pm$ (such as \texttt{galprop} \cite{2015ICRC...34..492M} and \texttt{rx-dmfit} \cite{2017JCAP...09..027M}) use a uniform diffusion coefficient. For modeling the relatively large region of interest for our analysis, this assumption is insufficient. We develop a numerical solution that allows for radial dependence in all transport coefficients -- including the diffusion coefficient. Using our numerical method, we solve for the spherically averaged electron-positron phase space density and compute the radio emission from this phase space density and an axisymmetric model of the magnetic field. We then set exclusion limits on the annihilation cross section of dark matter, while varying the diffusion coefficient normalization. 

The remainder of this paper is organized as follows. In Section~\ref{sec:data}, we describe the radio observations of M31 used in the analysis. Section~\ref{sec:DM_production} describes the spectrum and morphology of electrons and positrons injected into M31 through the annihilation of dark matter. We construct our models of the magnetic fields, interstellar radiation field (ISRF), and thermal gas density using a variety of relevant data in Section~\ref{sec:M31model}. 
The transport of electrons and positrons within M31 is described in Section~\ref{sec:propagation}. In this section, we include a discussion of the physics of charged particle transport in a galaxy and our numerical method for solving the transport equation for systems with position-dependent energy loss and diffusion. In Section~\ref{sec:synchrotron}, we calculate the intensity and morphology of the resulting synchrotron emission. Our statistical method for determining a data-driven background model and setting exclusion limits is described in Section~\ref{sec:SM}. In Section~\ref{sec:limits}, we present our results. Finally, we conclude in Section~\ref{sec:C}.

\section{Radio Observations of M31} \label{sec:data}

To constrain dark matter annihilation into high-energy electrons and positrons in M31, we use the non-thermal radio flux per unit frequency per beam $dS/d\nu$ from a survey of $\nu = 8.35 \unit{GHz}$ emission in M31 \cite{2020A&A...633A...5B} using the Effelsberg 100-m telescope. Our data has the thermal emission subtracted, along with 38 point sources which are not associated with M31. This is the highest frequency, and therefore highest resolution, intensity map of M31 measured by the Effelsberg telescope. The frequency bandwidth is $\Delta \nu = 1.1\unit{GHz}$, while the half-power beam-width (HPBW) is $1.5'$ (corresponding to a physical size of $0.34\, \unit{kpc}$ at the distance of M31). The root-mean-squared (rms) noise of the data is $\sigma_{\rm rms} = 0.25\,\unit{mJy/\rm{beam}}$ in the inner $9.13\,\unit{kpc} \times 9.13\,\unit{kpc}$ region and $\sigma_{\rm rms} = 0.3\,\unit{mJy/\rm{beam}}$ elsewhere. 
\begin{figure*}[th]
\includegraphics[width=2\columnwidth]{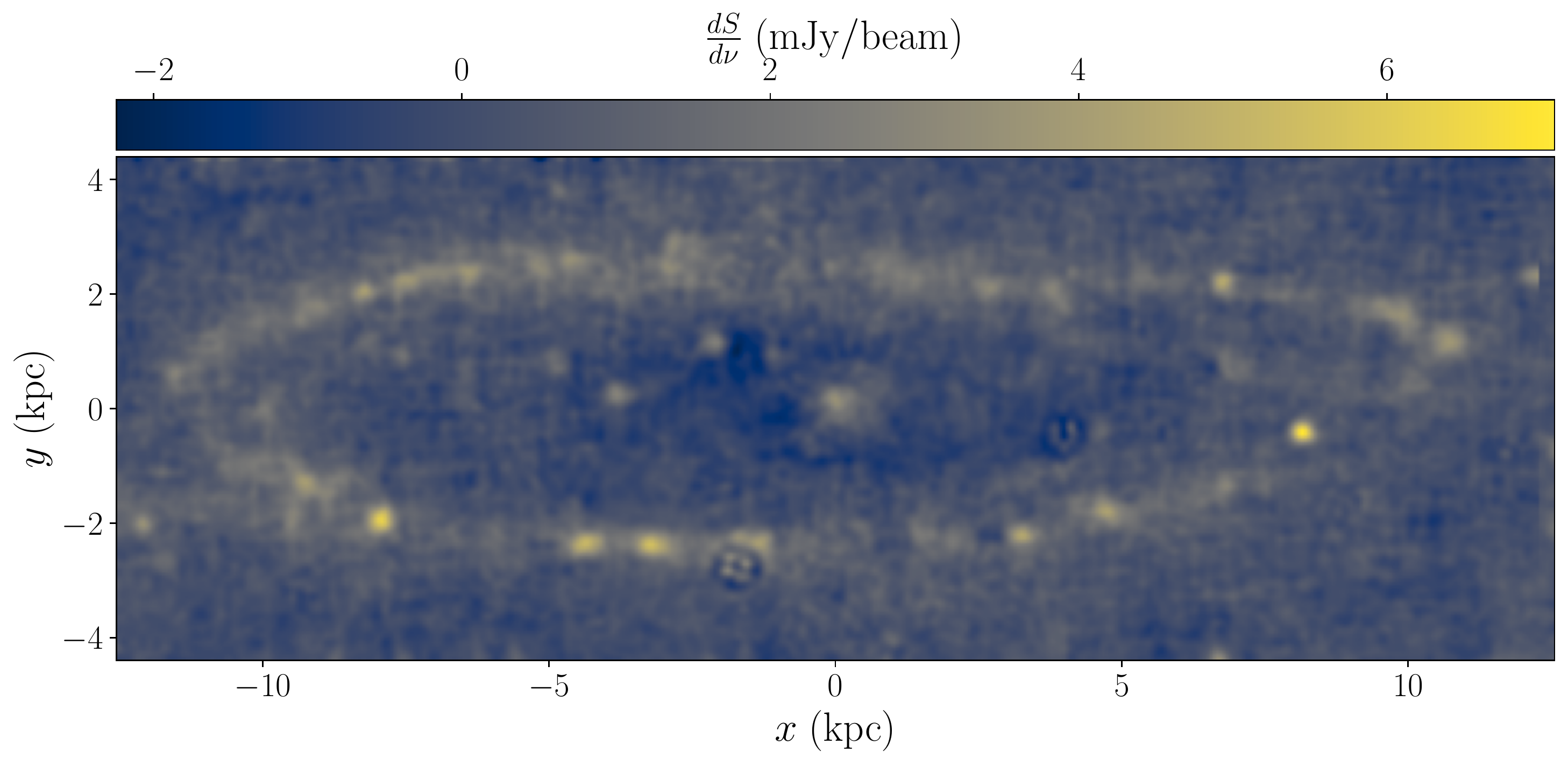}
\caption{Smoothed non-thermal radio intensity map of M31 from Ref.~\cite{2020A&A...633A...5B}, showing the flux per unit frequency per beam averaged over a frequency bandwidth of $1.1\unit{GHz}$. The HPBW projected into the plane of M31 is $0.340 \unit{kpc}$ and the rms noise is given by $\sigma_{\rm rms} = 0.25 \unit{mJy/\rm{beam}}$ in the inner $9.13 \unit{kpc} \times 9.13 \unit{kpc}$ region and $\sigma_{\rm rms} = 0.3 \unit{mJy/\rm{beam}}$ in the rest of the map. Digitized data for this figure was provided by the authors of Ref.~\cite{2020A&A...633A...5B}.}
 \label{fig:intensity_map}
 \end{figure*}
 
In Figure~\ref{fig:intensity_map}, we show the intensity map of the data.\footnote{Note that our vertical axis in Figure~\ref{fig:intensity_map} is inverted compared to Figure~9 of Ref.~\cite{2020A&A...633A...5B}.} The reported intensity at each pixel is the radio emission measured by the Effelsberg telescope from that location on the sky; this corresponds to the true differential flux convolved with the frequency band and the angular beam centered on that location. Our $x$ and $y$ coordinates are oriented so that the $x$ axis is aligned with the semimajor axis of M31, converting angular coordinates to lengths assuming a distance to M31 of 785~kpc \cite{2005MNRAS.356..979M}.

We note that the observations of M31 have significant negative values, well in excess of statistical expectations given the rms noise. Most notably, the data has a large negative excursion located near the center of M31, at $(x,y)\sim (-2,1)\,\unit{kpc}$. This may be due to over-subtraction of one of the point sources identified by Ref.~\cite{2020A&A...633A...5B}
These negative values suggest that pixels labeled as having a flux of 0 mJy/beam actually may have a significant (unknown) positive flux. This in part motivates our choice to set limits using {\it morphology} of the expected dark matter-induced radio signal, rather than overall intensity.

Like the Milky Way, M31 is a spiral galaxy with approximate axisymmetry around a rotating stellar disk. We adopt cylindrical coordinates with the origin at the center of M31, the cylindrical radius $R$ and the height away from the midplane of the disk $z$. The assumption of axisymmetry implies there is no dependence on the angle around the disk $\phi$. We will refer to the spherical radius from the center of M31 as $r$. Note that we observe M31 at an angle of inclination given by $\beta = 77.5^\circ$ \cite{2001ChPhL..18.1420M}, and so the cylindrical $(R,z,\phi$) coordinates are projected on to the $x-y$ coordinate system of Figure~\ref{fig:intensity_map}. 

\section{Dark Matter Production of $e^\pm$ in M31} \label{sec:DM_production}

Dark matter annihilation to unstable Standard Model particles such as $b$-quarks, $\tau$ leptons, or $W$ bosons will result in cascade decays involving large numbers of leptons and QCD bound states, many of which themselves will decay into prompt photons and $e^\pm$. Though the exact spectrum of stable final-state particles that result from these decay showers of course depends on the initial Standard Model pair produced in the annihilation, there are also broad similarities regardless of the progenitors. As explicit calculation of the showers show (using a particle hadronization and decay package such as \texttt{pythia8} \cite{2022arXiv220311601B}), the final state particles will have energies in the ${\cal O}(0.1-10\,\unit{GeV})$ range for dark matter with the weak-scale masses typically expected for thermal relics. In this section, we calculate the injection morphology and spectrum of electrons and positrons produced in M31 due to dark matter annihilation, assuming weak-scale masses and annihilation into $b\bar{b}$ pairs. 

While the flux on Earth of prompt photons from dark matter annihilation involves the integration of the dark matter density squared along the line of sight, electrons and positrons generated far from the Earth do not propagate to detectors here. Instead, we must track the evolution of the $e^\pm$ phase space density as the particles diffuse and energy is lost -- a task we will take up in Section~\ref{sec:propagation}. For now, we will quantify the rate of production of the $e^\pm$ with a source term, which depends on the local dark matter density, $\rho_\chi(\boldsymbol{x})$, at every location within M31 and the particle physics model of the dark matter candidate. The source term or injection density rate of $e^\pm$ due to dark matter self-annihilation is given by
\begin{equation} \label{eq:injection}
Q_e(\boldsymbol{x},E) = \frac{\langle\sigma v\rangle}{2 m_\chi^2}\frac{d N_e}{dE} \rho_\chi(\boldsymbol{x})^2,
\end{equation}
where $\langle\sigma v\rangle$ is the thermally averaged annihilation cross section, $m_\chi$ is the dark matter particle mass, $dN_e/dE$ is the injection spectrum of $e^\pm$ per annihilation in terms of the total energy\footnote{Here and throughout this work we will use units where $c = \hbar = 1$.} $E = (m_e^2 + p^2)^{1/2}$ of the $e^\pm$. Here and throughout this work, we have assumed that dark matter is its own antiparticle. If it is not and the abundance of dark matter particles and anti-particles is symmetric, there is an additional factor of $1/2$ in Eq.~\eqref{eq:injection}.

The energy spectrum of the $e^\pm$ source is determined by $dN_e/dE$ which is influenced by the dark matter mass and the annihilation channel. Our choices for these parameters are motivated by fits of dark matter annihilation to the gamma ray excess in the Milky Way's Galactic center \cite{2013PhRvD..88h3521G, 2014PhRvD..90b3526A, 2015JCAP...03..038C, 2016PDU....12....1D}. In the Milky Way, the dark matter candidates that best fit the gamma ray excesses have $m_\chi \sim 30-50 \; \unit{GeV}$ annihilating to $b\bar{b}$ or $m_\chi \sim 10 \; \unit{GeV}$ annihilating to $\tau^+ \tau^-$ \cite{2016PDU....12....1D, 2015JCAP...03..038C, 2014PhRvD..90b3526A, 2013PhRvD..88h3521G, 2015PhRvD..91j2001B}. A similar signal in M31 has a best fit of dark matter with mass $m_\chi \sim 10 \; \unit{GeV}$ annihilating to $b\bar{b}$, or $m_\chi \sim 5 \; \unit{GeV}$ annihilating to $b\bar{b}$ and $\tau^+\tau^-$ democratically \cite{2018PhRvD..97j3021M}. 

\begin{figure}
\centering
\includegraphics[width=0.9\columnwidth]{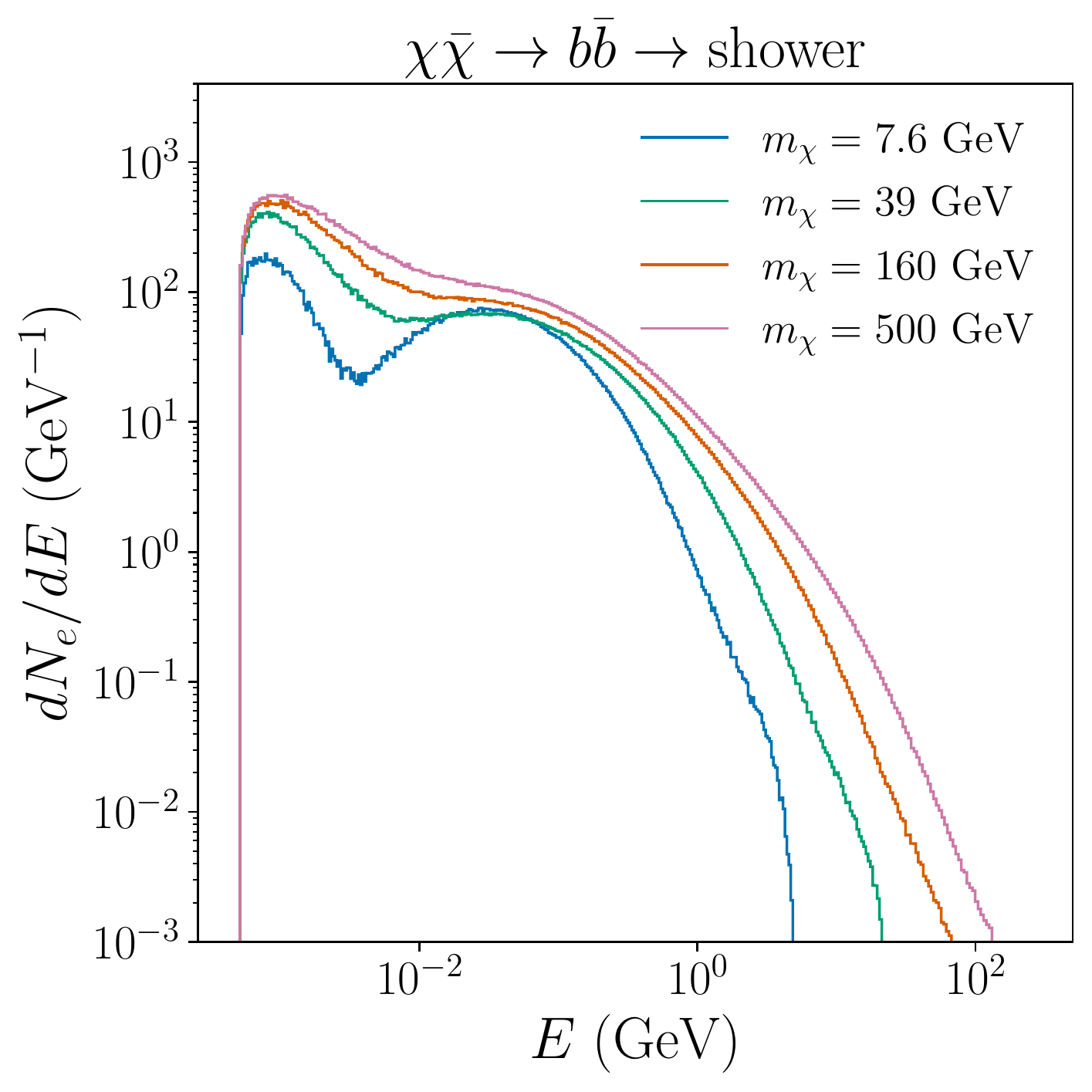}
\caption{The number of $e^\pm$ in final states per unit energy per annihilation of dark matter into $b\bar{b}$ for a representative sample of dark matter masses $m_\chi$. 
 }\label{fig:electron_spectrum}
 \end{figure}

In part motivated by these fits, in this work we consider dark matter annihilation into $b\bar{b}$ with $m_\chi$ in the range $6-500\,\unit{GeV}$. We scan in dark matter mass from $6-500\,\unit{GeV}$, and use \texttt{pythia8} to decay, shower, and hadronize the $b\bar{b}$ annihilations to calculate $dN_e/dE$ for each choice of $m_\chi$.\footnote{The \texttt{pythia} shower was modified to allow decays of Standard Model particles which are meta-stable on detector timescales, but whose decays would be astrophysically relevant, e.g., $\mu$, $\pi^0$, $\pi^\pm$, and neutrons.} We show in Figure~\ref{fig:electron_spectrum} the resulting $e^\pm$ spectra for a sample of representative dark matter masses. All spectra are cut off at low energy by $E=m_e$. 

The morphology of the source is determined by the dark matter distribution of M31. This can be fit by a modified Navarro-Frenk-White (NFW) profile \cite{1996ApJ...462..563N, 1997ApJ...490..493N, 1996MNRAS.278..488Z}
\begin{equation} \label{eq:NFW}
\rho_\chi = \frac{\rho_0}{\left(r/r_s\right)^{\gamma_{\rm NFW}}\left(1 + r/r_s\right)^{3-\gamma_{\rm NFW}}}
\end{equation}
 where $\gamma_{\rm NFW}$ is the logarithmic inner-slope, $\rho_0$ is the scale density and $r_s$ is the scale radius. In the Milky Way, the gamma ray excess favors an inner slope of $\gamma_{\rm NFW} \sim 1.25$ \cite{2016PDU....12....1D, 2015JCAP...03..038C, 2014PhRvD..90b3526A, 2013PhRvD..88h3521G}. This slope is adopted by Ref.~\cite{2018PhRvD..97j3021M} for the dark matter distribution of M31. However, there is considerable uncertainty in the M31 dark matter distribution. In keeping with available kinematic fits to the rotation curve, we use a standard NFW with $\gamma_{\rm NFW} = 1$. For the scale density and the scale radius, we use the best-fit values from analysis of kinematic data \cite{2012A&A...546A...4T}:
 \begin{equation} \label{eq:DM_params}
     \begin{split}
        \rho_0 = & \,  (0.418 \pm 0.068)\,\unit{GeV/cm^3},  \\
        r_s = &  \,  (16.5\pm 1.5)\,\unit{kpc}. \\
    \end{split}
\end{equation}

\section{Astrophysical Model of M31} \label{sec:M31model}

The intensity and morphology of a radio signal which originates from electrons and positrons in a galaxy depends greatly on how the charged particles propagate. In M31, the most important propagation effects are diffusion and energy loss \cite{2018PhRvD..97j3021M}. To calculate the effects of these, we must first model the properties of the interstellar medium.

In this section, we present our models for the magnetic field, the interstellar radiation field (ISRF), and the various components of thermal gas within M31. As relativistic $e^\pm$ propagate, the interstellar magnetic field causes them undergo synchrotron energy loss, emitting photons at radio frequencies. Random fluctuations in the magnetic field also diffuse the charged particles through space. The ISRF causes the $e^\pm$ to lose energy through inverse-Compton scattering. Lastly, the various components of gas cause energy loss through bremsstrahlung and Coulomb scattering.

The measured distance to M31 has varied considerably in the literature. Early measurements found a value of $690 \unit{kpc}$ \cite{1963AJ.....68..435B}, but modern techniques and measurements prefer $760-785\unit{kpc}$ \cite{2005MNRAS.356..979M, 2010A&A...511A..89C, 2021ApJ...920...84L}. As a result, the references we used to construct our astrophysical models assume a variety of distances to the galaxy. 
Throughout this work, we use a distance to M31 of 785~kpc \cite{2005MNRAS.356..979M, 2010A&A...511A..89C} obtained using measurements of tip of the red giant branch stars. When necessary, we scale the results of previous works used in our model of M31.  


\subsection{Magnetic Fields of M31}\label{sec:magnetic field}

The magnetic field $\boldsymbol{B}$ in M31 is turbulent, with fluctuations on many length-scales, ranging from the size of the galaxy down to far below the resolution limit of our experimental probes. 
The details of small-scale fluctuations are not observationally accessible, but the expectation value of $\boldsymbol{B}^2$ and the power spectrum of fluctuations of the magnetic field at different locations in the galaxy can be measured though radio polarization and cosmic ray propagation observations, respectively. These observationally accessible features of the galactic magnetic field suffice for our purposes.

We write the magnetic field as the product of an axi-symmetric field magnitude and a random dimensionless vector-field that depends on location $\boldsymbol{x}$: 
\begin{equation}
    \boldsymbol{B}(\boldsymbol{x}) = \bar{B}(R,z)\boldsymbol{b}(\boldsymbol{x}).
\end{equation}
The vector $\boldsymbol{b}$ contains the local fluctuations in the field, and 
\begin{equation}\label{eq:second moment}
\bar{B}(R, z)^2 \equiv \langle\boldsymbol{B}(\boldsymbol{x})^2\rangle
\end{equation}
is the expectation value of the magnitude of the field squared, which we assume is independent of $\phi$. Formally, this is an ensemble average with respect to the probability distribution that the magnetic field is sampled from.

As is conventional \cite{1998APh.....9..227C, 1998ApJ...509..212S, 2007ARNPS..57..285S, 2012ApJ...752...68V, 2018PhRvD..97j3021M}, we characterize the magnetic field fluctuations in terms of a power spectrum normalized as 
\begin{equation} \label{eq:power spec}
\langle \boldsymbol{b}(\boldsymbol{x})^2 \rangle  = \int\displaylimits_{k_0}^\infty dk P_b(k) = 1,
\end{equation}
where $k_0$ is the minimum wavenumber for which the power spectrum applies.
The length-scale $1/k_0$ is typically assumed to be a factor of $\mathcal{O}(10-100)$ smaller than of the characteristic length-scale of the galaxy \cite{1998APh.....9..227C}. For M31, this implies $\mathcal{O}(0.1 \unit{kpc}) \lesssim 1/k_0 \lesssim \mathcal{O}(1 \unit{kpc})$ (though when setting limits we will vary this parameter over a much more conservative range). Observations of the propagation of cosmic rays in the Galaxy \cite{1995ApJ...441..209H} find a diffusion coefficient of the form $D \propto E^\delta$ with $\delta\simeq 1/3$. This is consistent with magnetic field fluctuations that follow a Kolmogrov spectrum \cite{1941DoSSR..30..301K}, $P_b \propto k^{-5/3}$. The diffusion of charged particles moving in a magnetic field with fluctuations obeying a Kolmogrov spectrum will be discussed in detail in Section~\ref{sec:diff coef}.

\begin{figure}[th]
\includegraphics[width=0.9\columnwidth]{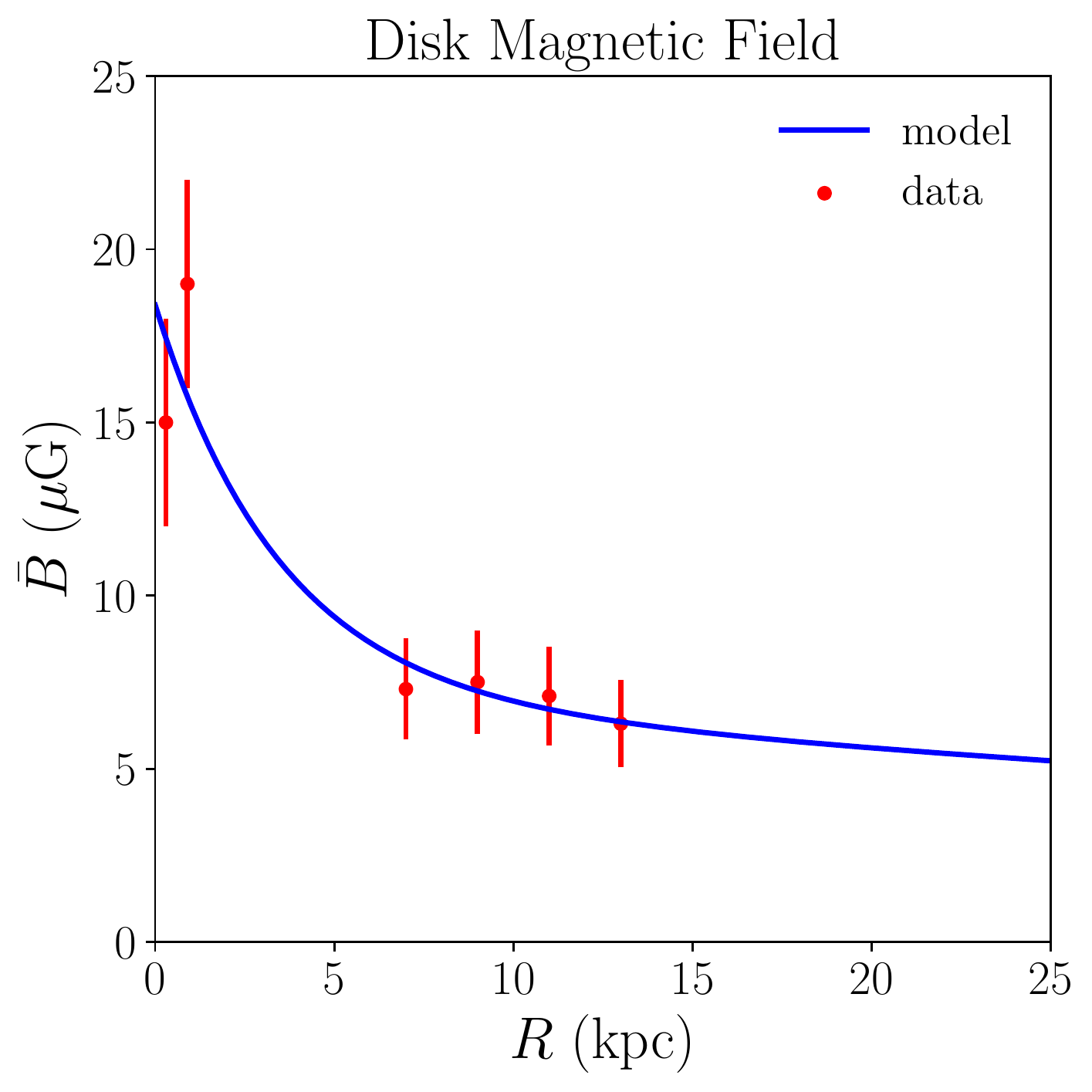}
\caption{RMS Magnetic field strength in the disk of M31, as measured by Refs.~\cite{1998IAUS..184..351H,2004A&A...414...53F} (red). Our double-exponential fit Eq.~\eqref{eq:magnetic field model} (with the parameters of Table~\ref{tab:magneticfield_parameters}) is shown in blue.}
 \label{fig:magnetic_field_disk}
 \end{figure}

We determine the $R$ dependence of $\bar{B}$ from measurements of the M31 magnetic field (taken to be the root-mean-squared (RMS) field strength) in the disk in three  regions: within the inner $1\unit{kpc}$ \cite{1998IAUS..184..351H}, in the range $6-14 \unit{kpc}$ \cite{2004A&A...414...53F}, and in intergalactic space \cite{2019A&A...622A..16O}. 
The measurements from Refs.~\cite{1998IAUS..184..351H, 2004A&A...414...53F} are shown in Figure~\ref{fig:magnetic_field_disk}.
The intergalactic magnetic field has been measured to be at most $0.3 \unit{\mu G}$ \cite{2019A&A...622A..16O}, which we take to be the $1\sigma$ upper bound of the field strength outside of M31. To require our fit to the M31 field strength to agree with this upper bound, we include $\bar{B} = 0.15 \pm 0.15 \unit{\mu G}$ at $R = 300 \unit{kpc}$ (though it is not shown in Figure~\ref{fig:magnetic_field_disk}) in addition to the measurements from Refs.~\cite{1998IAUS..184..351H, 2004A&A...414...53F}.  

We fit the RMS magnetic field measurements to the functional form
\begin{equation}
\label{eq:magnetic field model}
\bar{B}(R, z) = \left(B_0 e^{-R/R_{B,0}} + B_1 e^{-R/R_{B, 1}} \right)e^{-|z|/h_B(R)}.
\end{equation}
which has sufficient flexibility to fit the available data. Our best-fit parameters of Eq.~\eqref{eq:magnetic field model} are shown in Table~\ref{tab:magneticfield_parameters}.

\begin{table}[h!]
\centering
 \begin{tabular}{|c||c|} 
 \hline\hline
 \multicolumn{2}{c}{Magnetic Field} \\
 \hline\hline
 $B_0 \unit{(\mu G)}$ & $11.2 \pm 2.9$ \\ 
 $B_1 \unit{(\mu G)}$ & $7.2  \pm 1.9$\\
 $R_{B,0} \unit{(kpc)}$ & $3.5  \pm 2.6$ \\
 $R_{B,1} \unit{(kpc)}$ & $77.6  \pm 21$ \\ [1ex] 
 \hline
 \end{tabular}
 \caption{Parameter values of Eq.~\eqref{eq:magnetic field model}, fit to the data at $|z|=0$ (shown in Figure~\ref{fig:magnetic_field_disk}).}\label{tab:magneticfield_parameters}
\end{table}

Assuming equipartition between the cosmic ray and magnetic field energy density, the vertical scale height of the magnetic field is approximately four times the scale height of the disk of synchrotron emission \cite{2012SSRv..166..133H}, which itself is approximated by the scale height of the HI disk \cite{2004A&A...414...53F}. 
The HI disk scale height as a function of $R$ was measured by Ref.~\cite{1991ApJ...372...54B}; rescaling from the distance to M31 assumed in that work ($690\unit{kpc}$), the magnetic field scale height is
\begin{equation}\label{eq:B scale height}
\begin{split}
    h_B(R)=& 4h_{\rm syn} = 4 h_{\rm HI}\\ 
    =& (0.83 \pm 0.17 \,\unit{kpc}) + (0.064 \pm 0.012)\times R.
\end{split}
\end{equation}

\subsection{Interstellar Radiation Fields of M31}\label{sec:isrf}

The ISRF of M31 has contributions from the CMB, starlight, and infrared emission from dust. We model each component, and sum the results to obtain the total energy density of radiation within M31. Most straightforward is the energy density of the CMB, which is (for our purposes) uniform and given by
\begin{equation}\label{eq:isrf_cmb}
    \rho_{\rm CMB} = \frac{\pi^2(k_B T_{\rm CMB})^4}{15} = 0.26 \unit{eV/cm^3},
\end{equation}
where $k_B T_{\rm CMB} = 2.3\times 10^{-4} \unit{eV}$ \cite{2009ApJ...707..916F}.

We determine the energy density from stars and dust from the measured luminosity distribution of M31. The energy density of stellar radiation $\rho_*$ in M31 is related to the bolometric luminosity density $Q_*$ by
\begin{equation}\label{eq:energy_density_star}
\rho_*(\boldsymbol{x}) = \frac{1}{4\pi}\int d^3 x' \frac{Q_*(\boldsymbol{x'})}{|\boldsymbol{x}-\boldsymbol{x'}|^2}.
\end{equation}
In Ref.~\cite{2011A&A...526A.155T}, the extinction-corrected stellar luminosity distribution of M31 was modeled for five different structural components (bulge, disk, nucleus, young disk and stellar halo). As these distributions are extinction corrected, they describe the luminosities that would be observed without dust absorbing stellar emission.
Assuming that the dust is in equilibrium with the starlight,
the luminosity it emits in IR is equal to the luminosity that it absorbs. Therefore, the extinction-corrected stellar luminosity distribution integrated over all wavelengths approximates the bolometric luminosity from stars and dust combined. 

Ref.~\cite{2011A&A...526A.155T} models the luminosity density $Q_j$ as
\begin{equation}\label{eq:lum_density_model}
Q_j(\boldsymbol{x}) = Q_{0,j} \exp{\left[\left(\frac{\sqrt{R^2+(z/q_j)^2}}{A_{0,j}}\right)^{1/{N_j}}\right]},
\end{equation}
where $j = ($bulge, disk, nucleus, young disk, stellar halo) indexes the various components of the luminosity distribution and $(Q_{0,j},q_j,A_{0,j},N_j)$ are parameters determined separately for each component. As Ref.~\cite{2011A&A...526A.155T} finds that the M31 luminosity is dominated by the bulge and disk components, 
we only include those in our model of the ISRF.

Ref.~\cite{2011A&A...526A.155T} fits the parameters $q_j,A_{0,j},N_j$ to data, and provides the extinction corrected total luminosity of each M31 structural component in each of the $ugriz$ filter bands. These luminosities are defined as
\begin{equation}
    L_{a, j} \equiv\left.\left(\lambda \frac{dL_j}{d\lambda}\right)\right|_{\lambda_a},
\end{equation}
where $a=(u,g,r,i,z)$ represents the spectroscopic band of the measurement, and $\lambda_a$ is the central wavelength of the relevant band.

The $Q_{0,j}$ depend on the total bolometric luminosity of each component $L_{{\rm bol},j}$:
\begin{equation}\label{eq:luminosity_norm}
    L_{{\rm bol},j} = \int d^3 x' Q_j(\boldsymbol{x'}).
\end{equation} 
The bolometric luminosity of the $j$ component can be found by integrating
\begin{equation}
    L_{\mathrm{bol}, j} = \int d\lambda \frac{dL_j}{d\lambda}
\end{equation}
Therefore, we need the luminosity per unit wavelength -- or spectral energy distribution (SED) -- of each component. 

\begin{figure*}
\centering
\subfigure[]{\label{fig:bulge_SED}\includegraphics[width=0.9\columnwidth]{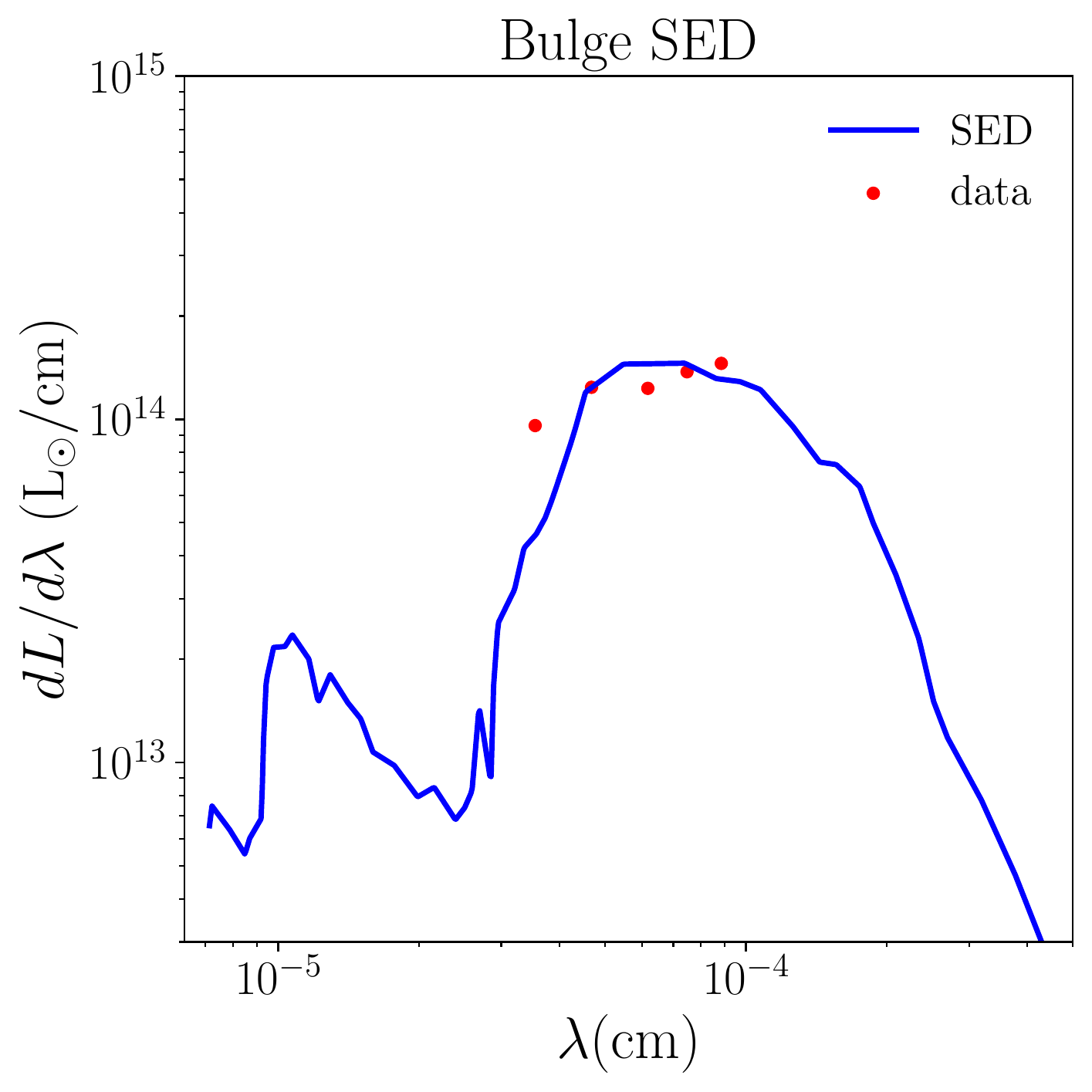}}
\subfigure[]{\label{fig:disk_SED}\includegraphics[width=0.9\columnwidth]{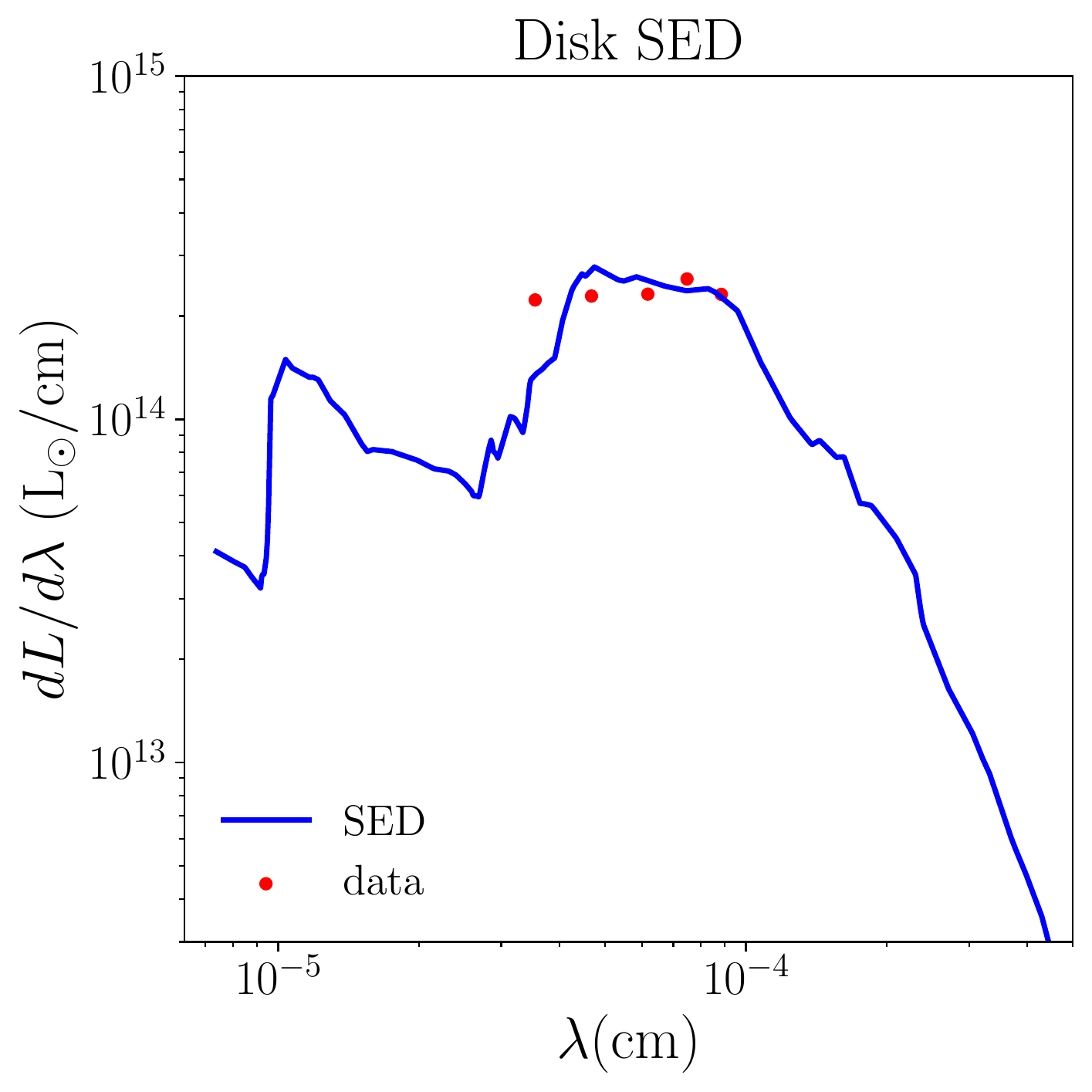}}
\caption{Best-fit rescaled SED models \cite{2012MNRAS.426..892G, 2012IAUS..284..112G} and observed differential luminosities in $ugriz$ filters \cite{2011A&A...526A.155T} for (a) the bulge and (b) the disk.}
\label{fig:SEDs}
\end{figure*}

The SED is only available in the inner 1~kpc of M31 \cite{2012MNRAS.426..892G}, but M31's bulge is concentrated in the inner $1-2 \unit{kpc}$ \cite{2011A&A...526A.155T}. Therefore, we use the SED from the inner 1~kpc of M31 \cite{2012MNRAS.426..892G} as a template for the SED of the bulge, renormalized so that it accounts for the luminosity of the whole bulge. To chose the appropriate normalization we digitize and smoothly interpolate the extinction corrected SED of the inner 1~kpc of M31 from Figure 1 of Ref.~\cite{2012MNRAS.426..892G}, and fit the proportionality constant between the inner kpc and the entire bulge by minimizing the $\chi^2$ between the re-normalized SED and the bulge luminosity in each band, derived from the fits of Ref.~\cite{2011A&A...526A.155T}. Our best fit rescaled SED and the measured luminosities in the $ugriz$ bands for the bulge are shown in Figure~\ref{fig:bulge_SED}. Using the best fit rescaled SED, we then integrate $dL_{\rm bulge}/d\lambda$ over wavelengths $\lambda$ to obtain the bolometric luminosity of the bulge.

For the functional form of the SED for the disk, we subtract our best-fit SED for the bulge from the extinction corrected SED for the whole galaxy (given in Ref.~\cite{2012IAUS..284..112G}). Using this functional form, we repeat the procedure that we used for the bulge: we perform a $\chi^2$ minimization to find the proportionality constant that gives the best agreement between the rescaled SED and the luminosity values for the disk, and integrate $dL_{\rm disk}/d\lambda$ over $\lambda$ to get the bolometric luminosity of the disk. Our best fit rescaled SED and the measured luminosities in the $ugriz$ bands for the disk are shown in Figure~\ref{fig:disk_SED}. 

\begin{figure}[th]
\includegraphics[width=0.9\columnwidth]{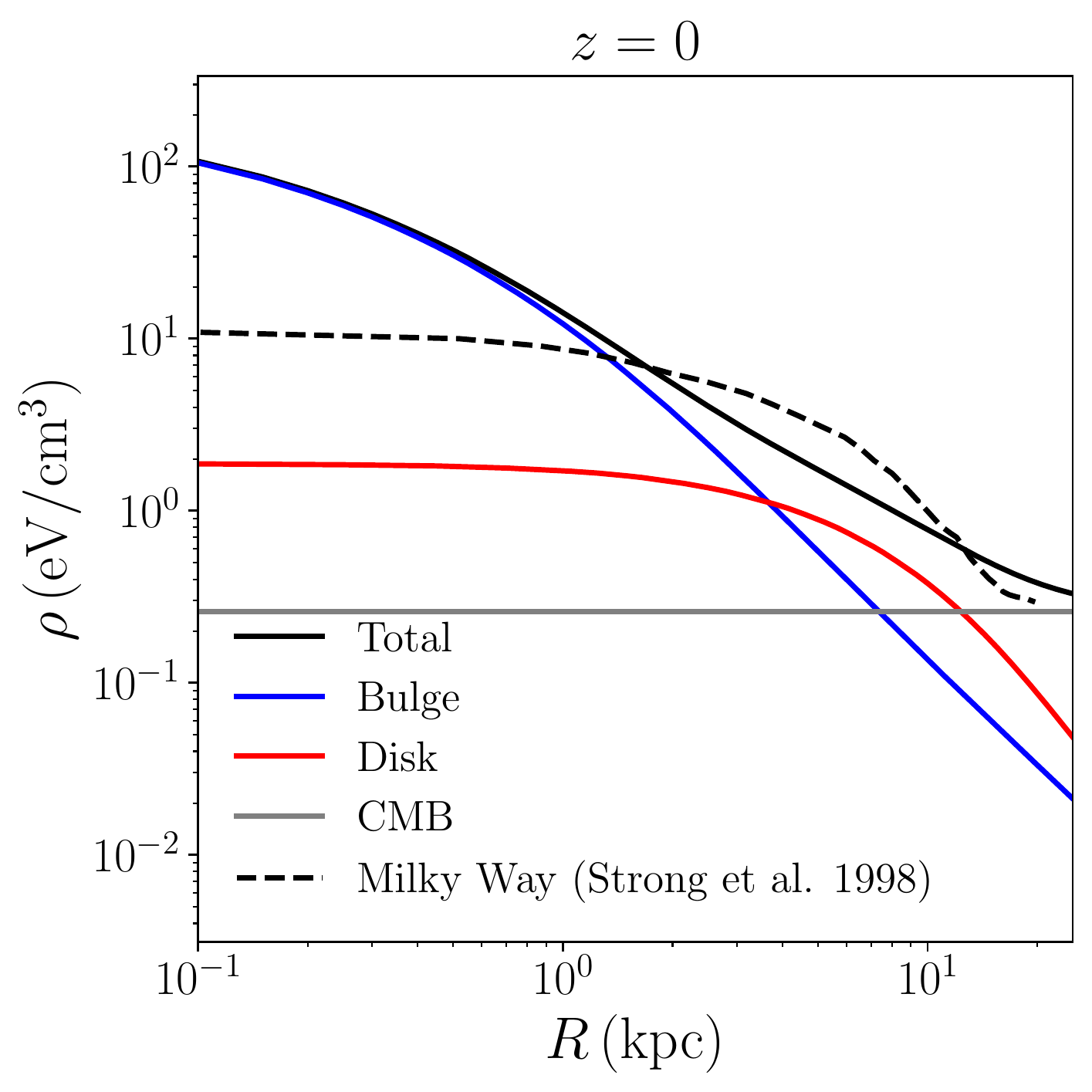}
\caption{The ISRF radiation density for M31 along the disk ($z=0$). The CMB result is given in Eq.~\eqref{eq:isrf_cmb}. The bulge and disk components come from replacing $Q_*$ with $Q_{\rm bulge}$ and $Q_{\rm disk}$, respectively in Eq.~\eqref{eq:energy_density_star}. The Milky Way ISRF is digitized from Figure~1 of Ref.~\cite{2000ApJ...537..763S}.}
\label{fig:ISRF}
\end{figure}

We use our bolometric luminosity values and Eq.~\eqref{eq:luminosity_norm} to calculate $Q_{0, j}$ for each component. We show the values of the structural parameters $(Q_{0,j},q_j,A_{0,j},N_j)$ and luminosities $(L_{u,j}, L_{g, j}, L_{r,j}, L_{i, j}, L_{z, j}, L_{\mathrm{bol}, j})$ in Table~\ref{tab:radiation_parameters} (for $j={\rm bulge},{\rm disk}$).  We add our results for the luminosity density of the disk and bulge to get $Q_*(\boldsymbol{x})$ and numerically integrate Eq.~\eqref{eq:energy_density_star} to obtain $\rho_*$. As we derived our luminosity density distributions from extinction corrected luminosities \cite{2011A&A...526A.155T} and SEDs \cite{2012MNRAS.426..892G, 2012IAUS..284..112G}, our result for $\rho_*$ contains contributions from starlight and dust. Our model for the ISRF $\rho_{\rm tot} = \rho_* + \rho_{\rm CMB}$ as well as the radiation density from individual components in the plane of the disk are given in Figure~\ref{fig:ISRF}.

It is important to note that our model for the starlight luminosity of the innermost regions of M31 is significantly larger than the equivalent for the Milky Way \cite{2000ApJ...537..763S} (shown with the dashed curve in Figure~\ref{fig:ISRF}). Previous dark matter studies of radio emission from the center of M31 used a starlight model scaled with an $\mathcal{O}(1)$ factor from the Milky Way \cite{2018PhRvD..97j3021M, 2013PhRvD..88b3504E, 2021MNRAS.501.5692C, 2022PhRvD.106b3023E}. The higher luminosity in the center of the galaxy that we find in our model leads to greater energy losses into X-rays from  $e^\pm$ inverse Compton scattering with the starlight photons. As a consequence of this increased energy-loss mechanism, the radio signature of dark matter-produced electrons and positrons in the galactic center is reduced. 

At distances further from the center ($\gtrsim 1\,\unit{kpc}$), our starlight model more closely matches those previously assumed for M31. As we will describe in detail, our dark matter constraints are obtained from radio emission in this region rather then from the center itself. Consequentially, we are less sensitive to differences in the starlight in the core of M31.

\begin{table}[h!]
\centering
 \begin{tabular}{|c||c|c|}
 \hline\hline
 \multicolumn{3}{c}{Radiation Field}\\
 \hline\hline
 Parameter & Bulge Value & Disk Value \\ [0.5ex] 
 \hline
 \multicolumn{3}{c}{Structural Parameters}\\
 \hline
 $Q_0\,(10^{10}\,L_\odot/\rm{kpc}^3)$ & $1.4 \times 10^{2}$ & $1.9 \times 10^{-2}$ \\
 $A_0\,\unit{(kpc)}$ & $4.6\times 10^{-3}$ & $2.6$ \\
 $q$ & 0.72 & 0.17 \\
 $N$ & 2.7 & 1.2 \\
 \hline
 \multicolumn{3}{c}{Observed Luminosities}\\
 \hline
 $L_u\, (10^{10}\,L_\odot)$ & 0.34 & 0.78 \\
 $L_g\, (10^{10}\,L_\odot)$ & 0.57 & 1.1 \\
 $L_r\, (10^{10}\,L_\odot)$ & 0.75 & 1.4 \\
 $L_i\, (10^{10}\,L_\odot)$ & 1.0 & 1.9 \\
 $L_z\, (10^{10}\,L_\odot)$ & 1.3 & 2.0 \\ 
 $L_{\rm bol}\, (10^{10}\,L_\odot)$ & 2.0 & 3.0 \\ [1ex]
 \hline
 \end{tabular} 
  \caption{Top: best-fit parameters to the extinction-corrected luminosity distribution Eq.~\eqref{eq:lum_density_model}. Bottom: observed extinction-corrected luminosities in $ugriz$ filter bands followed by our derived bolometric luminosities.  The bulge values are in the second column while the disk values are in the third column. All values except for $L_{\rm bol}$ and $Q_0$ are taken from Ref.~\cite{2011A&A...526A.155T}, see text for details of our calculations of $L_{\rm bol}$ and $Q_0$.}
  \label{tab:radiation_parameters}
\end{table}

\subsection{Gas in M31}\label{sec:gas}

The thermal gas of M31 can be split into ionized gas and neutral gas, each of which play different roles in the energy loss of relativistic electrons and positrons. Elastic collisions between the ionized gas and the $e^\pm$ result in Coulomb losses in the $e^\pm$. This leads to a net transfer of energy out of the $e^\pm$ and into the gas. Interactions between $e^\pm$ and ionized gas also result in bremsstrahlung losses due to inelastic collisions. Neutral gas only causes bremsstrahlung losses. As the rate of energy loss depends on the properties of the ionized and neutral gas, we must model HI, H$_2$, and $^4$He gas separately.

\subsubsection{Ionized Gas}
Due to the difficulty of observing M31 as compared to our own Galaxy, the gas model of M31 is motivated by that of the Milky Way. Following Ref.~\cite{2001AJ....122..908G}, we model the ionized gas density as
\begin{equation}\label{eq:ionized gas}
\langle n_{\rm ion} \rangle \equiv \bar{n}_{\rm ion}(R, z) = \bar{n}_{\mathrm{ion},0} \sech^2{\left(R/R_{\rm ion}\right)} \sech^2{\left(z/z_{\rm ion}\right)}
\end{equation}
where $\langle n_{\rm ion} \rangle$ is the ion density averaged over scales that are small compared to the galaxy but large compared to fluctuations in the ion density.  The parameters $\bar{n}_{\rm ion}$, $R_{\rm ion}$, and $z_{\rm ion}$ will be fit to M31 measurements of the ion density from H$\alpha$ emission \cite{1994ApJ...431..156W} and Faraday rotation \cite{2004A&A...414...53F}, as we will discuss below. 

We first extract the ion density at the mid-plane of M31 at $R\simeq 9\unit{kpc}$ from measurements of H$\alpha$ emission \cite{1994ApJ...431..156W}. Here, the observable is the emission measure (${\cal E}$) along the line of sight, which is related to the ion density by
\begin{equation} \label{eq:EM}
    {\cal E} = \int d\ell \, \langle n_{\rm ion}^2\rangle,
\end{equation}
Ref.~\cite{1994ApJ...431..156W} measures the value of $\cal E$ along a line of sight at $R=9 \unit{kpc}$ and then projects to the result that would be observed if M31 were viewed face-on, finding $\mathcal{E} = 6-15 \unit{pc \,cm^{-6}}$. 

The ionized gas inhabits the galaxy in clumps that are small compared to astrophysical scales. These clumps make up a fraction of the volume of the galaxy given by the fill factor $\phi$, which Ref.~~\cite{1994ApJ...431..156W} assumes to be the Milky Way value of $0.2$. Under this assumption (and taking the maximal value of $\mathcal{E} = 15 \unit{pc\, cm^{-6}}$ as well as their median scale height of $\hat{z}_{\rm ion} = 500 \unit{pc}$), Ref.~~\cite{1994ApJ...431..156W} calculates a mid-plane ion density of $\hat{n}_{\rm ion} = 0.39 \unit{cm^{-3}}$ \cite{1994ApJ...431..156W}. This is the density in a gas clump corresponding to a density field in the neighborhood of $R=9 \unit{kpc}, z = 0 \unit{kpc}$ of 
\begin{equation}\label{eq:filling function}
    n_{\rm ion} = \mathcal{F}(\boldsymbol{x}) \hat{n}_{\rm ion}
\end{equation}
where $\mathcal{F}$ is the filling function which varies over short length-scales and is defined to be $1$ inside a clump of ionized gas and $0$ outside. Its spatial average over astrophysical length-scales is equal to the fill factor $\phi$.
The result obtained by Ref.~\cite{1994ApJ...431..156W} for $\hat{n}_{\rm ion}$ is likely an overestimate since they obtained it using their maximal value of $\cal E$. Therefore, we modify it to the result they would have obtained if they had used their central value of $\cal E$.

Given Eqs.~\eqref{eq:EM}\&\eqref{eq:filling function}, the mid-plane ion density scales with the emission measure and the scale height as
\begin{equation}
    \hat{n}_{\rm ion} \propto \sqrt{\frac{\mathcal{E}}{\hat{z}_{\rm ion}}}.
\end{equation}
We can use this dependence to re-scale the results of Ref.~\cite{1994ApJ...431..156W} for $\hat{n}_{\rm ion}$, as well as propagate uncertainties based on the measured value of $\mathcal{E} \simeq (10 \pm 5) \unit{pc\, cm^{-6}}$ and the range of assumed values for the scale height $\hat{z}_{\rm ion} \simeq (500 \pm 300) \unit{pc}$. After doing so, our re-scaled result for the mid-plane ionized gas density becomes $\hat{n}_{\rm ion} = (0.32 \pm 0.20) \unit{cm^{-3}}$. Using Eq.~\eqref{eq:filling function}, and averaging over a small neighborhood around $R=9 \unit{kpc}, z = 0 \unit{kpc}$ leads to our inferred value of $\bar{n}_{\rm ion}$ in this neighborhood:
\begin{equation}
    \bar{n}_{\rm ion}^{\rm obs} = \phi \hat{n}_{\rm ion} = (0.063 \pm 0.039) \unit{cm^{-3}}.
\end{equation}

By observing rotation measures (RM) from Faraday rotation and assuming magnetic field equipartition, Ref.~\cite{2004A&A...414...53F} determines $\bar{n}_{\rm ion}$ in the upper layers of the thermal disk (between $0.3 - 1 \unit{kpc}$ from the mid-plane) at three different values of $R$ between $8-14 \unit{kpc}$. We take the distance of these measurements from the midplane to be the midpoint of the upper layers of the thermal disk ($z=0.65\unit{kpc}$). Errors were not reported for these results; we make the conservative choice to use errors of $50\%$ of the measured value.

\begin{table}[h!]
    \centering
    \begin{tabular}{|c|c|c|c|}
        \hline\hline
        \multicolumn{4}{c}{Ionized Gas Measurements} \\
            \hline\hline
            $R \; \rm (kpc)$ & $z \; \rm (kpc)$ & $\bar{n}_{\rm ion} \; \rm (cm^{-3})$ & Ref.  \\ [0.5ex] 
            \hline
            9 & 0 & $0.063$ & {\cite{1994ApJ...431..156W}}\\
            9 & 0.65 & $8\times 10^{-3}$ & {\cite{2004A&A...414...53F}} \\
            11 & 0.65 & $7\times 10^{-3}$ & {\cite{2004A&A...414...53F}} \\
            13 & 0.65 & $4\times 10^{-3}$ & {\cite{2004A&A...414...53F}} \\
            \hline
    \end{tabular}
    \caption{Values of the ionized gas density derived from observations of H$\alpha$ emission \cite{1994ApJ...431..156W} and Faraday rotation \cite{2004A&A...414...53F}. These derived values are used to fit our model of the ionized gas density given in Eq.~\eqref{eq:ionized gas}.}
    \label{tab:ionized_gas_data}
\end{table}

We summarize the ionized gas observations from Faraday rotation \cite{2004A&A...414...53F} and our derived value at the midplane from the measurement of emission measure \cite{1994ApJ...431..156W} in Table~\ref{tab:ionized_gas_data}. Using these observations and our derived value, we perform a $\chi^2$ fit to Eq.~\eqref{eq:ionized gas}; we show our results in Table~\ref{tab:gas_parameters}.

\begin{table}[h!]
\centering 
 \begin{tabular}{|c||c|} 
 \hline\hline
 \multicolumn{2}{c}{Gas Density} \\
 \hline\hline
 Parameter & Value \\ [0.5ex] 
 \hline
 \multicolumn{2}{c}{Ionized} \\
 \hline
 $\bar{n}_{\mathrm{ion},0} \unit{(cm^{-3})}$ & $0.14 \pm 0.15 $ \\
 $R_{\rm ion} \unit{(kpc)}$ & $9.4 \pm 5.7$ \\
 $z_{\rm ion} \unit{(kpc)}$ & $0.39 \pm 0.09$ \\
 \hline
 \multicolumn{2}{c}{$\rm HI$} \\
 \hline
 $h_{\mathrm{HI},0} \unit{(kpc)}$ & $0.21 \pm 0.04$ \\
 $S$ & $0.016 \pm 0.003$\\
 \hline
 \multicolumn{2}{c}{$\rm H_2$} \\
 \hline
 $h_{\mathrm{H_2}, 0} \unit{(kpc)}$ & $0.15 \pm 0.075$ \\
 $R_{\rm H_2} \unit{(kpc)}$ & $13.2 \pm 6.6$ \\
 \hline
 \end{tabular}
 \caption{Parameter values for our models of interstellar gas in M31. The top panel of the table has the best-fit parameter values for the ionized gas density in M31. The next two panels list parameters for the HI and $\rm H_2$ distributions along the $z$ coordinate given by Eqs.~\eqref{eq:HI scale} and \eqref{eq:H2 scale height M31}, respectively.}\label{tab:gas_parameters}
\end{table}

\subsubsection{HI Gas}

\begin{figure*}
\centering
\subfigure[]{\label{fig:HI_gas}\includegraphics[width=0.9\columnwidth]{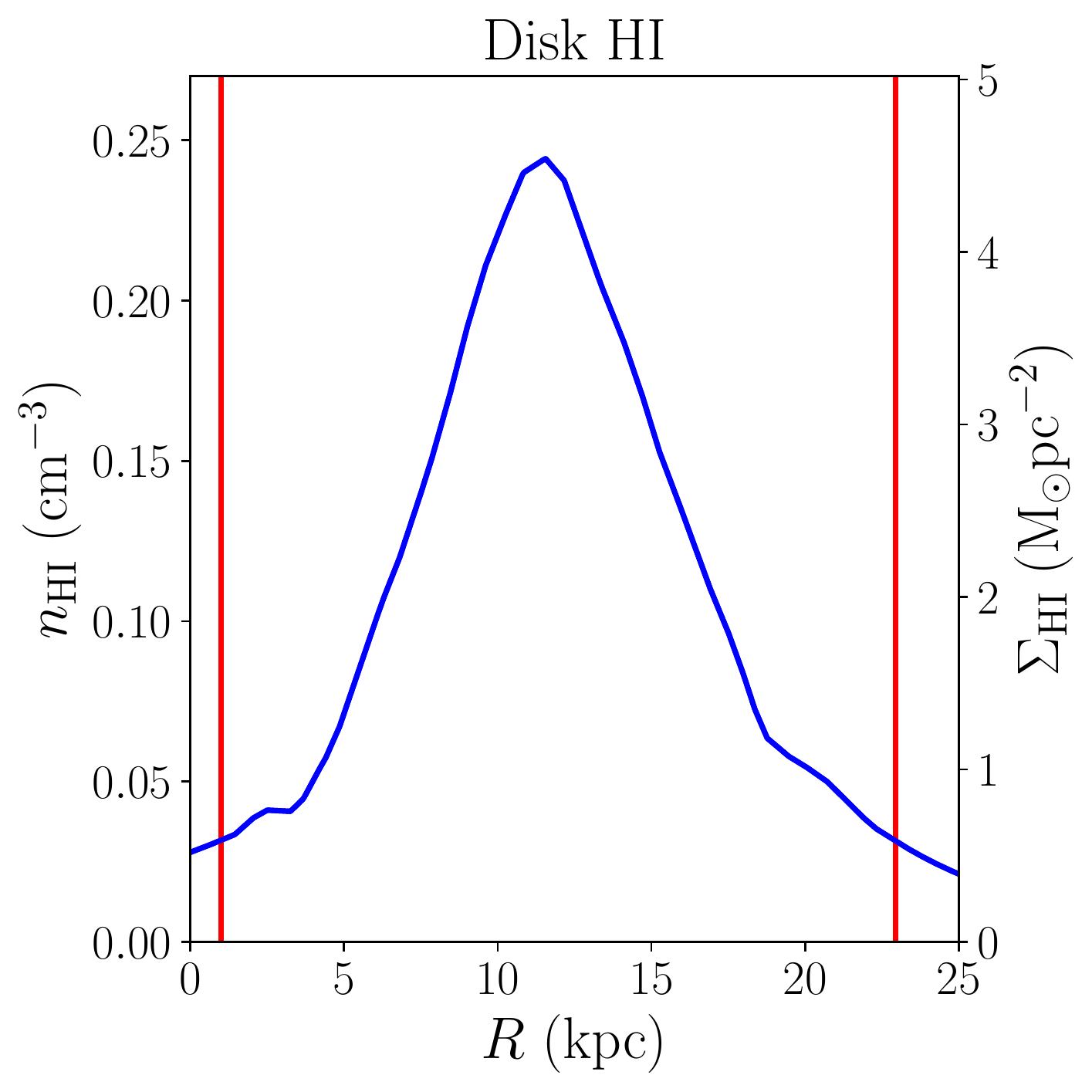}}
\subfigure[]{\label{fig:H2_gas}\includegraphics[width=0.9\columnwidth]{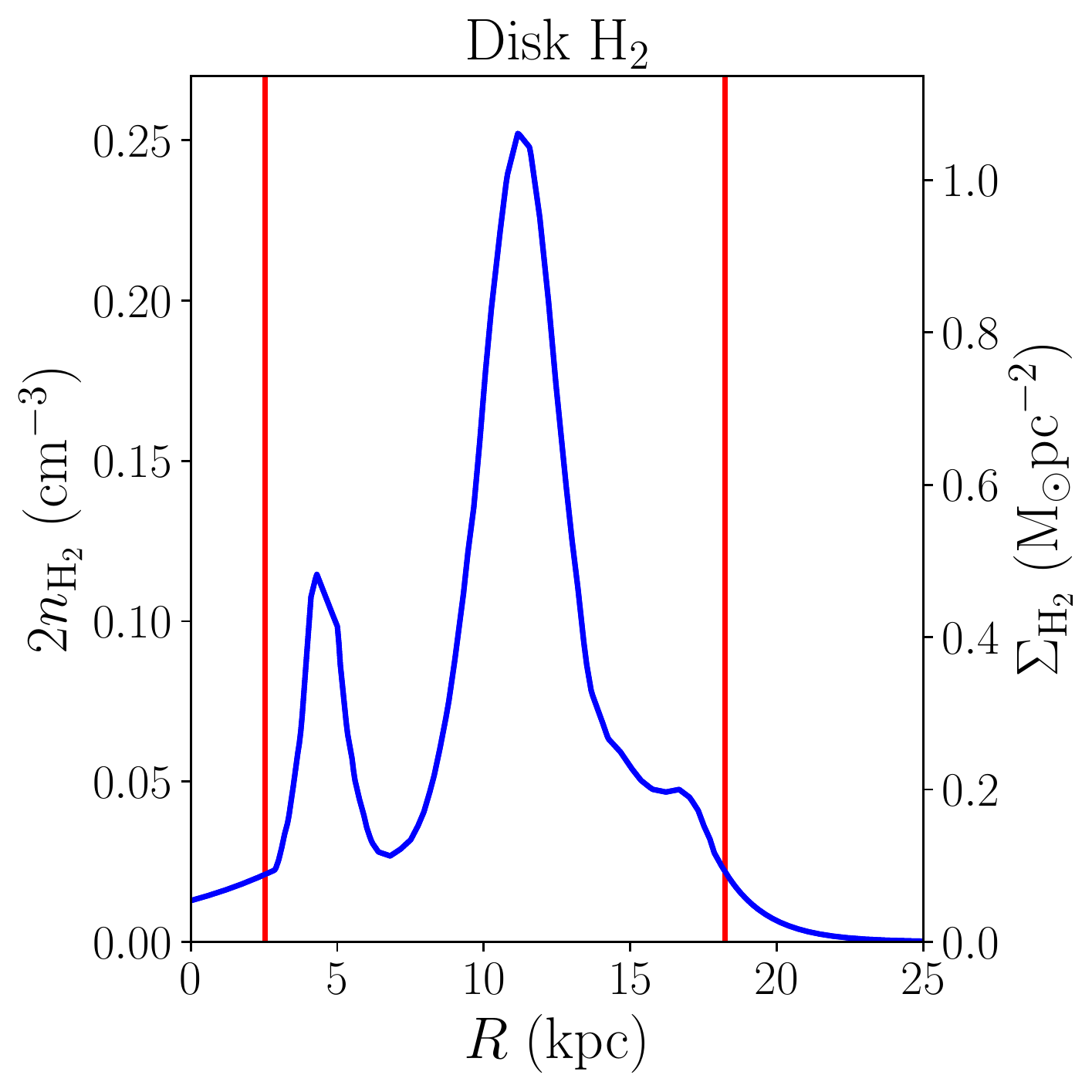}}
\caption{Number density (left axis) and surface density (right axis) of (a) HI and (b) $\rm H_2$ gas in the plane of the disk. The digitized and interpolated distributions from Ref.~\cite{2009A&A...505..497Y} are within the two vertical red lines. Outside these regions, we fit exponential extrapolations, matching the function values and first derivatives at the boundaries.
}

\label{fig:H_gas}
\end{figure*}

For the HI gas density, we digitize the radial column density distribution provided by Ref.~\cite{2009A&A...505..497Y} in the range $R\in [1, 23] \,  \unit{kpc}$. We then interpolate and extrapolate the digitized column density to get a function valid for all $R\geq 0$. For the extrapolation, we use a growing exponential at small $R$ and a decaying exponential at large $R$, such that the value and slope of the function are continuous at each boundary. Our interpolation and extrapolation of the fit to the column density of HI within the disk of M31 \cite{2009A&A...505..497Y} is shown in Figure~\ref{fig:HI_gas} (see the vertical axis on the right). 

We assume that the gas density on the disk is proportional to the column density. For the $z$-dependence of the HI density we assume a decaying exponential with a scale-height of $h_{\rm HI}(R)$ taken from Ref.~\cite{1991ApJ...372...54B}, as previously discussed in Section~\ref{sec:magnetic field}:
\begin{equation}\label{eq:HI scale}
   h_{\rm HI}(R) = h_{\mathrm{HI},0} + S\times R,
\end{equation}
where $h_{\mathrm{HI},0}$ and $S$ are reported in Table~\ref{tab:gas_parameters}. We normalize the HI density to the total HI mass reported in Table~1 of \cite{2009A&A...505..497Y}. The resulting gas density on the disk is shown in Figure~\ref{fig:HI_gas} (see the vertical axis on the left).

\subsubsection{H$_2$ Gas}

For the $\rm H_2$ density, we digitize the radial column density distribution provided by Ref.~\cite{2009A&A...505..497Y} in the range $R\in [2.5, 18 ]\, \unit{kpc}$. We then repeat our interpolation/extrapolation procedure to obtain a model of the column density of $\rm H_2$ on the disk. The results are shown in Figure~\ref{fig:H2_gas}. We again assume that the density of $\rm H_2$ on the disk is proportional to the column density. 

\begin{figure}
    \centering
    \includegraphics[width=0.9\columnwidth]{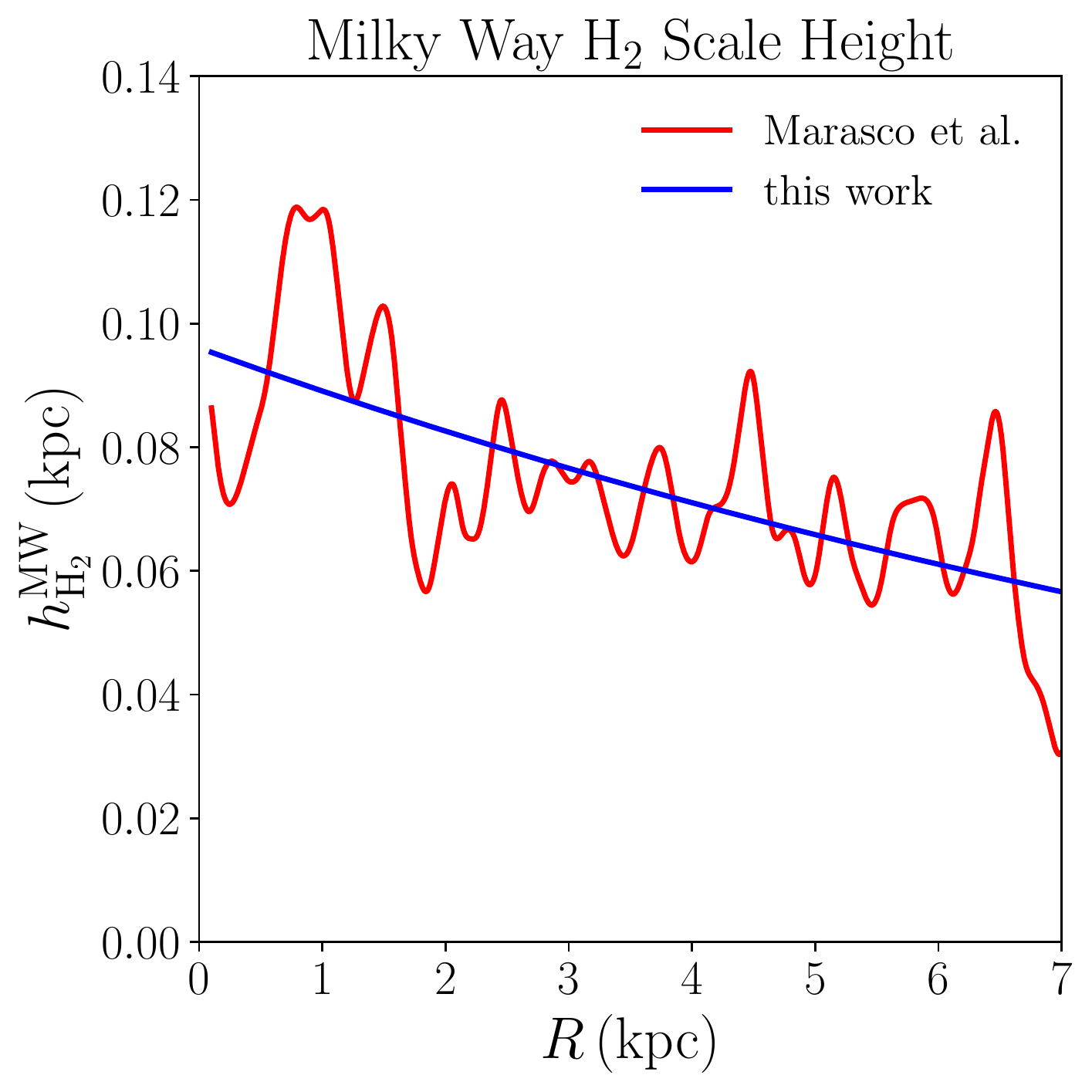}
    \caption{Fit to the Milky Way H$_2$ scale height \cite{2017A&A...607A.106M} (red curve). The parameterized fit of Eq.~\eqref{eq:H2 scale height MW} is shown in blue.}
    \label{fig:H2_scaleheight_MW}
\end{figure}

We assume a decaying exponential for the $z$-dependence, however there is little observational data available for the scale height of $\rm H_2$ in M31. We therefore use the $\rm H_2$ scale height in the Milky Way (derived from the data in Ref.~\cite{2017A&A...607A.106M}), and re-scale to M31 using a comparison of the HI scale heights in the Milky Way and M31. We digitize the fit to the $\rm H_2$ scale height data in Figure~10 of Ref.~\cite{2017A&A...607A.106M}, 
and smooth-out the fluctuations by fitting the digitized version of the fit to the functional form 
\begin{equation}\label{eq:H2 scale height MW}
h_{\rm H_2}^{\rm MW} = h_{\mathrm{H_2}, 0}^{\rm MW} e^{-R/R_{\rm H_2}}.
\end{equation}
We compare our model for the scale height in the Milky Way to that of Ref.~\cite{2017A&A...607A.106M} in  Figure~\ref{fig:H2_scaleheight_MW}.

To obtain a scale height for H$_2$ gas in M31 from the scale height of H$_2$ in the Milky Way, we take the ratio of the average HI scale height for M31 (Eq.~\eqref{eq:HI scale}) to that of the Milky Way (see Figure~6 of Ref.~\cite{2017A&A...607A.106M}) in the region $R \in [0, 7]\,\unit{kpc}$. We find this ratio is $1.55$, and assume this ratio holds for the H$_2$ gas:
\begin{equation}\label{eq:H2 scale height M31}
h_{\rm H_2} = 1.55 \times h_{\rm H_2}^{\rm MW} = h_{\mathrm{H_2}, 0} e^{-R/R_{\rm H_2}}.
\end{equation}
The resulting values of $h_{\mathrm{H_2}, 0}$ and $R_{\rm H_2}$ are given in Table~\ref{tab:gas_parameters}. The errors in these two parameters are dominated by the systematic errors of converting from their values in the Milky Way so we conservatively set their errors to $50\%$ of their values.

Again, we normalize our distribution for the ${\rm H}_2$ density based on the measured value for the total $\rm H_2$ gas mass reported in Table 1 of Ref.~\cite{2009A&A...505..497Y}. The resulting ${\rm H}_2$ density on the disk is shown in Figure~\ref{fig:H2_gas}.

\subsubsection{$^4$He Gas}

The last significant component of neutral gas that we have to model is $\prescript{4}{}{\mathrm{He}}$. The mass fraction of $\prescript{4}{}{\mathrm{He}}$ in simulated spiral galaxies was found to vary spatially over the range $Y_{\rm He} \simeq 0.25 - 0.3$ \cite{2019A&A...630A.125V}. The lower limit is the primordial value set by Big Bang Nucleosynthesis, and higher values are due to stellar production.
For simplicity, we approximate that $\prescript{4}{}{\mathrm{He}}$ production from stars is negligible, so the ratio of $\prescript{4}{}{\mathrm{He}}$ to H is set by Big Bang Nucleosynthesis:
\begin{equation}
    N_{\rm He} = \frac{N_H Y_{\rm He}}{4(1-Y_{\rm He})} \simeq N_H/12.
\end{equation}
The errors in our model for the Helium density from using this approximation are $<25\%$. As the effects of bremsstrahlung from $\prescript{4}{}{\mathrm{He}}$ are  subdominant compared to the other components in our propagation model, the errors in our final results from this approximation will be negligible as well.


Further, we assume that the $\prescript{4}{}{\mathrm{He}}$ density has the same morphology as the total hydrogen density. The local density for $^4$He is then derived from the HI and $\rm H_2$ gas:
\begin{equation}\label{eq:He density}
    n_{\rm He} = \left(\frac{1}{12}\right)\times n_H = \left(\frac{1}{12}\right)\times (n_{\rm HI} + 2n_{\rm H_2}).
\end{equation}

\section{Propagation of $e^\pm$ in M31} \label{sec:propagation}

The production of electrons and positrons by dark matter annihilation provides a source $Q_e$ for the phase space-density $f_e$ within a galaxy. From their initial locations, the $e^\pm$ will diffuse in turbulent magnetic fields and undergo energy loss from synchrotron radiation as well as inverse Compton, bremsstrahlung and Coulomb scattering. The synchrotron losses lead to the radio signal we will use to constrain dark matter annihilation, but all forms of energy loss must be tracked to determine the evolution of $f_e$ in energy and position.

The evolution of the phase space density $f_e$ is controlled by the diffusion-loss equation:
\begin{equation}\label{eq:diff loss}
\frac{\partial f_e}{\partial t} = \partial_i[\mathcal{D}_{ij}(\boldsymbol{x}, E)\partial_j f_e] + \frac{\partial}{\partial E}[b(\boldsymbol{x}, E) f_e] + Q_e(\boldsymbol{x}, E).
\end{equation}
where $f_e(\boldsymbol{x}, E) = dn_e/dE$ is the phase space density of electrons at position $\boldsymbol{x}$ and energy $E$, $\mathcal{D}_{ij}(\boldsymbol{x}, E)$ is the diffusion matrix, and $b(\boldsymbol{x}, E)$ is the energy loss parameter. The position-dependent diffusion matrix depends on both the RMS magnetic field, $\bar{B}$ and the turbulent fluctuations of the field at small scales. The loss parameter depends on $\bar{B}^2$, the ISRF, and the densities of the various gas components, which have been modeled in Section~\ref{sec:M31model}.

\subsection{Diffusion Matrix} \label{sec:diff coef}

Fluctuations in the magnetic field cause the relativistic $e^\pm$ to exhibit diffusive motion. In a uniform magnetic field with strength $B$, the motion of $e^\pm$ is helical, with a Larmor radius of
\begin{equation} \label{eq:larmor}
r_L = \frac{E \sin \alpha}{eB} = (1.1 \times 10^{-7} \unit{pc}) \left(\frac{E}{1 \unit{GeV}}\right) \left(\frac{10 \unit{\mu G}}{B}\right).
\end{equation}
Here, $E$ is the particle energy and $\alpha$ is the pitch angle, which is defined as the angle of the velocity with respect to the direction of the magnetic field. For a changing magnetic field, the Larmor radius can still be defined, provided the field variations are small over the distance traversed by the particle in the time it takes for its phase to change by $\mathcal{O}(1)$. For the ${\cal O}(10\,\unit{\mu G})$ magnetic fields in the disk of M31 (see Section~\ref{sec:magnetic field}), electrons and positrons of the energies expected from the annihilation of dark matter have Larmor radii of $\lesssim \mathcal{O}(10^{-7}\,\unit{pc})$. As the fluctuations of the field follow a power law distribution on length-scales smaller than $1/k_0\sim \mathcal{O}(1 \unit{kpc})$, the magnetic field is dominated by Fourier modes that are much larger than the Larmor radius. Modes that are of order the Larmor radius and smaller can be treated perturbatively.

The motion of an electron or positron under the influence of the large-scale magnetic field fluctuations is well-described by the adiabatic approximation: 
the particle exhibits helical motion about the local field with an axis that gradually changes as the particle moves along the slowly changing magnetic field \cite{2011hea..book.....L}. The magnetic field fluctuations with length-scales below the Larmor radius perturb this adiabatic motion, causing pitch angle scattering which leads to diffusion along the axis of the local magnetic field \cite{2017ApJ...837..140S}. Since the direction of the large-scale magnetic field slowly changes over space and time, the diffusion is shared evenly in all directions leading to isotropic diffusion \cite{2017ApJ...837..140S}. 

With these approximations and assuming magnetic field fluctuations characterized by the Kolmogrov spectrum, introduced in Section~\ref{sec:magnetic field}, the diffusion matrix is \cite{1998APh.....9..227C, 2015MNRAS.448.3747R}
\begin{widetext}
\begin{equation}\label{eq:diff coef}
\mathcal{D}_{ij} \simeq \left(1.5 \times 10^{28} \unit{cm^2/s}\right) \delta_{ij} \left(\frac{d_0}{1 \unit{kpc}}\right)^{2/3}\left(\frac{10 \unit{\mu G}}{\bar{B}}\right)^{1/3}\left(\frac{E}{1\unit{GeV}}\right)^{1/3} \equiv D_0 \delta_{ij} \left(\frac{10 \unit{\mu G}}{\bar{B}}\right)^{1/3}\left(\frac{E}{1\unit{GeV}}\right)^{1/3},
\end{equation}
\end{widetext}
where $d_0 = 1/k_0$ is largest length-scale over which the Kolmogrov spectrum of magnetic field fluctuations is valid. Since the diffusion matrix is isotropic, it can be written as 
\begin{equation}
  \mathcal{D}_{ij} = \delta_{ij}D.
\end{equation}
It is conventional to refer to $D$ the diffusion coefficient. We absorb the uncertainties in the prefactor of Eq.~\eqref{eq:diff coef} and $d_0$ into the constant $D_0$. A range of possible values for $D_0$ (in both the Milky Way and M31) have been suggested in the literature. We review these briefly here.

Ref.~\cite{2013MNRAS.435.1598B} infers the diffusion coefficient for $e^\pm$ near star-forming regions in M31 from measurements of non-thermal radio emission at $\nu = 1.4 \unit{GHz} $ and one higher frequency using two methods. The first method infers the diffusion coefficient from the difference in morphology between the two frequencies. The second method uses the difference between non-thermal emission and thermal emission at each of the frequencies, assuming that the thermal emission has a similar morphology to the source distribution of cosmic ray electrons.
These methods allow Ref.~\cite{2013MNRAS.435.1598B} to extract the diffusion coefficient at two electron and positron energies:  $4.1$ and $7.5 \unit{GeV}$. 

 Rescaling Ref.~\cite{2013MNRAS.435.1598B}'s results to the magnetic field parameters of M31, we obtain $D_0 \simeq 1.1 \times 10^{28} \unit{cm^2/s}$ using the first method and $D_0 \simeq 3.5\times 10^{27} \unit{cm^2/s}$ using the second. Neither method fully models the propagation of cosmic rays, and the variation between the two results makes it difficult to identify either value as our default $D_0$ value. 

For further guidance about the value of $D_0$ in M31, we review studies of propagation in the Milky Way. The \texttt{galprop} cosmic ray propagation model \cite{1998ApJ...509..212S} uses observations of cosmic rays in the Milky Way \cite{2001ApJ...563..768Y, 2009ApJ...698.1666G} to determine its best-fit diffusion coefficients \cite{2011ApJ...729..106T}. 
The assumed propagation model includes a uniform diffusion coefficient of the form $D = \tilde{D_0} \left(E/4 \unit{GeV}\right)^\delta$ for which the best-fit parameters were found to be 
 $\tilde{D}_0 = (8.3 \pm 1.5) \times 10^{28} \unit{cm^2/s}$ and $\delta = 0.31 \pm 0.02$ \cite{2011ApJ...729..106T}. 
 Comparing the model of Ref.~\cite{2011ApJ...729..106T} to our form for the diffusion coefficient, Eq.~\eqref{eq:diff coef}, and assuming that the cosmic rays studied were subject to a constant $10\unit{\mu G}$ magnetic field,
 these measurements imply $D_0=(5.2 \pm 0.9) \times 10^{28} \unit{cm^2/s}$. 
 Ref.~\cite{2021PhRvD.104h3005G} constructed 
 MIN, MED and MAX propagation models for $1 \unit{GeV}$ cosmic ray energies in the Milky Way, which can be interpreted as a range of $D_0 = \left[5.9\times 10^{27} - 2.0\times 10^{28}\right] \unit{cm^2/s}$, assuming the relevant magnetic field is a constant $10\unit{\mu G}$.

Underestimating the diffusion coefficient would under-predict how far particles will move before emitting most of their energy, leading to an over-prediction of the signal in regions of high dark matter density. A large value of $D_0$ will likewise result in a smaller signal flux for a given cross section. Therefore, to set conservative limits on the cross section of dark matter annihilation, we must avoid assuming too small a value for $D_0$. To set our conservative upper limit on $D_0$, we select a maximum value of $d_0\equiv 1/k_0$ in Eq.~\eqref{eq:diff coef} by setting $k_0$ equal to the wavenumber of a fluctuation with wavelength of $R_{B,1} = 77.6 \unit{kpc}$ (the longest scale-length in our magnetic field model). This leads to $d_0 \lesssim R_{B,1}/(2\pi) \simeq 12.5 \unit{kpc}$, implying $D_0\lesssim 8 \times 10^{28} \unit{cm^2/s}$. 

Given the variation in diffusion coefficients in the Milky Way and M31, as well as our conservative upper bound, we consider $D_0$ in the range
\begin{equation} \label{eq:diff coef range}
    3\times 10^{27}\unit{cm^2/s}\leq D_0\leq 8\times 10^{28}\unit{cm^2/s},
\end{equation}
We select as a default value $D_0 = 1 \times 10^{28} \unit{cm^2/s}$.

\begin{figure}
    \centering
    \includegraphics[width=0.9\columnwidth]{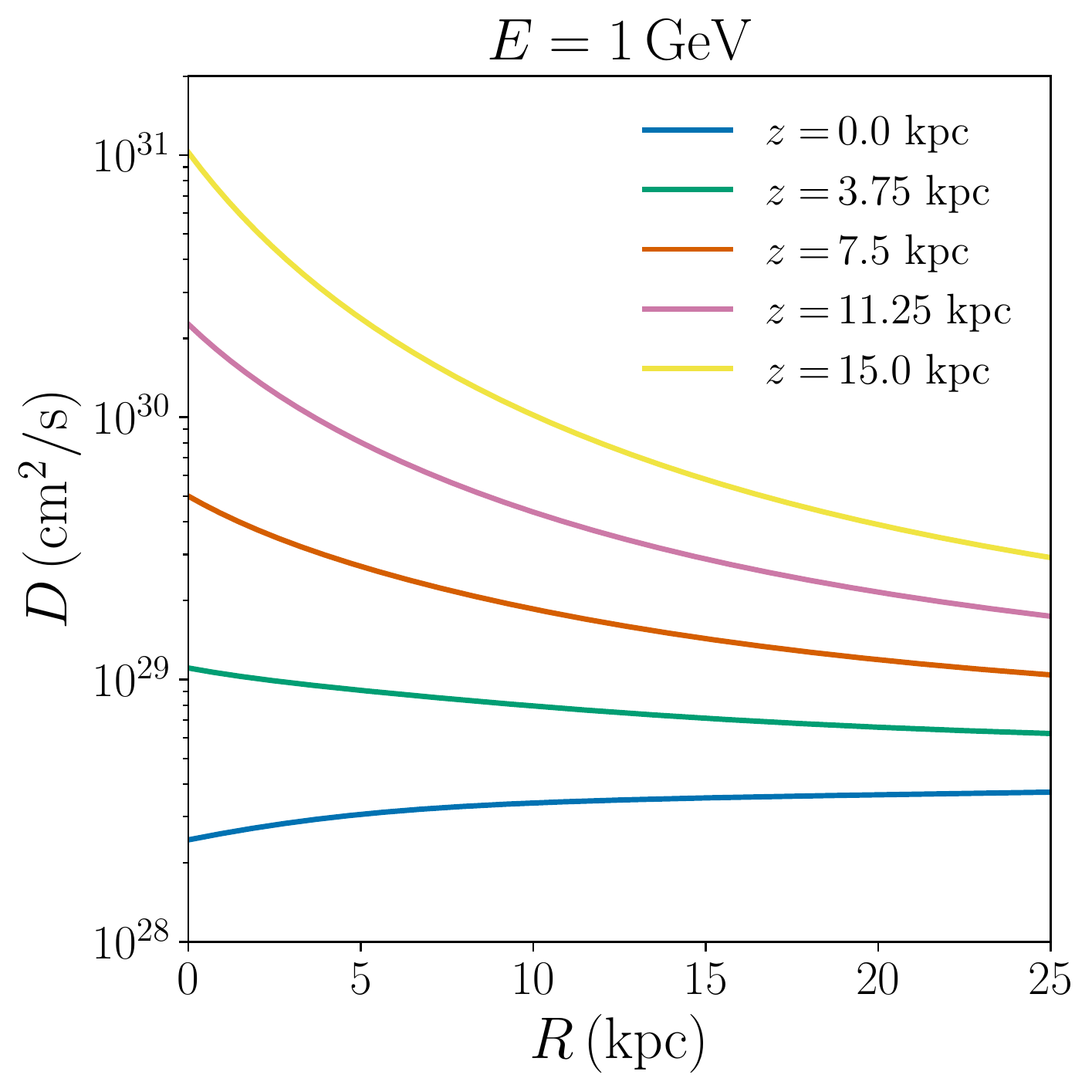}
    \caption{Diffusion coefficient as a function of $R$ for $E=1 \unit{GeV}$ and various values of $z$, with $D_0 = 10^{28}$~cm$^2$/s.
    }
    \label{fig:diffusion_coef}
\end{figure}

Though $D_0$ is position-independent, the diffusion coefficient $D$ explicitly depends on the magnetic field, and our model for the magnetic field is position dependent (see Section~\ref{sec:data}). As a result, the diffusion coefficient also depends on location within M31. We show the dependence on location within M31 in Figure~\ref{fig:diffusion_coef} for our default diffusion coefficient normalization ($D_0=1\times 10^{28} \unit{cm^2/s}$) and $E=1 \unit{GeV}$. The diffusion coefficient varies more rapidly with $z$ when $R$ is small, as a result of the magnetic field scale height increasing with $R$. 

Prior studies of radio emission from dark matter annihilation in M31 make the approximation that the diffusion coefficient is zero \cite{2013PhRvD..88b3504E, 2016PhRvD..94b3507C, 2021MNRAS.501.5692C} or position independent \cite{ 2018PhRvD..97j3021M, 2022PhRvD.106b3023E}. This latter assumption is sufficient when the region of interest is small and the diffusion coefficient is nearly constant over the region. 
Over the length scales of interest, the variations in the diffusion coefficient must be taken into account, and we develop a numerical method for calculating the evolution of phase space density of the charged particles which can accommodate a position-dependent diffusion coefficient. We describe our numerical solution in Section~\ref{sec:solving diff loss}, after introducing the energy-loss terms that enter into Eq.~\eqref{eq:diff loss} in the next subsection. To our knowledge this is the first study that uses a position-dependent diffusion coefficient to set limits on dark matter annihilation in M31 via radio emission. 

\subsection{Energy Loss due to Radiative Processes} \label{sec:energy loss coef}

As the electrons and positrons diffuse through the ISM they lose energy through radiative processes. This energy loss is encoded in Eq.~\eqref{eq:diff loss} by the loss parameter $b$. In M31, the relevant losses are inverse Compton (IC), synchrotron, bremsstrahlung, and Coloumb interactions:
\begin{equation}\label{eq:loss param}
\begin{split}
-\frac{dE}{dt} \equiv & \, b(\boldsymbol{x}, E) = b_{\rm IC}(\boldsymbol{x}, E) + b_{\rm sync}(\boldsymbol{x}, E) + \\
 & b_{\rm brem}(\boldsymbol{x}, E) + b_{\rm C}(\boldsymbol{x}, E).
\end{split}
\end{equation}
We treat each of these terms in turn.

The inverse Compton scattering between $e^\pm$ and the ambient starlight, rescattered light from dust, and CMB emission will transfer energy from the charged particles into the photons, at a rate \cite{2011hea..book.....L}
\begin{equation}\label{eq:IC loss}
b_{\rm IC} = b_{\rm IC}^{(0)} \left(\frac{\rho_{\gamma}(\boldsymbol{x})}{10 \unit{eV/cm^3}}\right) \left(\frac{E}{1 \unit{GeV}}\right)^2,
\end{equation}
where $b_{\rm IC}^{(0)} = 1.0 \times 10^{-15} \unit{GeV/s}$ and $\rho_\gamma$ is the total radiation energy density, derived in Section~\ref{sec:isrf}. 

Synchrotron emission occurs due to the acceleration of charged particles in galactic magnetic fields. As described in Section~\ref{sec:magnetic field}, the magnetic fields of M31 do not change appreciably over the Larmor radius of the relevant $e^\pm$. Additionally, due to pitch angle scattering, the pitch angles are approximately uniformly occupied. Therefore, the energy loss due to synchrotron emission can be determined by assuming a locally constant magnetic field and averaging the energy loss over all pitch angles. The expression for the loss due to synchrotron is given by \cite{2011hea..book.....L}
\begin{equation}\label{eq:loss sync}
b_{\rm sync}= b_{\rm sync}^{(0)} \left(\frac{\bar{B}(R, z)}{10\unit{\mu G}}\right)^2\left(\frac{E}{1 \unit{GeV}}\right)^2
\end{equation}
where $b_{\rm sync}^{(0)} = 2.5 \times 10^{-16} \unit{GeV/s}$. 

The third term in Eq.~\eqref{eq:loss param} is the contribution to the loss from bremsstrahlung emission due to $e^\pm$ scattering with neutral hydrogen, neutral helium, and ionized gas:
\begin{equation}\label{eq:brem loss splitup}
b_{\rm brem} = b_{\rm H}(\boldsymbol{x},E) + b_{\rm He}(\boldsymbol{x},E) + b_{\rm ion}(\boldsymbol{x},E).
\end{equation} 
The expressions for these three components of the bremsstrahlung loss are given by \cite{1970RvMP...42..237B, 2000ApJ...537..763S}
\begin{eqnarray}
b_{\rm H} & = & b_{\rm H}^{(0)} \left(\frac{n_{\rm H}(\boldsymbol{x})}{1 \unit{cm^{-3}}}\right) \left(\frac{E}{1 \unit{GeV}}\right), \nonumber \\
b_{\rm He} & = & b_{\rm He}^{(0)}  \left(\frac{n_{\rm He}(\boldsymbol{x})}{1 \unit{cm^{-3}}}\right) \left(\frac{E}{1 \unit{GeV}}\right), \label{eq:brem loss} \\
b_{\rm ion} & = & b_{\rm ion}^{(0)} \left(\frac{n_{\rm ion}(\boldsymbol{x})}{1 \unit{cm^{-3}}}\right) \left(\frac{E}{1 \unit{GeV}}\right)\left[1 + \frac{1}{7.94}\ln{\left(\frac{E}{1 \unit{GeV}}\right)}\right], \nonumber
\end{eqnarray}
where $b_{\rm H}^{(0)} = 1.22 \times 10^{-16} \unit{GeV/s}$,  $b_{\rm He}^{(0)} = 3.61 \times 10^{-16} \unit{GeV/s}$ and $b_{\rm ion}^{(0)} = 1.74 \times 10^{-16} \unit{GeV/s}$. Our models for the density of each gas component were presented in Section~\ref{sec:gas}.

Lastly, the fourth contribution to the loss parameter is from Coulomb interactions with ionized gas and is given by \cite{1972Phy....60..145G, 1988SoPh..115..313S}
\begin{equation}\label{eq:C loss}
b_{\rm C} = b_{\rm C}^{(0)} \left(\frac{n_{\rm ion}(\boldsymbol{x})}{1 \unit{cm^{-3}}}\right)\left[1 +\frac{1}{82}\ln{\left(\frac{E}{1 \unit{GeV}}\frac{1 \unit{cm^{-3}}}{n_{\rm ion}}\right)}\right],
\end{equation}
where $b_{\rm C}^{(0)} = 6.2 \times 10^{-16} \unit{GeV/s}$. Note that Coulomb losses are not radiative processes involving the loss of energy from the charged particles into photons, but rather are due to an energy transfer from the relativistic $e^\pm$ to non-relativistic ions in the interstellar plasma.

\begin{figure*}
    \centering
    \subfigure[]{\label{fig:loss_coef_ER}\includegraphics[width=0.9\columnwidth]{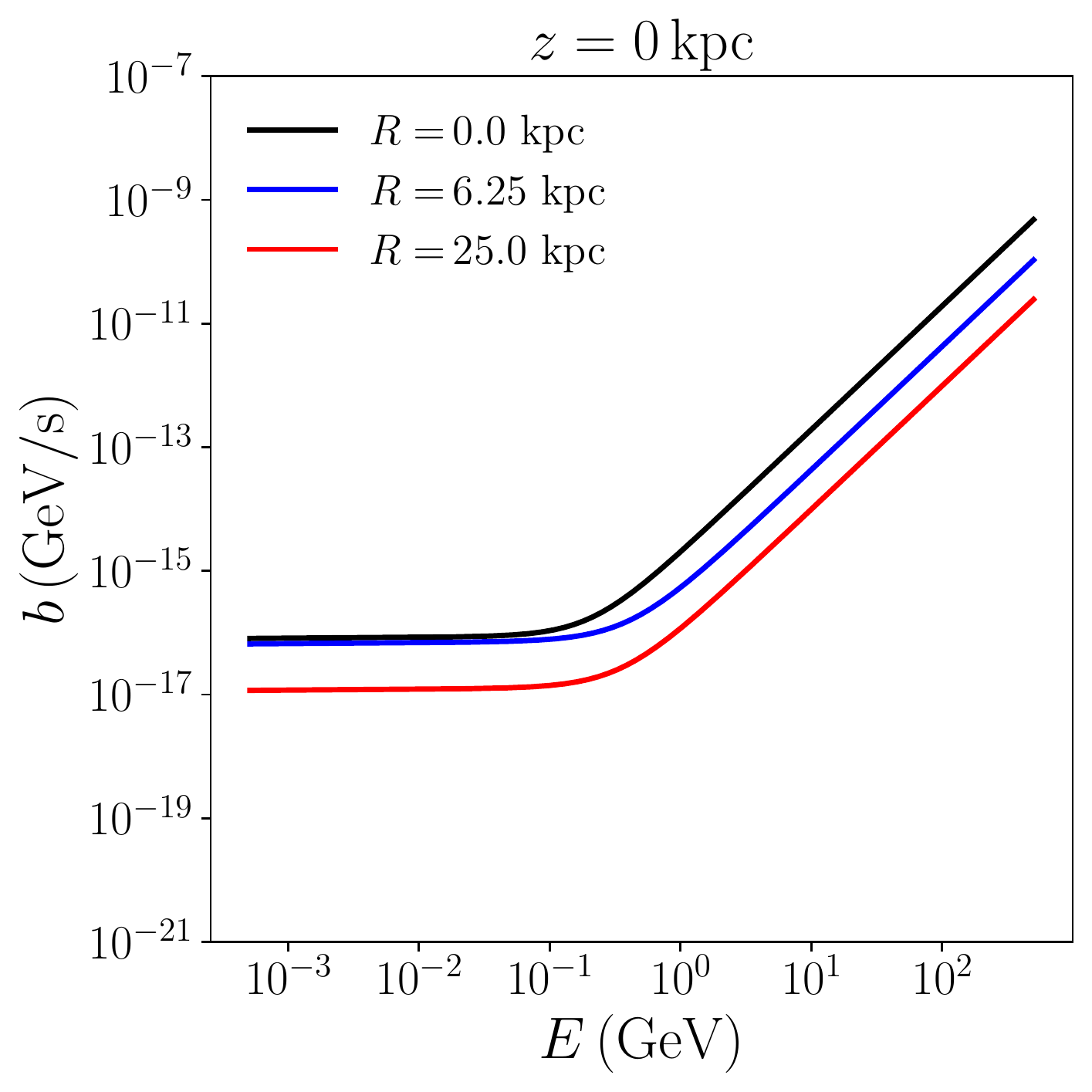}}
    \subfigure[]{\label{fig:loss_coef_Etype}\includegraphics[width=0.9\columnwidth]{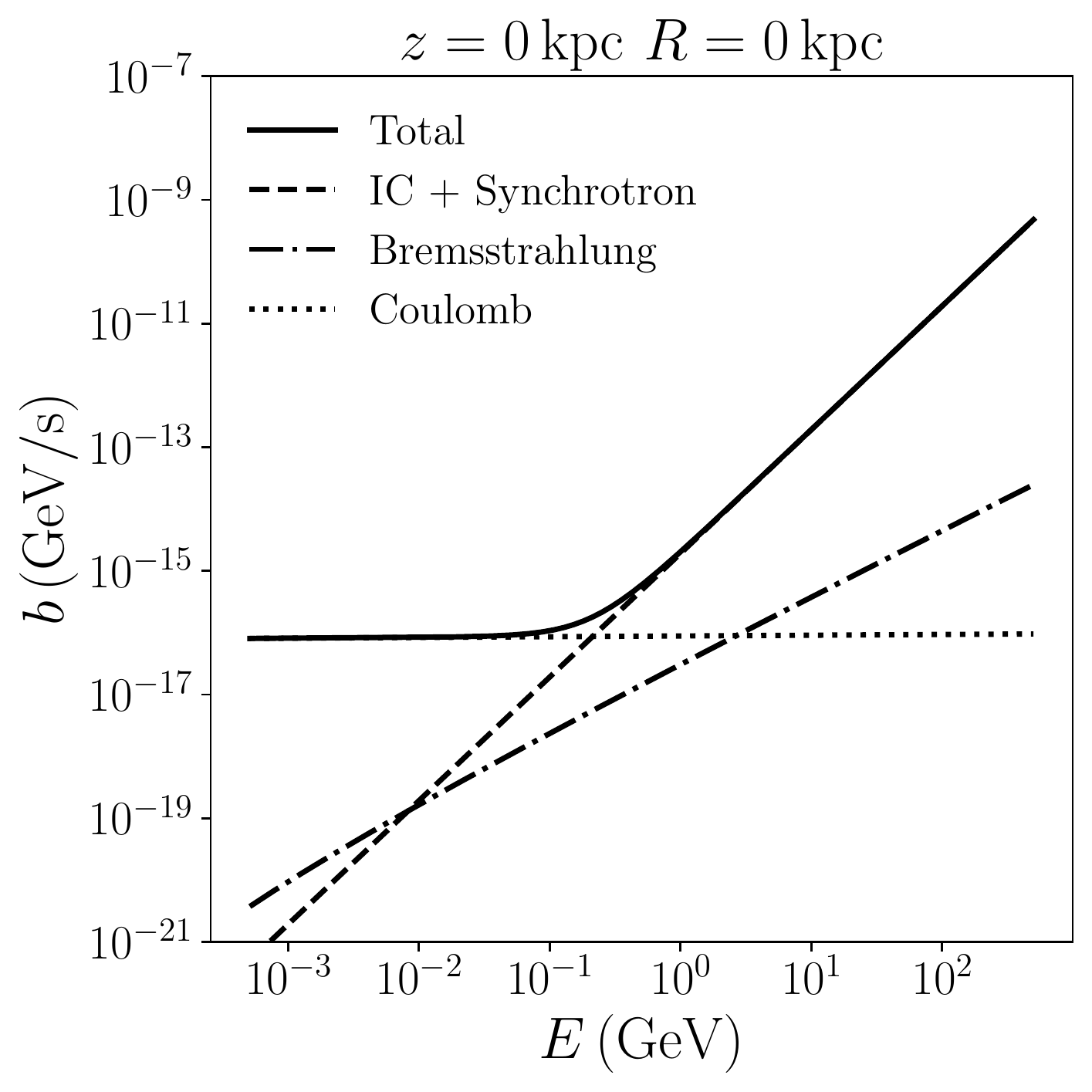}}
    \caption{(a) The energy dependence of the loss coefficient for various values of $R$ at $z=0$. 
    (b) The energy dependence of the total loss coefficient (solid line) and its subcomponents at $R=z=0$. Inverse Compton and synchrotron losses, which have the same energy dependence, are shown as the dashed line, bremsstrahlung as dot-dashed, and Coulomb losses as the dotted line.
    }
\label{fig:loss_coef}
\end{figure*}

We show the resulting loss coefficient as a function of energy in Figure~\ref{fig:loss_coef}. Figure~\ref{fig:loss_coef_ER} shows the total loss coefficient given by Eq.~\eqref{eq:loss param} for various values of $R$ on the disk. Figure~\ref{fig:loss_coef_Etype} shows the total loss coefficient at the origin and the contributions to it from the individual processes discussed in this subsection. Coulomb losses dominate at low energy, inverse Compton and synchrotron losses dominate at high energy, and bremsstrahlung only becomes marginally important at intermediate energies for $R \simeq 10\,\unit{kpc}$ due to the large concentration of interstellar gas in the ring-like structure (discussed in Section~\ref{sec:ring background}).

\subsection{Solving the Diffusion Loss Equation}\label{sec:solving diff loss}

We now turn to the numerical solution to the diffusion loss equation (Eq.~\eqref{eq:diff loss}) in M31, assuming the electron and positron injection from dark matter from Section~\ref{sec:DM_production} and the astrophysical model of M31 from Section~\ref{sec:M31model}.

To motivate our approach, it is useful to consider the two dynamic time scales which characterize the diffusion ($\tau_D$) and  energy loss ($\tau_b$), defined implicitly through rewriting Eq.~\eqref{eq:diff loss} as
\begin{equation}\label{eq:approx diff loss}
\frac{\partial f_e}{\partial t} = -\frac{f_e}{\tau_D} - \frac{f_e}{\tau_b} + Q_e(\boldsymbol{x}, E).
\end{equation}
These timescales depend on $R$, $z$, $E$ and derivatives of $f_e$ over $f_e$. As a result, $\tau_b$ and $\tau_D$ are independent of the overall magnitude of $f_e$. 

The timescale for diffusion is
\begin{equation}
   \tau_D^{-1} = -\left(\partial_i D\right)\frac{\partial_i f_e}{f_e} - D\frac{\nabla^2 f_e}{f_e}.
\end{equation}
In the approximation that $f_e$ depends on position only as a power-law in $r$, 
\begin{eqnarray}
    \tau_D & \sim & \frac{r^2}{D}\left[1 + z \frac{\partial_z D}{D} + R\frac{\partial_R D}{D} \right]^{-1} \equiv \frac{L(\boldsymbol{x})^2}{D}\nonumber \\
    & \sim & (2 \times 10^{16} \unit{s}) \left(\frac{L(\boldsymbol{x})}{5\unit{kpc}}\right)^2 \left(\frac{1 \times 10^{28} \unit{cm^2/s}}{D}\right). \label{D timescale}
\end{eqnarray}
where $L(\boldsymbol{x})$ is a length-scale that determines the rate that diffusion causes the phase space density to change. Assuming $f_e$ is a power law in $E$, the characteristic timescale for energy loss 
can be approximated by
\begin{equation}\label{b timescale}
\tau_b \simeq \frac{E}{b} = (1\times 10^{16} \unit{s}) \left(\frac{E}{1 \unit{GeV}}\right)\left(\frac{1 \times 10^{-16} \unit{GeV/s}}{b}\right).
\end{equation}

\begin{figure*}
\centering
\subfigure[]{\label{fig:tauDvsba}\includegraphics[width=0.9\columnwidth]{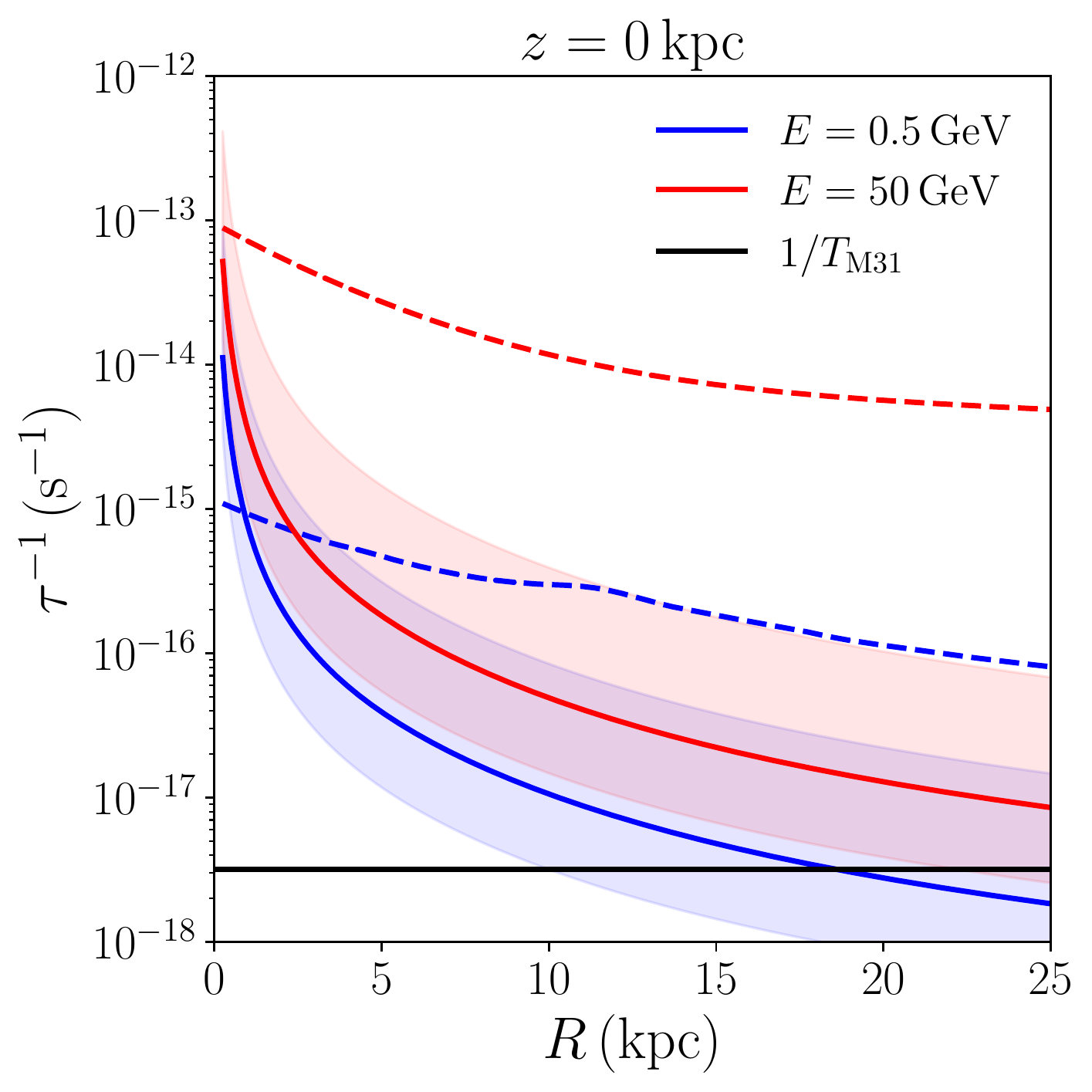}}
\subfigure[]{\label{fig:tauDvsbb}\includegraphics[width=0.9\columnwidth]{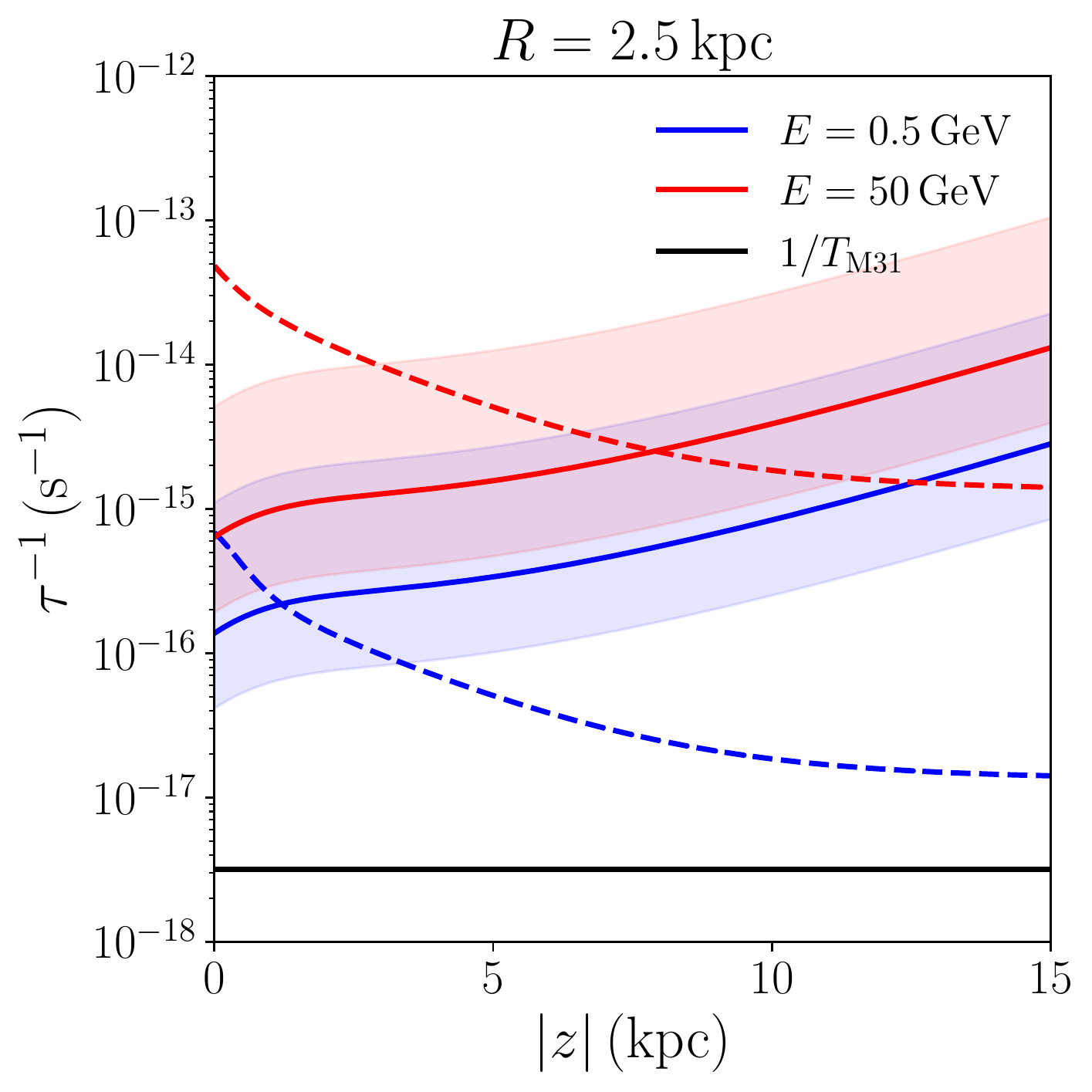}}
\caption{The inverse timescales for diffusion (solid lines) and loss (dashed lines), Eqs.~\eqref{D timescale} and \eqref{b timescale}. We show for comparison the inverse of the age of M31 (solid black), $T_{\rm M31} = 10^{10}\,\unit{years}$. The shaded regions around each solid line shows the variation of the inverse timescales for diffusion as $D_0$ is varied within the range given in Eq.~\eqref{eq:diff coef range}.}
\label{fig:tauDvsb}
\end{figure*}

The propagation is dominated by diffusion when  $\tau_D \ll \tau_b$ and dominated by energy loss when $\tau_D\gg \tau_b$. 
Both diffusion and loss will dominate at different values of $E$ and $\boldsymbol{x}$. In Figure~\ref{fig:tauDvsb} we plot the inverse timescales for diffusion and loss over a range of $R$ and $z$. Along the disk, loss tends to dominate (Figure~\ref{fig:tauDvsba}), whereas diffusion becomes the more important term off of the disk (Figure~\ref{fig:tauDvsbb}).

\begin{figure*}[th!]
\includegraphics[width=1.5\columnwidth]{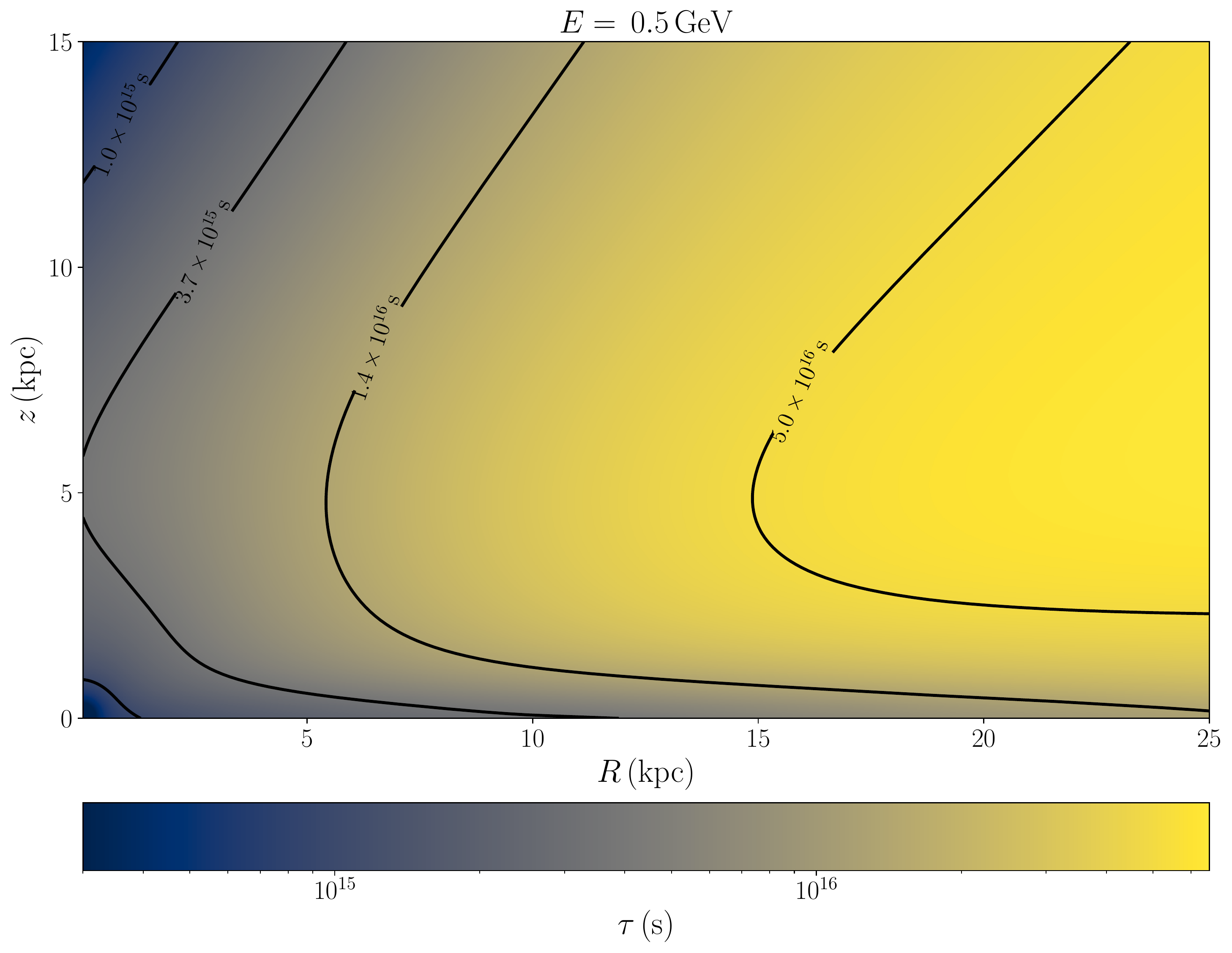}
\caption{Dynamic timescale $\tau$ of M31 as a function of $R$ and $z$ for $E = 0.5 \, \rm GeV$ (the minimum energy contributing significantly to the $8.35\,\unit{GHz}$ synchrotron signal) and $D_0 = 3 \times 10^{27} \rm cm^2/s$ (the lower bound on the diffusion coefficient). 
}
\label{fig:tau}
\end{figure*}

In regions of phase space where $\tau \equiv (\tau_b^{-1}+\tau_D^{-1})^{-1} \ll T_{\rm M31}$ (where $T_{\rm M31} \simeq 3\times 10^{17}\,\unit{s}$ is the approximate age of M31), the phase space density $f_e$ today will be well-approximated by the equilibrium density. In  Figure~\ref{fig:tau}, we show $\tau$ for $E=0.5 \, \unit{GeV}$, $R\in [0, 25] \unit{kpc}$ and $z\in [0, 15] \unit{kpc}$ assuming our lowest diffusion coefficient normalization, $D_0=3\times 10^{27} \rm cm^2/s$. $E = 0.5 \unit{GeV}$ is a lower bound on the range of energies that contribute significantly to $\nu =8.35 \unit{GHz}$ radio emission in M31. Due to the energy dependence of the diffusion and loss coefficients, $\tau$ decreases as $E$ increases for $E\gtrsim 0.1 \unit{GeV}$. As larger $D_0$ also makes $\tau$ smaller, the combination of $D_0$ and $E$ shown in Figure~\ref{fig:tau} provides an upper bound on $\tau$.

As can be seen in Figure~\ref{fig:tau}, within $R<25\,\unit{kpc}$ and $|z|<15\,\unit{kpc}$, we find $\tau < T_{\rm M31}$. Near the center of the galaxy and for higher $e^\pm$ energies, $\tau$ decreases. Though some regions at large $R$ have timescales comparable to the age of M31, these regions are far from the inner part of the galaxy where the Effelsberg radio data will be used to set limits. 
We are therefore justified in following the general approach of the literature \cite{2013PhRvD..88b3504E, 2016PhRvD..94b3507C, 2018PhRvD..97j3021M, 2021MNRAS.501.5692C} by approximating the phase space density $f_e$ of $e^\pm$ in M31 today as the equilibrium density.
 
If $b$ and $D$ do not depend on $\boldsymbol{x}$, a semi-analytic solution exists for the equilibrium density (see e.g., Ref.~\cite{2017JCAP...09..027M}). When the region of interest is small, homogeneous coefficients can be obtained by averaging the diffusion and loss coefficients over the relevant volume \cite{2017JCAP...09..027M, 2018PhRvD..97j3021M}. However, our goal in this paper is to compute the synchrotron distribution over the field of view of the radio data in Figure~\ref{fig:intensity_map}, that is, most of the galactic disk of M31. Based on the astrophysical models (described in Section~\ref{sec:M31model}), the diffusion and loss coefficients will vary significantly over this region. We must therefore solve Eq.~\eqref{eq:diff loss} in the case of non-homogeneous coefficients.

While the source term is spherically symmetric, the diffusion and loss coefficients are axially symmetric, implying that the solution to Eq.~\eqref{eq:diff loss} depends on $R$, $z$ and $E$. However, a fully axially symmetric numeric solution is intractable given our numeric approach. 
To overcome this problem, we average Eq.~\eqref{eq:diff loss} over solid angle $\Omega$:
\begin{equation}\label{eq:average diff loss}
\frac{\partial \langle f_e\rangle}{\partial t} = \frac{\partial}{\partial r}\langle D\left(\partial_r f_e \right)\rangle + \frac{2}{r}\langle D\left(\partial_r f_e\right)\rangle+\frac{\partial}{\partial E}\langle b f \rangle + Q_e,
\end{equation}
where (for an arbitrary function $g(E,\boldsymbol{x})$),
\begin{equation}
\langle g\rangle (E, r) \equiv \frac{1}{4\pi}\int d\Omega g(E, \boldsymbol{x}).
\end{equation} 
Spherically averaging Eq.~\eqref{eq:average diff loss}, we find
\begin{equation}\label{eq:sph sym diff loss}
\frac{\partial \langle f_e\rangle}{\partial t} = \left[\frac{\partial}{\partial r}+\frac{2}{r}\right]\left(\bar{D}\partial_r \langle f_e\rangle \right)+ \frac{\partial}{\partial E}\left(\bar{b} \langle f_e\rangle \right) + Q_e
\end{equation}
where
\begin{equation}\label{eq:sph ave coefs}
\bar{D} \equiv \frac{\langle D \partial_r f_e\rangle}{\langle \partial_r f_e \rangle}, \quad \bar{b} \equiv \frac{\langle b f_e\rangle}{\langle f_e \rangle}.
\end{equation}

\begin{figure*}
\centering
\subfigure[]{\label{fig:weighted_vs_unweightedD}\includegraphics[width=0.9\columnwidth]{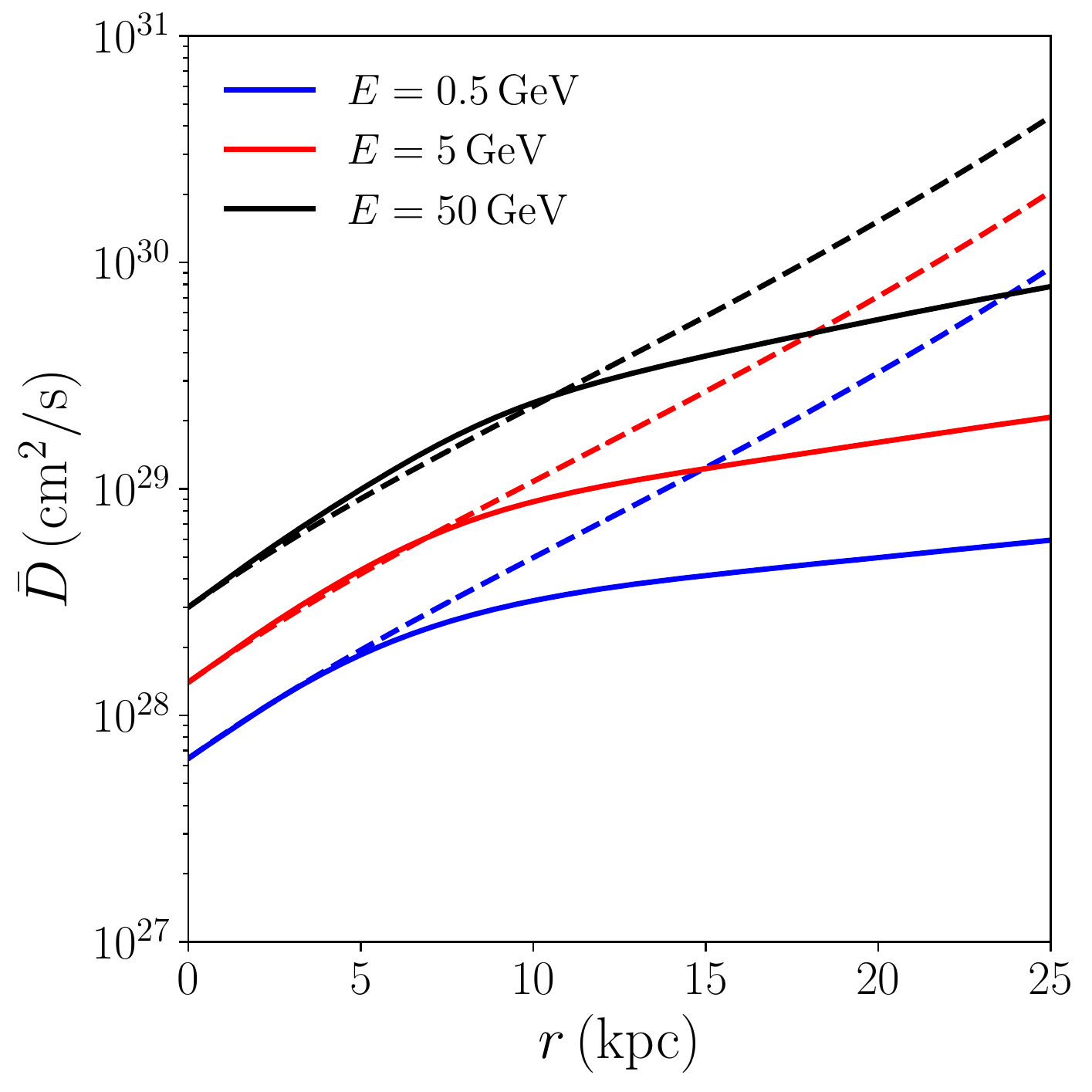}}
\subfigure[]{\label{fig:weighted_vs_unweightedb}\includegraphics[ width=0.9\columnwidth]{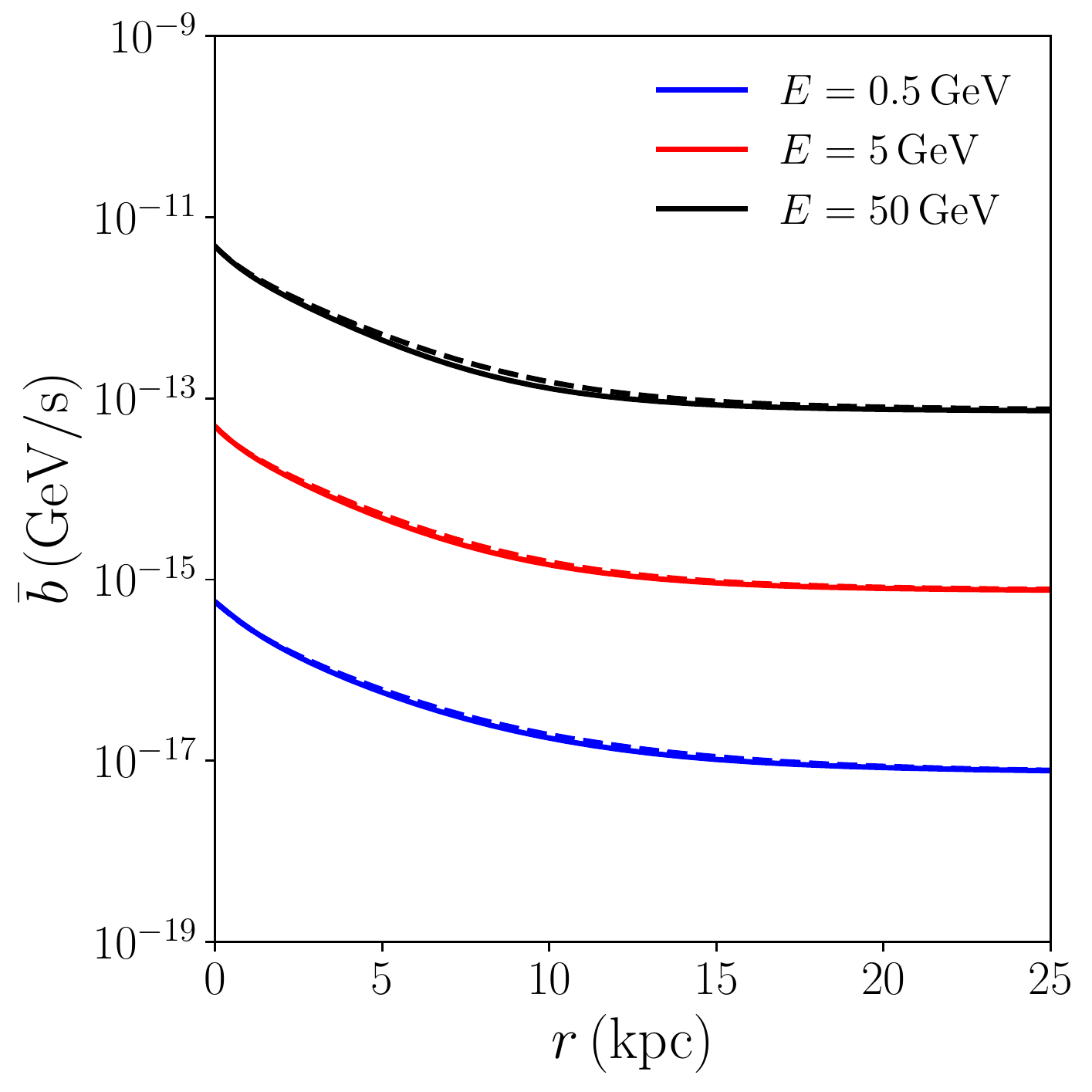}}
\caption{Spherically averaged diffusion coefficient (left) and loss coefficient (right) using the unweighted average (dashed) and the weighted average (solid) for a range of energies. We use our default value of $D_0=1\times 10^{28} \unit{cm^2/s}$.}
\label{fig:weighted_vs_unweighted}
\end{figure*}

The averaged coefficients $\bar{D}$ and $\bar{b}$ required to solve for $\langle f_e \rangle$  in Eq.~\eqref{eq:sph sym diff loss} themselves depend on $f_e$. To calculate $\bar{D}$ and $\bar{b}$, we use approximate solutions for $f_e$, then use these averaged coefficients to numerically solve the spherically averaged diffusion loss equation for $\langle f_e \rangle$. 

We calculate these approximate solutions for $\bar{D}$ and $\bar{b}$ in two different ways: first by assuming that $f_e$ is approximately spherically symmetric, and in the second approach taking into account approximate deviations from spherical symmetry. For the region of M31 of interest for our analysis of the radio data (namely, the region inside $\sim 10\,\unit{kpc}$), the resulting solutions for $\langle f_e \rangle$
(and the resulting synchrotron emission) are similar regardless of our assumptions.

Our first approximate solution -- the ``unweighted" solution -- assumes that deviations from spherical symmetry for $f_e$ are small. If this is the case,
\begin{equation}\label{eq:sph ave coefs 2}
\begin{split}
\bar{D} &\simeq \langle D\rangle\\
\bar{b} &\simeq\langle b\rangle,
\end{split}
\end{equation}
and we can numerically average over solid angles our models for $D$ and $b$ given in Sections \ref{sec:diff coef} and \ref{sec:energy loss coef}.

\begin{figure*}
\centering
\subfigure[]{\label{fig:electron_phasespaceE_mx}\includegraphics[width=0.9\columnwidth]{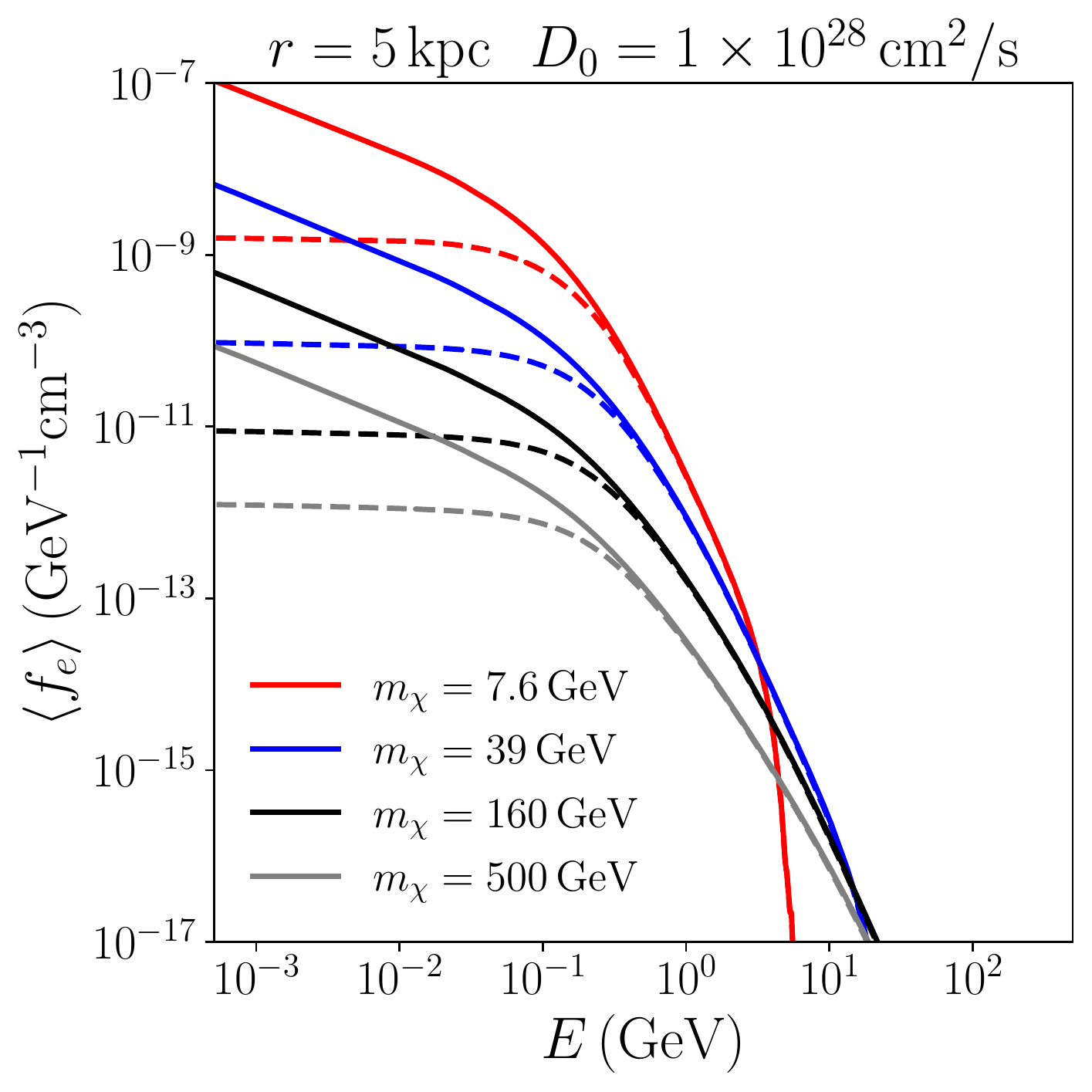}}
\subfigure[]{\label{fig:electron_phasespacer_D0}\includegraphics[width=0.9\columnwidth]{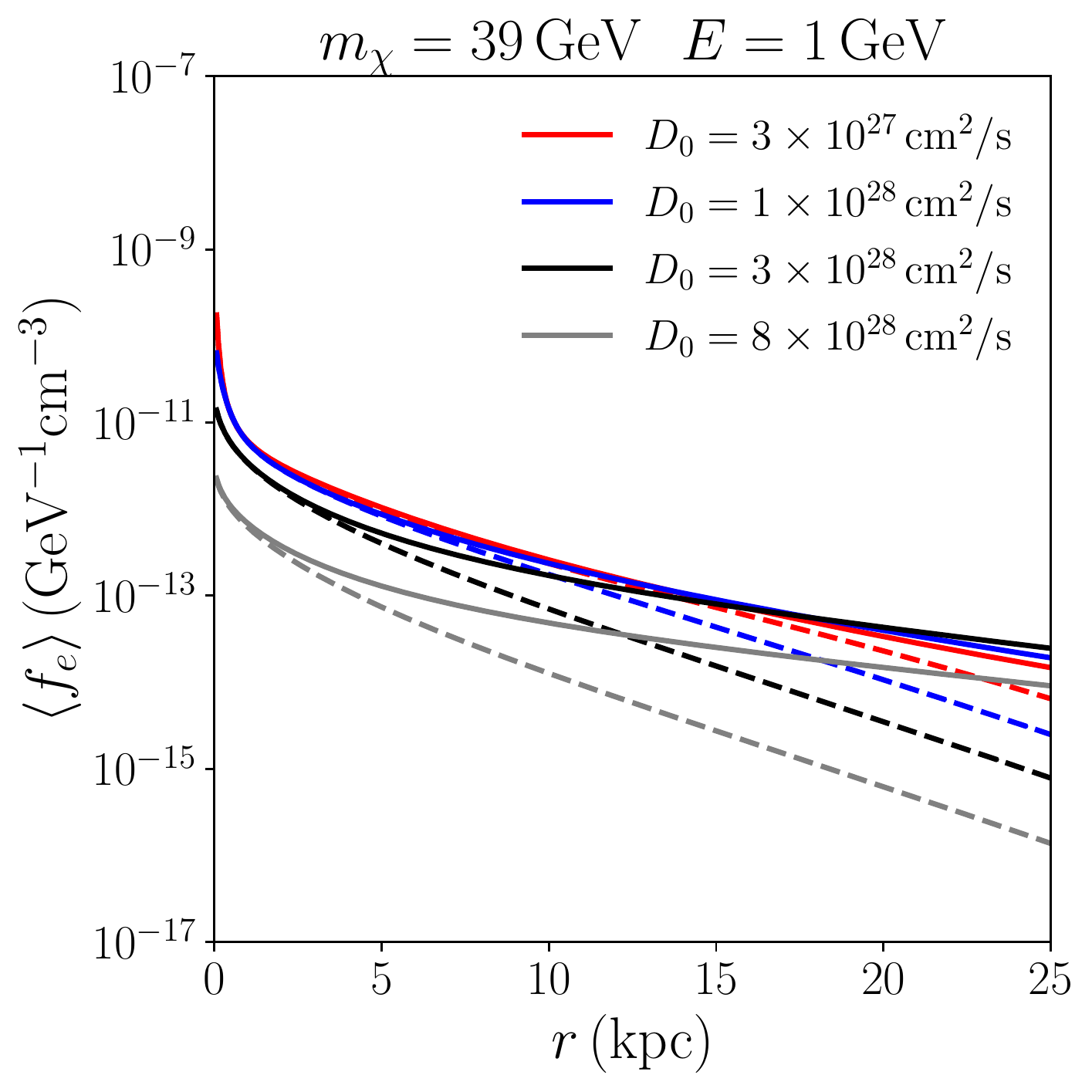}}
\caption{Spherically averaged equilibrium phase space density of $e^\pm$ as a function of (a) $E$ and (b) $r$ from dark matter with an annihilation cross-section of $\langle \sigma v \rangle = 2.2 \times 10^{-25}\rm cm^3/s$. In (a) we keep $r$ and $D_0$ constant for various values of $m_\chi$. In (b) we hold $E$ and $m_\chi$ constant and vary $D_0$. Dashed and solid lines are as in Figure~\ref{fig:weighted_vs_unweighted}.}
\label{fig:e distrib}
\end{figure*}

For our second solution for the diffusion and loss coefficients, we
calculate the spherically-averaged $\bar{D}$ and $\bar{b}$ parameters by substituting into Eq.~\eqref{eq:sph ave coefs} the approximate equilibrium solution for $f_e$ obtained from solving Eq.~\eqref{eq:approx diff loss} with $\partial f_e/\partial t=0$:
\begin{equation}\label{eq:fe approx}
f_e \simeq Q_e \tau,
\end{equation}
where $\tau^{-1} \equiv \tau_b^{-1}+\tau_D^{-1}$. We refer to this as the ``weighted'' solution, as $\bar{D}$ and $\bar{b}$ are obtained as averages of $D$ and $b$ weighted by $\partial_r f_e$ and $f_e$, respectively.

Figure~\ref{fig:weighted_vs_unweighted} shows the spherically averaged diffusion and loss coefficients for the unweighted (dashed) and weighted (solid) averaging schemes. The two methods give very similar results for the loss coefficient across all of phase space and galactic radii. For the diffusion coefficient, the two calculations agree for the inner part of M31, $r \lesssim 10\unit{kpc}$. As we will show, the disagreement at large radii in the diffusion coefficients does not result in significant differences in the predicted synchrotron emission from the region of M31 that we will use to set limits. As a result, the constraints derived from radio observations are robust across these different solutions.

Using either the weighted or unweighted solutions for $\bar{D}$ and $\bar{b}$, we must solve the diffusion loss equation for $\langle f_e\rangle$. Defining $u\equiv r\langle f_e \rangle$, Eq.~\eqref{eq:sph sym diff loss} becomes
\begin{equation}\label{eq:diff loss eq u}
\frac{\partial u}{\partial t} = \frac{\partial}{\partial r}\left[\bar{D}(r, E)\frac{\partial u}{\partial r}\right] -\frac{\partial \bar{D}}{\partial r}\frac{u}{r} + \frac{\partial}{\partial E}\left[\bar{b}(r, E)u\right]+rQ(r, E)
\end{equation} 
Under this redefinition, the boundary condition at $r=0$ can be easily written as $u(0, E) =0$, as long as $f_e$ does not diverge faster than $1/r$. This is satisfied if the inner slope of the dark matter density has a power law index $\gamma_{\rm NFW}<1.5$. The other required boundary conditions are $u(49.9\unit{kpc}, E) = 0$ and $u(r, m_\chi) = 0$. To solve Eq.~\eqref{eq:diff loss eq u}, we discretize $r$, $E$ and $t$ and use finite differences to approximate the derivatives. This leads to a recursive equation for $u$ at the next time-step in $t$, given its value at the current $t$. 

Forward difference schemes for solving
Eq.~\eqref{eq:diff loss eq u} are only stable if the time-step satisfies $\Delta t \lesssim (\Delta r)^2/D$ over the whole domain \cite{1992nrfa.book.....P}, where $\Delta r$ is the grid-spacing.
Given the approximate age of M31 and our grid-spacing $\Delta r = 62 \,\unit{pc}$, ${\cal O}(10^{7})$ time-steps would be needed to reach the equilibrium solution using a forward difference method. Backward differences, on the other hand, are unconditionally stable \cite{1992nrfa.book.....P} for any size of time-step. We therefore use backward differences to approximate the derivatives on the right-hand-side, leading to an implicit equation for $u$ at the next time-step, which can be solved with a sparse matrix method. We choose the time-step to be much larger than the maximum timescale in the problem to minimize the number of iterations required. Further details about our numerical method for solving the diffusion-loss equation are provided in Appendix \ref{sec:numerical method}.

The results for the equilibrium solutions of $\langle f_e\rangle$ are shown in Figure~\ref{fig:e distrib}. Figure~\ref{fig:electron_phasespaceE_mx} shows the energy dependence of $\langle f_e \rangle$ at $r = 5 \unit{kpc}$ for a representative set of values of $m_\chi$ and our default value of $D_0$. Figure~\ref{fig:electron_phasespacer_D0} shows the dependence on $r$ for each value of $D_0$ and $E=1 \unit{GeV}$. The results become more sensitive to changes in $D_0$ for $D_0\gtrsim 3 \times 10^{28} \unit{cm^2/s}$. In both panels, the solid curves represent the weighted solution while the dashed curves represent the unweighted solution.

\section{Synchrotron Spectrum and Morphology} \label{sec:synchrotron}

Relativistic electrons and positrons in M31 accelerate in the galactic magnetic field, leading to synchrotron emission. The power emitted per unit frequency from an electron or positron at pitch angle $\alpha$ and energy $E$ is \cite{1998clel.book.....J}
\begin{equation}\label{eq:power per frequency}
\frac{dP}{d\nu}(\nu, \alpha, E) = 2 \pi \sqrt{3}e^2 \gamma \nu_0x \int\displaylimits_{x/\sin{\alpha}}^{\infty} d\xi K_{5/3}(\xi),
\end{equation}
where $\nu_0 \equiv e\bar{B}/(2\pi \gamma m_e)$, $x=2\nu/(3\gamma^3 \nu_0)$, $\gamma$ is the Lorentz factor and $K_n$ is the $n^{\rm th}$-order modified Bessel function of the second kind. 
The differential flux can therefore be obtained by averaging Eq.~\eqref{eq:power per frequency} over uniformly distributed pitch angles, and convolving with the spherically-averaged phase space density of electrons leading to:
\begin{equation}\label{eq:flux integs}
\begin{split}
\frac{d^2S}{d\Omega d\nu} =& \frac{1}{4\pi}\int\displaylimits_{\rm los}dl\int\displaylimits_{m_e}^\infty dE \langle f_e\rangle (r(l, \Omega), E)\langle dP/d\nu\rangle_\alpha \\
\langle dP/d\nu\rangle_\alpha \equiv & \frac{1}{2} \int\displaylimits_{-1}^1 d (\cos \alpha)\frac{dP}{d\nu},
\end{split}
\end{equation}
where $\Omega=(\theta,\phi)$ is the location on the sky.

\begin{figure}
\centering
\includegraphics[width=0.9\columnwidth]{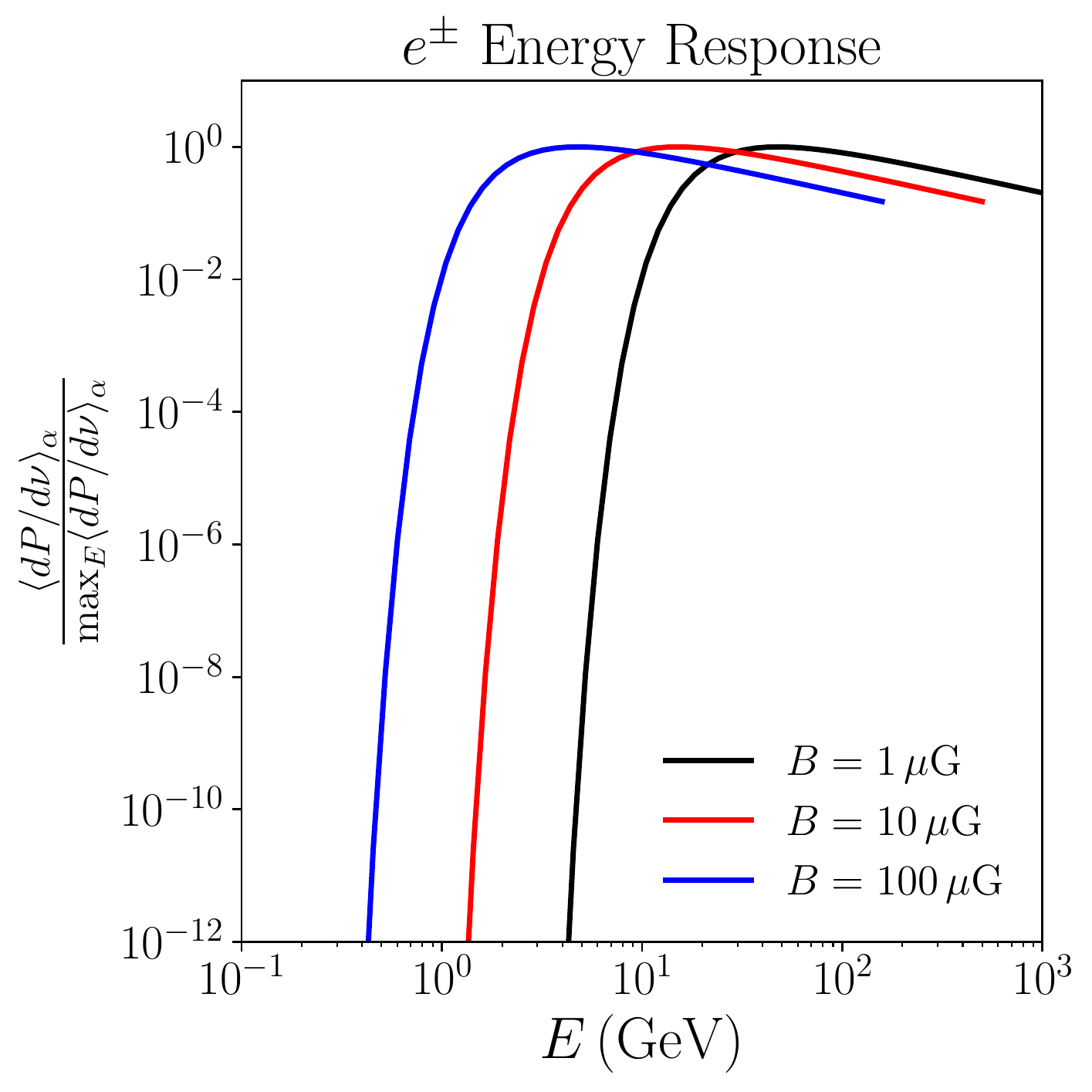}
\caption{Energy response of $e^\pm$ producing synchrotron emission of frequency $\nu = 8.35 \unit{GHz}$ for a variety of magnetic field strength values. 
}
 \label{fig:power}
 \end{figure}

For $\sin \alpha \sim \mathcal{O}(1)$ and $x\gg 1$, $dP/d\nu$ is exponentially suppressed at low energies \cite{1998clel.book.....J}, so most of the power is radiated by $e^\pm$ with energies satisfying
\begin{equation}\label{eq:E_cut}
    E\gtrsim 10 \unit{GeV}\left(\frac{\nu}{8.35 \unit{GHz}}\right)^{1/2}\left(\frac{10 \unit{\mu G}}{\bar{B}}\right)^{1/2}.
\end{equation}
As the Effelsberg radio telescope data used in this study is at frequencies around $8.35\unit{GHz}$, we are most interested in the $e^\pm$ produced through dark matter annihilation with energies of $\sim 10 \unit{GeV}$ and higher. This is shown in Figure~\ref{fig:power} where we plot the dependence of $\langle dP/d\nu\rangle_\alpha$ on $E$ for a variety of fixed values of $\bar{B}$.

\begin{figure*}
\centering
\includegraphics[width=2\columnwidth]{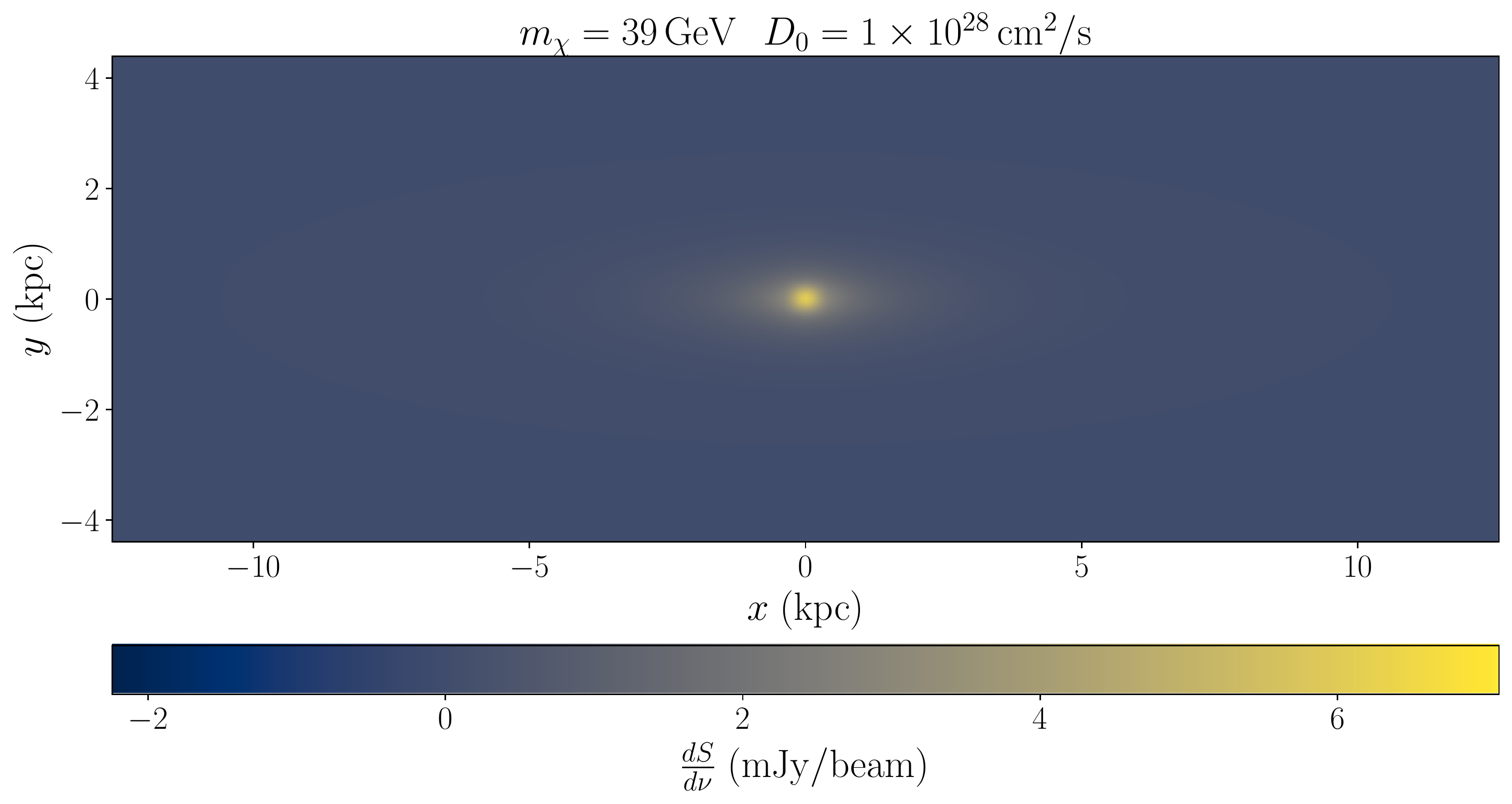}
\caption{Predicted synchrotron emission at a frequency of $\nu = 8.35 \unit{GHz}$ from dark matter with $m_\chi = 39 \unit{GeV}$ annihilating with a cross-section of $\langle \sigma v \rangle = 2.2\times 10^{-25} \, \rm cm^3/s$. In calculating this synchrotron map, we used our default value of $D_0$ and our weighted averaging scheme.}

\label{fig:signal}
\end{figure*}

\begin{figure*}
\centering
\subfigure[]{\label{fig:signal_mx}\includegraphics[width=0.9\columnwidth]{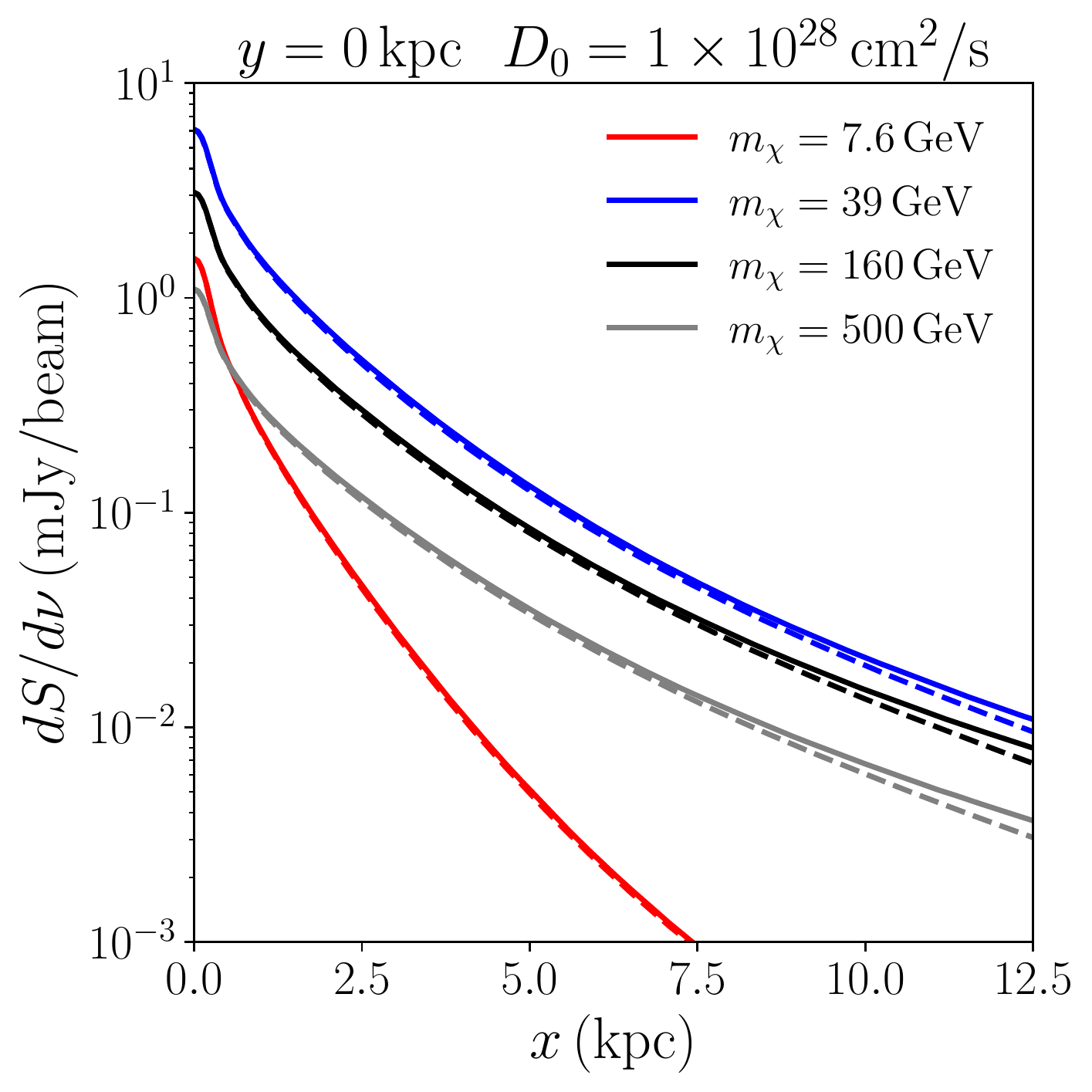}}
\subfigure[]{\label{fig:signal_D0}\includegraphics[width=0.9\columnwidth]{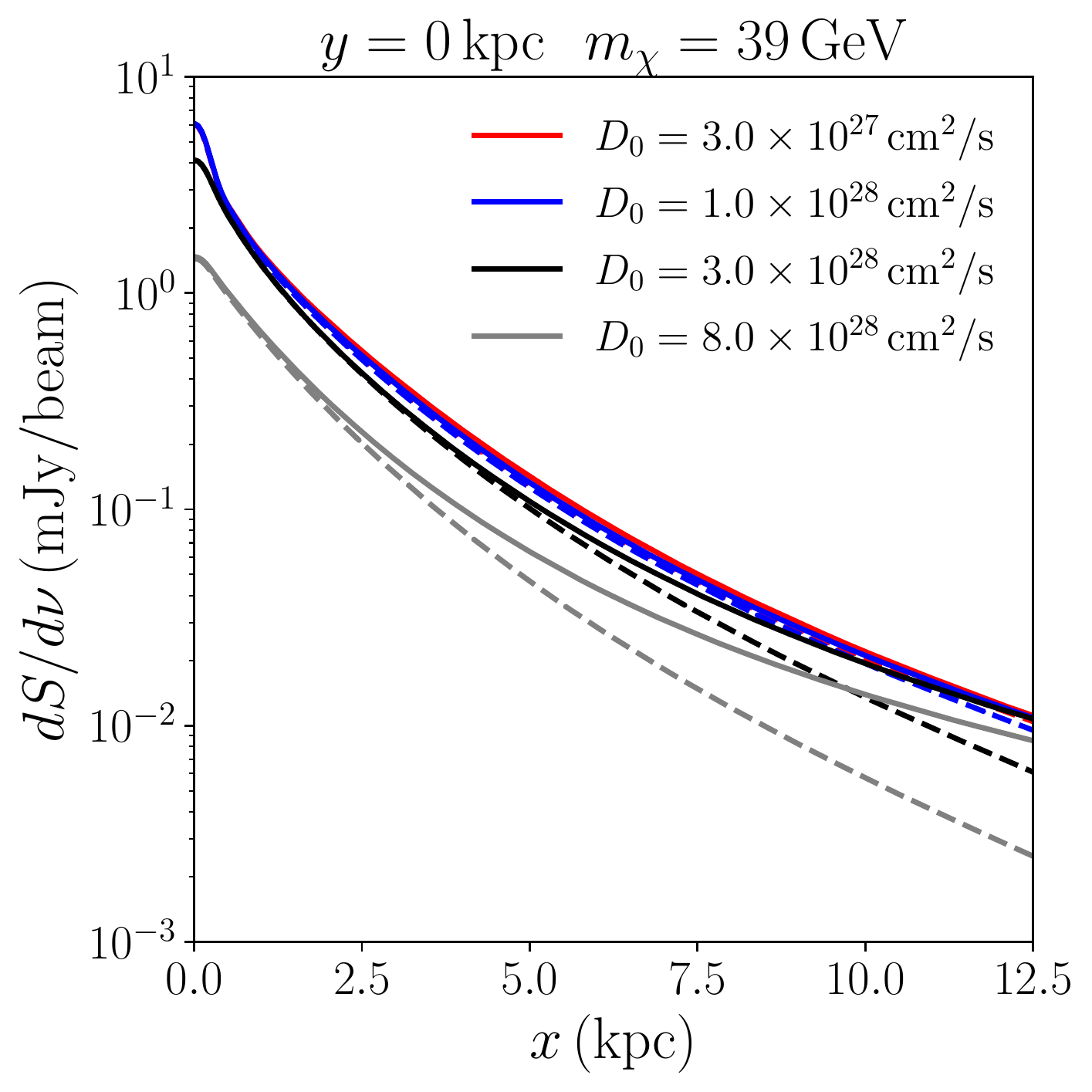}}
\caption{Predicted synchrotron emission at a frequency of $\nu = 8.35 \unit{GHz}$ from dark matter annihilating with a cross-section of $\langle \sigma v \rangle = 2.2\times 10^{-25} \, \rm cm^3/s$. The emission is shown as a function of $x$, the distance from the center of M31 in the plane of the sky along the semi-major axis. The flux is integrated over the effective beam size of the data, $\Omega_{\rm{beam}} = 2.157 \times 10^{-7}\,\unit{sr}$. In (a) we fix $D_0$ to the default value of $1\times 10^{28} \unit{cm^2/s}$ and vary $m_\chi$. In (b) we set $m_\chi = 39 \unit{GeV}$ and vary $D_0$. The solid curves show the emission using our weighted numerical solutions, dashed curves use the unweighted approach.
\label{fig:signal_slices}}
\end{figure*}

In Figure~\ref{fig:signal}, we show the $8.35 \unit{GHz}$ radio emission resulting from dark matter of mass $m_\chi = 39 \unit{GeV}$ annihilating with a cross section of $\langle \sigma v \rangle = 2.2\times 10^{-25}\unit{cm^3/s}$, assuming $D_0 = 1 \times 10^{28} \unit{cm^2/s}$. 
In Figure~\ref{fig:signal_slices}, we show the signal along the semi-major axis (a) for a variety of values of $m_\chi$, holding $D_0$ constant and (b) for a variety of values of $D_0$ holding $m_\chi$ constant. 

\section{Statistical Methodology} \label{sec:SM}

Having developed a numeric method to calculate the radio emission induced by dark matter annihilation in M31, we can now compare our predicted signal with data to set limits on the annihilation cross section to $\bar{b}b$ as a function of dark matter mass. Though the dark matter annihilation will be brightest in the center of the galaxy, this region also has significant baryonic sources whose intensities cannot be easily modelled. In addition, the flux near the center of M31 is sensitive to the value of $D_0$ for $D_0\gtrsim 1\times 10^{28} \unit{cm^2/s}$ (as seen in Figure~\ref{fig:signal_D0}). For these reasons we set limits using the expected {\it morphology} of the dark matter signal outside of the center, rather than the total intensity. This  also makes our constraints insensitive to possible mismeasurement of the overall zero-level of the radio data.

This approach requires data-driven modeling of the backgrounds within the galaxy. The background emission in M31 is complicated, with numerous point sources and a prominent ring feature (see Figure~\ref{fig:intensity_map}). None of these features are morphologically consistent with the expectations of dark matter annihilation, and can safely be attributed to baryonic physics. Even so, a multi-step process is required to define a search region and construct a background model within that region that does not risk fitting-away any potential signal.

In Section~\ref{sec:masking}, we first describe how we mask the point sources, the ring of radio emission in the disk, and the bright center of M31. This will allow us to define a search region interior to the ring, where the background can be approximated as the residual emission from the ring plus a constant. In the end, the radio emission from this search region will be used to set limits on the dark matter model.

Next in Section~\ref{sec:ring background}, we describe how we determine the background model within the search region using the data itself -- without absorbing dark matter emission (potentially present in the data) into the model. We introduce a background model with five free parameters: three morphological parameters $(\mu_1, \mu_2, \mu_3)$ controlling the shape of the residual background from the elliptical ring, and two coefficients $(w_1, w_2)$ which determine the intensity of each component of the background. The morphological parameters are fixed based on the data independent of the signal hypothesis, while the intensity coefficients are adjusted to their most likely values for each hypothesis.

Fixing the morphological parameters must be done carefully to avoid absorbing any signal present in the data into the background model. We leverage the fact that the signal peaks toward the center of M31, while the emission from the ring is dominant away from the center. We therefore can fix the morphological parameters by using the data away from the center of the intensity map (exterior to the dark matter-rich ``signal region''). The size of this signal region is determined by comparing fits of the morphological parameters of the background model assuming the presence or absence of a dark matter signal.

After defining the search region and fixing the morphology of our background model within this region, we set statistical limits on specific signal models. We use a $CL_{\rm s}$ test, described in Section~\ref{sec:CLs overview}. $CL_{\rm s}$ works by building distributions of test-statistics from synthetic observations generated from background-only and signal-plus background hypotheses. This test-statistic is sensitive to the morphology of the signal in addition to the amplitude, making this ideal for the distributed signal of dark matter in M31. Our full set of limits varying over astrophysical model parameters and using the methodology we describe here will be shown in Section~\ref{sec:limits}.

\subsection{Background Masks}\label{sec:masking}

The baryonic sources of radio emission in M31 are complicated and difficult to model from first principles. Overall, we expect relatively uniform background emission across the interior of the galaxy, overlaid with significant emission from the galactic center due to baryonic processes, as well as point sources throughout the galaxy. In addition, M31 contains a prominent elliptical ring-shaped structure in radio with a semi-major axis of approximately $10 \unit{kpc}$, due to significant star formation in this region \cite{2010A&A...517A..77T, 2016MNRAS.456.4128R}. All of these features can clearly be seen in the radio map of Figure~\ref{fig:intensity_map}. The location of the ring correlates with the highest densities of gas in our astrophysical models from Section~\ref{sec:M31model}.

Notably, other than the emission at the center of the galaxy, the spatial distribution of all of these sources of radio emission is inconsistent with emission sourced by dark matter. Rather than attempting to model these baryonic sources from first principles, we mask and remove them from our statistical analysis. As the emission at the center and the point sources are localized, we are able to completely remove them using masks. The ring of bright emission is broad enough that it cannot be removed completely. Instead, we model it as a Gaussian ring and mask its brightest emission.

\subsubsection{Point Source Masks} \label{sec:point_source_mask}

Ref.~\cite{2020A&A...633A...5B} has removed 38 point sources unrelated to M31. However, many point sources within the galaxy remain in the data. We locate point sources algorithmically by identifying circular regions (with a diameter of 0.75 times the HPBW of the beam) that are over-bright compared to the concentric annulus with inner and outer diameter of 2.25 and 2.75 times the HPBW, respectively. We classify a circular region centered on pixel $i$ a point-source at high ($\sim 4\sigma$) confidence if
\begin{equation}
    \langle d\rangle^{\rm (cir)}_i >  \langle d\rangle^{\rm (ann)}_i + 4\sigma_{\rm rms}
\end{equation}
where $\langle d\rangle^{\rm (cir)}_i$ and $\langle d\rangle^{\rm (ann)}_i$ are the flux per beam averaged over the circle and annulus centered at the pixel $i$, respectively (the noise $\sigma_{\rm rms}$ is defined in Section~\ref{sec:data}). For each pixel in the radio map that passes this criteria, we mask a circular region (of diameter 0.75 times the HPBW) centered on the pixel.

In addition to these conventional point sources, there is a feature (located near $x=4\unit{kpc}$, $y = 0\unit{kpc}$ in Figure~\ref{fig:intensity_map}) that is likely an artifact of the imaging process. As this feature does not have the intensity distribution of a point source, it was not identified by our point source algorithm, and we mask it by hand.\footnote{There is a similar feature near $x=-2 \unit{kpc}$, $y=-2.5 \unit{kpc}$. As this feature will not be in the search region (defined in Section~\ref{sec:annulus_mask}), we do not mask it manually.}

\subsubsection{Center Mask}\label{sec:center_mask}

The center of M31 is the brightest source of radio emission in the galaxy. While dark matter-induced emission would also peak in this region, much of the observed emission is likely due to difficult-to-model baryonic processes. Limits on annihilation can be set by using only this central emission \cite{2013PhRvD..88b3504E, 2021MNRAS.501.5692C, 2022PhRvD.106b3023E}, but the intensity of the dark matter signal here is sensitive to the diffusion parameter for $D_0\gtrsim 1 \times 10 ^{28}\unit{cm^2/s}$ (as shown in Figure~\ref{fig:signal_D0}). For these reasons, we also mask the center of M31 in our analysis and set limits on dark matter using the region outside the center. Here, the lower signal rate is off-set by the lower background, and the differing morphologies of the signal and background can be used to set limits less sensitive to uncertainties in the diffusion coefficient.

\begin{figure}
\centering
\includegraphics[width=0.9\columnwidth]{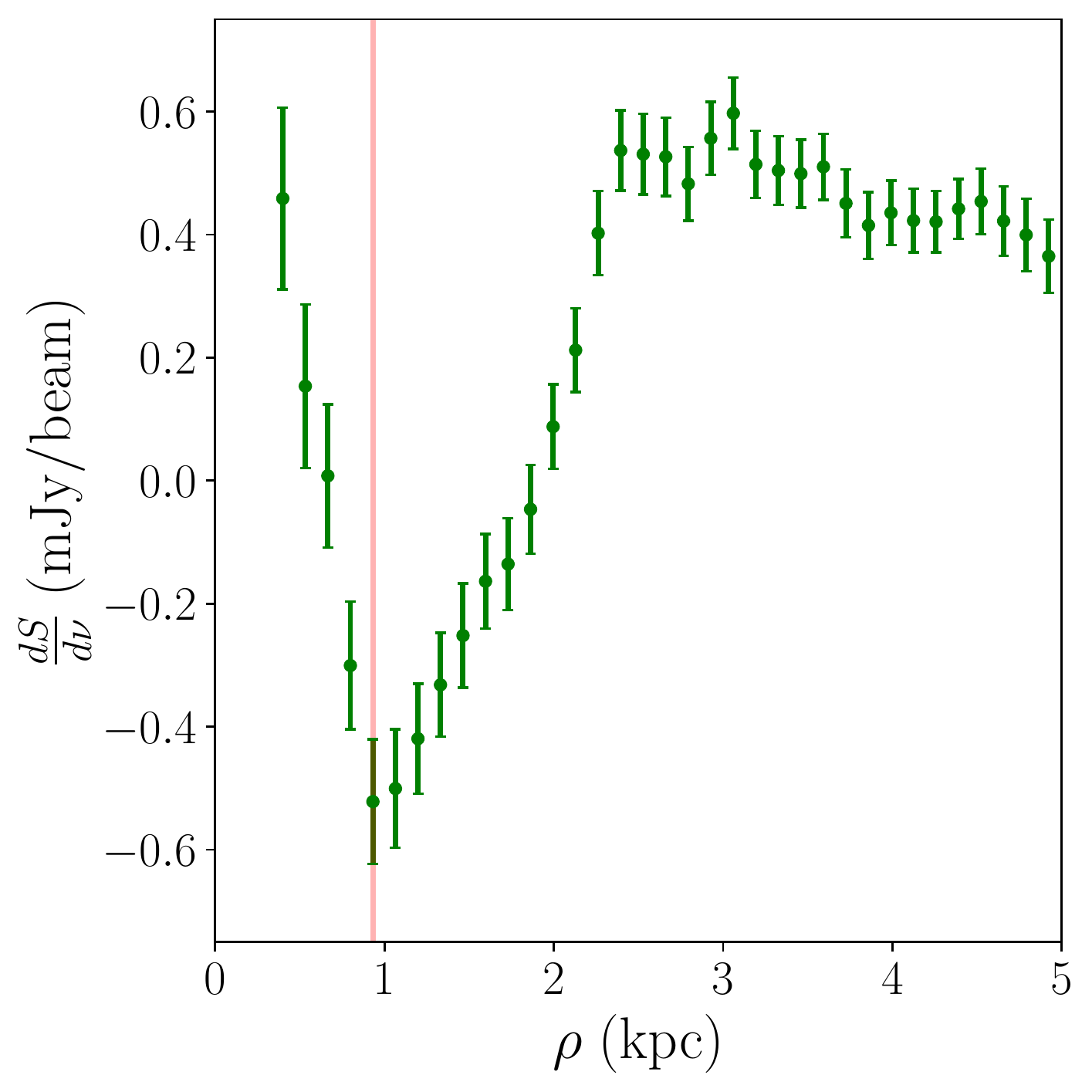}
\caption{Observed flux averaged over concentric circular annuli of radius $\rho$ in the plane of the sky, not including pixels that are in the point source mask. The errors in the annulus averaged flux in a particular bin are found by averaging the rms noise over the bin and dividing by the square root of the number of beams in the bin. The radius of the center circular mask is shown with a red vertical line.} \label{fig:annulus_averaged}
\end{figure}

\begin{figure*}
\centering
\subfigure[]{\label{fig:data_basemask}\includegraphics[width=1.8\columnwidth]{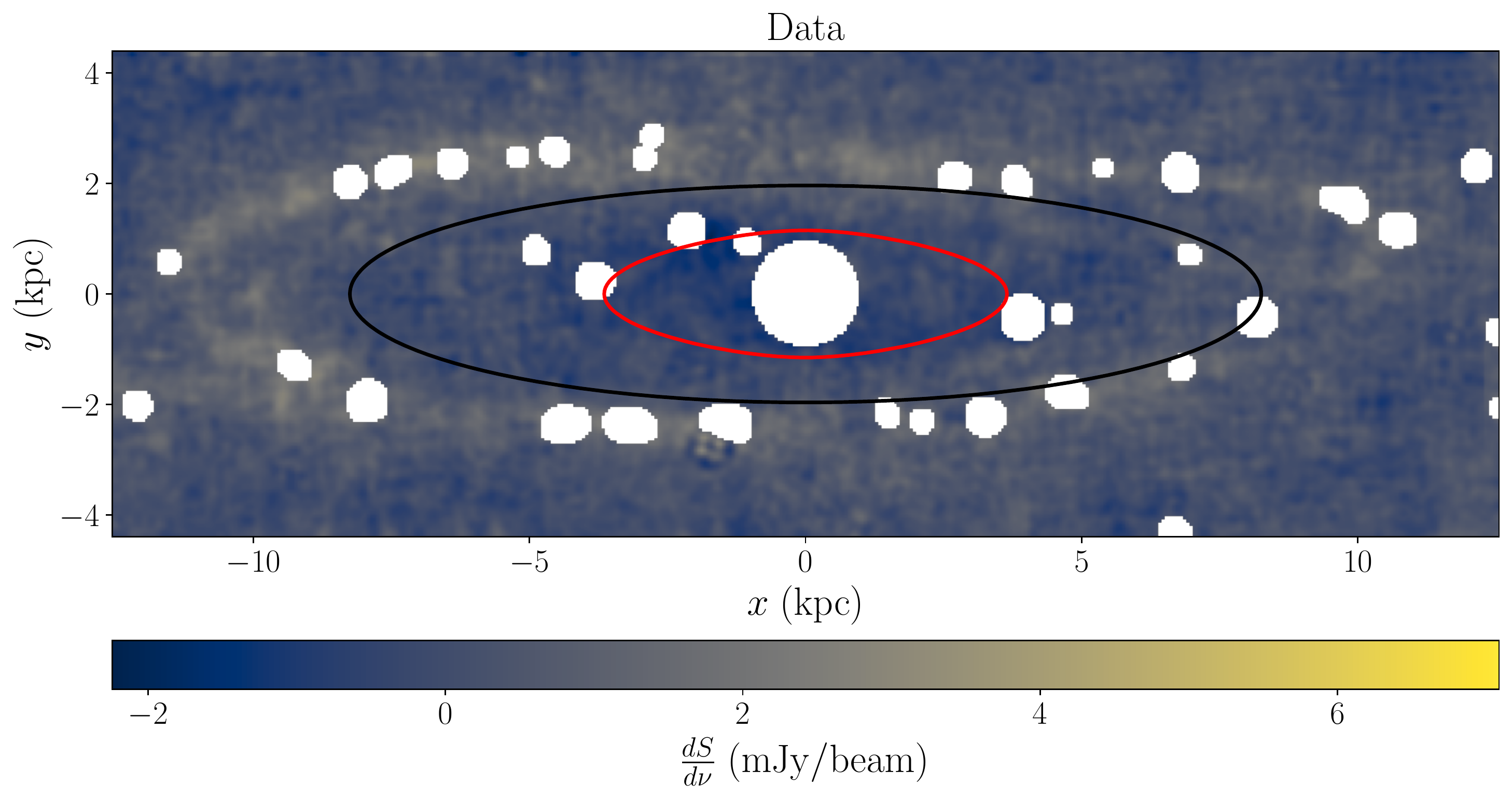}}
\subfigure[]{\label{fig:pseudodata_basemask}\includegraphics[width=1.8\columnwidth]{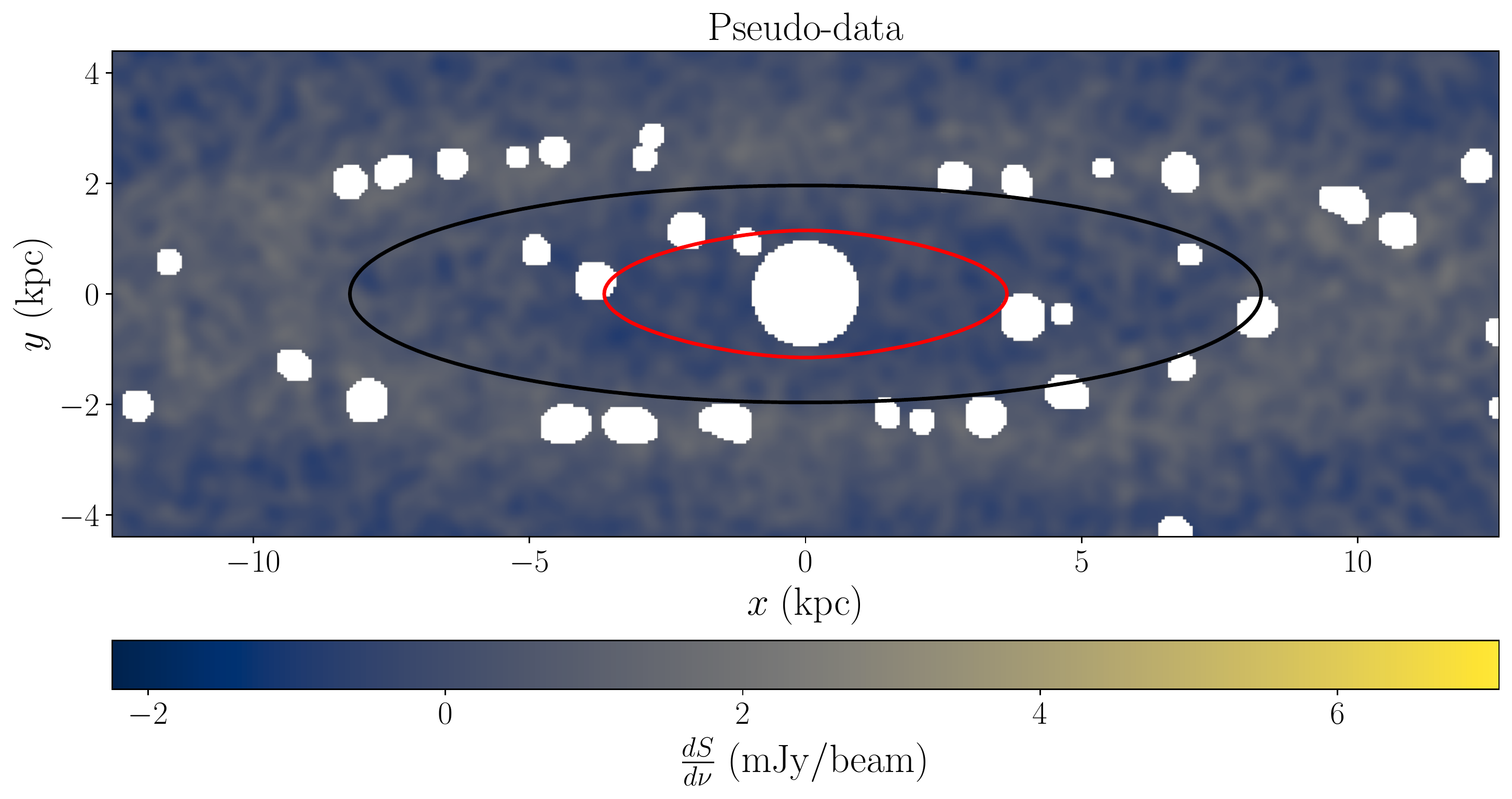}}
\caption{Intensity maps of the (a) radio data and (b) simulated pseudo-data using the globally-fit background model, with point source and center masks (described in Sections~\ref{sec:point_source_mask}  and \ref{sec:center_mask}, respectively) applied. The method of simulating the pseudo-data is described in Appendix~\ref{sec:simulating intensity maps}. The search region (used to set limits on dark matter annihilation, see Section~\ref{sec:annulus_mask}) consists of the unmasked pixels within the black contour. The signal region, masked when defining signal-independent background templates (see Section~\ref{sec:ring background}), is interior to the red contour.}
\label{fig:basemask}
\end{figure*}

Our point-source masking technique also identifies a source at the center of M31, but the default point-source mask is too small to cover the entire bright center region. To determine the size of the central circular mask, we plot the intensity, averaged over concentric circular annuli (excluding pixels in point-source masks), as a function of 2D radius $\rho = (x^2+y^2)^{1/2}$ in Figure~\ref{fig:annulus_averaged}. We mask the central region out to the minimum of this averaged flux, at $\rho=0.93\unit{kpc}$. The intensity map with point sources and the center masked is shown in Figure~\ref{fig:data_basemask}.

\subsubsection{Ring and Outside Masks}\label{sec:annulus_mask}

Finally, we must construct a mask for the elliptical ring of bright emission in the star forming region of the M31 disk \cite{2010A&A...517A..77T, 2016MNRAS.456.4128R}. Given the morphology of this feature, it cannot be due to dark matter annihilation. Masking it therefore does not risk removing a potential signal and setting overly-strong constraints. 

To construct the mask, we first fit the data to a sum of a uniform template and a Gaussian elliptical ring template which have the forms
\begin{equation}\label{eq:elliptical template}
\begin{split}
    \Phi^{u}(\boldsymbol{x}; w_1) = &\, w_1,\\
    \Phi^{r}(\boldsymbol{x}; w_2, \boldsymbol{\mu}) =&\,  w_2\exp{\left[ -\frac{\left(R_e(\boldsymbol{x}, \mu_1)-\mu_2\right)^2}{2 \mu_3^2}\right]},
\end{split}
\end{equation}
where 
\begin{equation}\label{eq:elliptical_radius}
R_e(\boldsymbol{x}, \mu_1) = \sqrt{x^2 + \mu_1^2 y^2}
\end{equation}
is the elliptical radius, $\boldsymbol{\mu} = (\mu_1,\mu_2,\mu_3)$ are free parameters of the model that control the shape and size of the ring, and $\boldsymbol{w} = (w_1,w_2)$  control the intensity of each component of the background. For the remainder of the paper, we will refer to $\boldsymbol{\mu}$ as the background morphological parameters and $\boldsymbol{w}$ as the background coefficients. The total background model is
\begin{equation} \label{eq:backgrond model}
        \Phi^b(\boldsymbol{x}; \boldsymbol{w}, \boldsymbol{\mu}) = \Phi^u(\boldsymbol{x}; w_1) + \Phi^r(\boldsymbol{x}; w_2, \boldsymbol{\mu})
\end{equation}

\begin{figure}
    \centering
    \includegraphics[width=0.9\columnwidth]{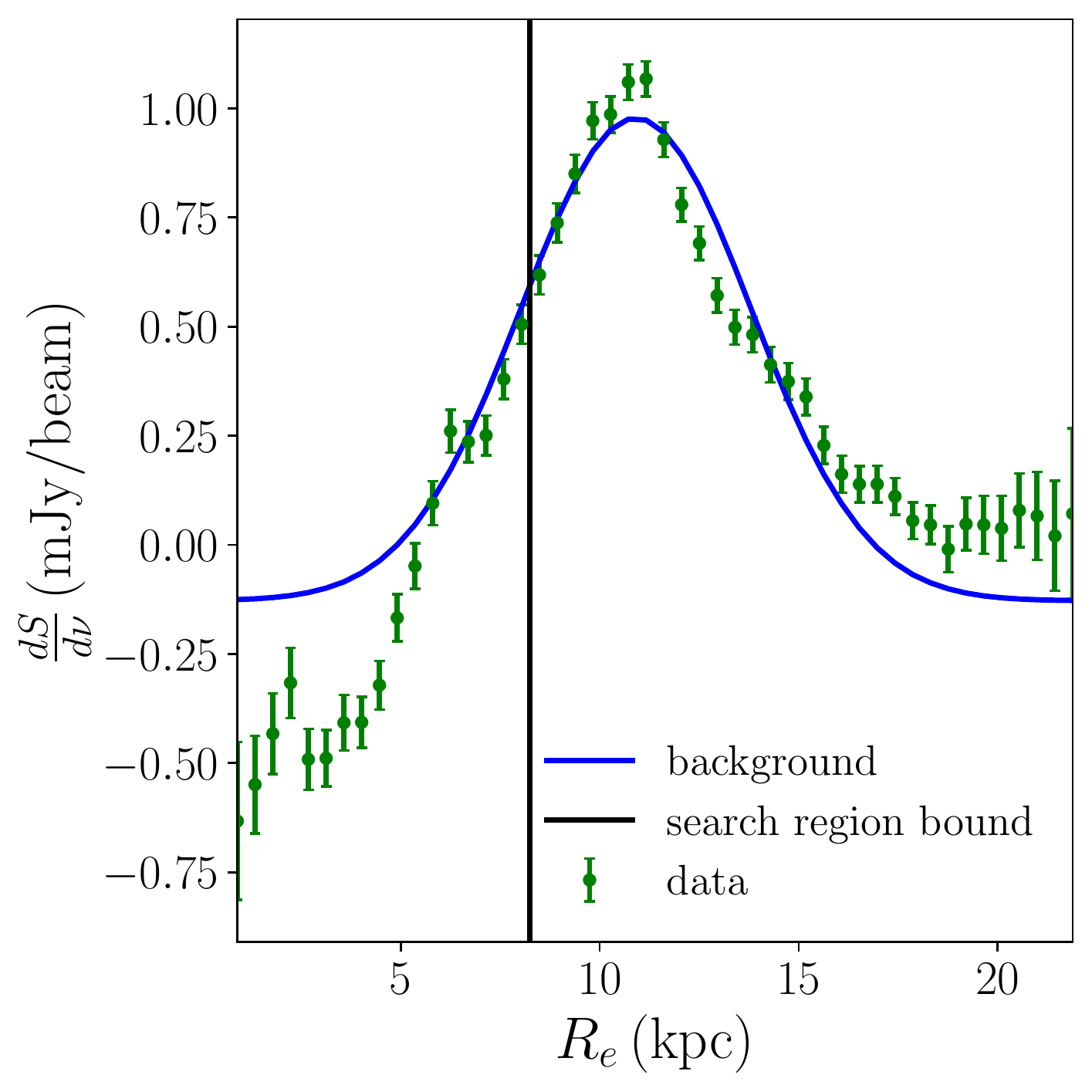}
    \caption{Synchrotron data and globally fit background model (parameters given in the ``Full Map Analysis'' section of Table~\ref{tab:annulusparameters}) averaged over elliptical annuli as a function of $R_e(\boldsymbol{x}, \mu_1)$ where $\mu_1$ is taken to be the globally fit value.}
    \label{fig:radial_flux}
\end{figure}

As the dark matter-induced annihilation signal is expected to be small at the radius of the ring, we can fit our model to the ring independent of the signal model.
We minimize the $\chi^2$ statistic between the observed flux ($d_i$ in pixel $i$ at location $\boldsymbol{x}_i$) and the ring plus uniform background model
\begin{equation}\label{eq:chi^2 ellipse}
\chi^2 = \sum\displaylimits_{i=1}^{N_{\rm pix}} \frac{\left[d_i-\Phi^b(\boldsymbol{x}_i; \boldsymbol{w}, \boldsymbol{\mu})\right]^2}{\sigma_{\rm{rms},\rm i}^2}.
\end{equation} 
with respect to all components of $\boldsymbol{w}$ and $\boldsymbol{\mu}$. The resulting best fit values for the morphological parameters are listed in Table~\ref{tab:annulusparameters} in the first section (labeled ``Full Map Analysis'') and second column (labeled ``Global Fit''). Figure~\ref{fig:radial_flux} shows the radio data (with point sources and galactic center masked) and the globally fit background model averaged over concentric elliptical annuli (with the same eccentricity as the globally fit ring model) as a function of $R_e$.

\begin{table}[t]
\centering 
 \begin{tabular}{|c||c||c|} 
 \hline
 Parameter & Global Fit & Signal-Region Masked \\ [0.5ex] 
 \hline \hline
 \multicolumn{3}{c}{Full Map Analysis} \\
 \hline
$\mu_1$ & $4.20$ & $4.28$\\
$\mu_2 (\unit{kpc})$ & $11.1$ & $11.1$\\
$\mu_3 (\unit{kpc})$ & $2.88$ & $2.42$ \\ 
\hline
\multicolumn{3}{c}{Right-Only Analysis} \\
 \hline
$\mu_1$ & $3.63$ & $3.63$\\
$\mu_2 (\unit{kpc})$ & $9.30$ & $9.17$\\
$\mu_3 (\unit{kpc})$ & $2.39$ & $2.15$ \\ 
\hline
 \end{tabular}
 \caption{Best-fit morphological parameters for the ring. The Global Fit has the parameter values fit to the data with the center and point sources masked, while the Signal-Region Masked fit is over data with the additional mask over the central signal-rich region applied. We separately show the parameters after fitting to the entire M31 data set (labeled ``Full Map Analysis''), and the data in the $x>0$ right-hand side of Figure~\ref{fig:intensity_map} (labeled ``Right-Only Analysis,'' see Section~\ref{sec:limits}).}\label{tab:annulusparameters}
\end{table}

We show a heatmap of the globally-fit background model superimposed with simulated errors in Figure~\ref{fig:pseudodata_basemask}. Our method for simulating random errors correlated over the beam size (which is much larger than the pixel size) is explained in Appendix~\ref{sec:simulating intensity maps}. 

It is clear from Figure~\ref{fig:radial_flux} that the emission outside of the ring ($R_e\gtrsim 15\unit{kpc}$) is significantly brighter than the emission inside ($R_e\lesssim 5 \unit{kpc}$). As the dark matter signal is expected to drop with distance from the center, it cannot be responsible for this excess emission outside the ring. Therefore, we mask exterior to the ring. The width of the best-fit Gaussian is too broad to completely mask the emission out to the level of statistical noise in the region interior to the ring. We instead mask the ring inward to 1 standard deviation from the peak of the Gaussian ring model.
That is, we mask all pixels satisfying
\begin{equation}\label{eq:elliptical masking critera}
   R_e(\boldsymbol{x}, \mu_1) > \mu_2 -\mu_3,
\end{equation}
using the globally-fit values for the morphological parameters $\boldsymbol{\mu}$.

The inner boundary of the ring mask is shown in black contours in each panel of Figure~\ref{fig:basemask}. The interior of this contour (minus the center and pixels masked as part of point sources) is the search region that will be used to constrain dark matter annihilation.

\subsection{Background Model of the Search Region}\label{sec:ring background}

Having selected our search region, we must now define our background model that we will use to construct our background-only and signal plus background hypotheses. The model has the functional form  of Eq.~\eqref{eq:backgrond model} (used to define the ring and outside-region mask in Section~\ref{sec:annulus_mask}), but as the intensity map has the potential to be signal-rich, we must fix the morphological parameters when calculating our limits more carefully than when we initially defined the search region.

If there was a (known) dark matter signal in the data, the parameters of the background model would be most accurately found by subtracting that signal from the data and then fitting our background model to the result. Alternatively, if there is no dark matter signal in the data, the parameters of the background model would be most accurately found by fitting the background model to the data itself. With only the point-source and circular center masks applied, the best fit parameters for the background model are sensitive to the (unknown) presence of signal in the data. We avoid the risk of the background model absorbing any signal present in the data by leveraging the morphology of the signal maps, which peak towards the center of M31. 

Unlike the procedure for constructing the globally-fit background model, if we fit the background model only using data outside the center of M31 (where dark matter contributes less to the radio flux), then fits with and without signal subtracted will be more in agreement. The level of statistical agreement will increase as we mask more of an assumed signal. To maximize the amount of signal masked for a given area masked, the mask should have the shape of a region bounded by a contour of constant signal intensity. We will call the inner signal-rich region the ``signal region," and the mask that covers it the ``signal-region mask."

Our strategy then is to mask the signal region, and fit the parameters of $\Phi^b$ to the data exterior to the mask (including data outside the search region). This fit will allow us to define
\begin{equation}\label{eq:data driven background model}
    \hat{\Phi}^b(\boldsymbol{x}; \boldsymbol{w}) = \Phi^b(\boldsymbol{x}; \boldsymbol{w}, \boldsymbol{\hat{\mu}})
\end{equation}
where $\boldsymbol{\hat{\mu}}$ are the morphological parameters, fit outside the signal region and fixed for the rest of the analysis, while $\boldsymbol{w}$ are free parameters which set the amplitude of the various components of the background. These free parameters will be fit to the data (or pseudo-data) in the search region when we set limits on the presence of a dark matter signal. The region exterior to the signal-region mask, used to find $\boldsymbol{\hat{\mu}}$, must be sufficiently signal-poor so that statistical tests that distinguish between signal plus background and background-only hypotheses obtain the same results regardless of whether $\boldsymbol{\hat{\mu}}$ is determined by assuming the presence or absence of signal in the data. 

To identify this signal-poor region, we use as a benchmark signal the flux from dark matter with $m_\chi = 38.6 \unit{GeV}$, $\langle \sigma v\rangle = 2.2\times 10^{-25} \unit{cm^3/s}$ and a diffusion normalization of $D_0 = 1\times 10^{28} \unit{cm^2/s}$.  As the leakage out of the masked signal region is minimal and the signal morphology depends only weakly on the choice of mass and diffusion parameters, the resulting fits can be applied to signals with other values of $m_\chi$ and $D_0$. The cross-section is chosen to be approximately an order of magnitude larger than the best fit value from the GCE \cite{2013PhRvD..88h3521G, 2014PhRvD..90b3526A, 2015JCAP...03..038C, 2016PDU....12....1D} and existing limits from dwarf galaxies \cite{2011PhRvL.107x1302A, 2011PhRvL.107x1303G, 2014PhRvD..89d2001A, 2015PhRvD..91h3535G, 2015ApJ...809L...4D, 2015PhRvL.115w1301A}. Our fitting procedure will ensure that the background model is not significantly influenced by the presence of dark matter signals of this intensity and weaker in the data.

We make a series of candidate signal-region masks that intersect the semi-major axis at $x$ values between $[1.0-8.8]\,\unit{kpc}$. For each of these masks, we fit our background model (Eq.~\eqref{eq:backgrond model}) to the remaining unmasked data with signal subtracted (defined as ``Fit A'') or without signal subtracted (``Fit B'') by minimizing Eq.~\eqref{eq:chi^2 ellipse} with respect to all components of $\boldsymbol{w}$ and $\boldsymbol{\mu}$. We take the sum in Eq.~\eqref{eq:chi^2 ellipse} to be over pixels not covered by the candidate signal region mask, the center mask, or point source masks.

For each candidate signal-region mask, we determine if Fits A and B of the morphological parameters lead to statistically indistinguishable results when testing for the presence of signal. To compare the background-only and signal plus background hypotheses, we introduce a test statistic, defined as
\begin{equation}\label{eq:TS main}
\begin{split}
\lambda_{\langle \sigma v\rangle, \boldsymbol{\theta}}(\{d_i\}) =  & \Delta \chi^2 = \chi_{s+b}^2 - \chi_{b}^2 \\ 
& = \sum_i\frac{\left[d_i - \hat{\Phi}^{s+b}_i(\langle \sigma v\rangle, \boldsymbol{\theta}, \boldsymbol{w^{s+b}})\right]^2}{\sigma_{\mathrm{rms}, i}^2}\\
& -\sum_i\frac{\left[d_i - \hat{\Phi}^{b}_i(\boldsymbol{w^{b}})\right]^2} {\sigma_{\mathrm{rms}, i}^2},
\end{split}
\end{equation}
This statistic will also be used in our methodology for setting limits on dark matter (see Section~\ref{sec:limits}).  In Eq.~\eqref{eq:TS main}, $\{d_i\}$ are the differential flux values in each pixel of the intensity map, and the sum runs over pixels $i$ that are in the search region (defined in Section~\ref{sec:masking}). $\boldsymbol{w^{b}}$ ($\boldsymbol{w^{s+b}}$) are the most-likely values of the background coefficients, $\boldsymbol{w} = (w_1, w_2)$, under the background-only (signal plus background) hypothesis and are determined analytically for each intensity map.
The test statistic is constructed such that higher test statistic values imply that the intensity map is more background-like. 

The test statistic depends on the signal hypothesis being tested through the signal plus background model 
\begin{equation}\label{eq:s+b model}
    \hat{\Phi}^{s+b}_i(\langle \sigma v\rangle, \boldsymbol{\theta}, \boldsymbol{w^{s+b}}) = \Phi_i^{s}(\langle \sigma v\rangle, \boldsymbol{\theta}) +  \hat{\Phi}^b_i(\boldsymbol{w^{s+b}}), 
\end{equation}
where the signal flux at the pixel centered at solid angle $\Omega_i$ is given by
\begin{equation}
    \Phi_i^{s}(\langle \sigma v\rangle, \boldsymbol{\theta}) \equiv \Omega_{\rm beam}\left.\frac{d^2 S}{d\Omega d\nu}\right|_{\Omega_i, \langle\sigma v\rangle, \boldsymbol{\theta}}
\end{equation}
using the differential flux calculated in Section~\ref{sec:synchrotron}. The signal hypothesis is parameterized by the cross section $\langle \sigma v\rangle$ and a vector $\boldsymbol{\theta}$, containing $m_\chi$, $D_0$ and all other default astrophysical parameters, given in Section~\ref{sec:M31model}.

For a given candidate signal-region mask, we calculate distributions of test statistics from ensembles of pseudo-data using background models with morphological parameters fixed by Fits A and B. These ensembles are drawn using the methods described in Appendix~\ref{sec:simulating intensity maps}, using the $\hat{\Phi}^b_i$ appropriate for each fit, summing over pixels in the search region for each calculation of $\lambda_{\langle \sigma v \rangle, \boldsymbol{\theta}}$. As the test statistic requires a choice of signal parameters, we use as our reference signal model $m_\chi = 38.6\,\unit{GeV}$, $D_0 = 1\times 10^{28}\,\unit{cm^2/s}$, and a cross-section for which the signal plus background hypothesis is easily distinguished from the background hypothesis: $\langle\sigma v\rangle = 1.1\times 10^{-25}\,\unit{cm^3/s}$. 
If -- for a given signal mask -- the distribution of test-statistics is indistinguishable between background Fits A and B, then the same will be true for the result of a statistical test for distinguishing signal plus background from background. This means the morphological parameters can be fit to the data using that signal region mask without absorbing potential signal from the data.

\begin{figure}
    \centering
    \includegraphics[width=0.9\columnwidth]{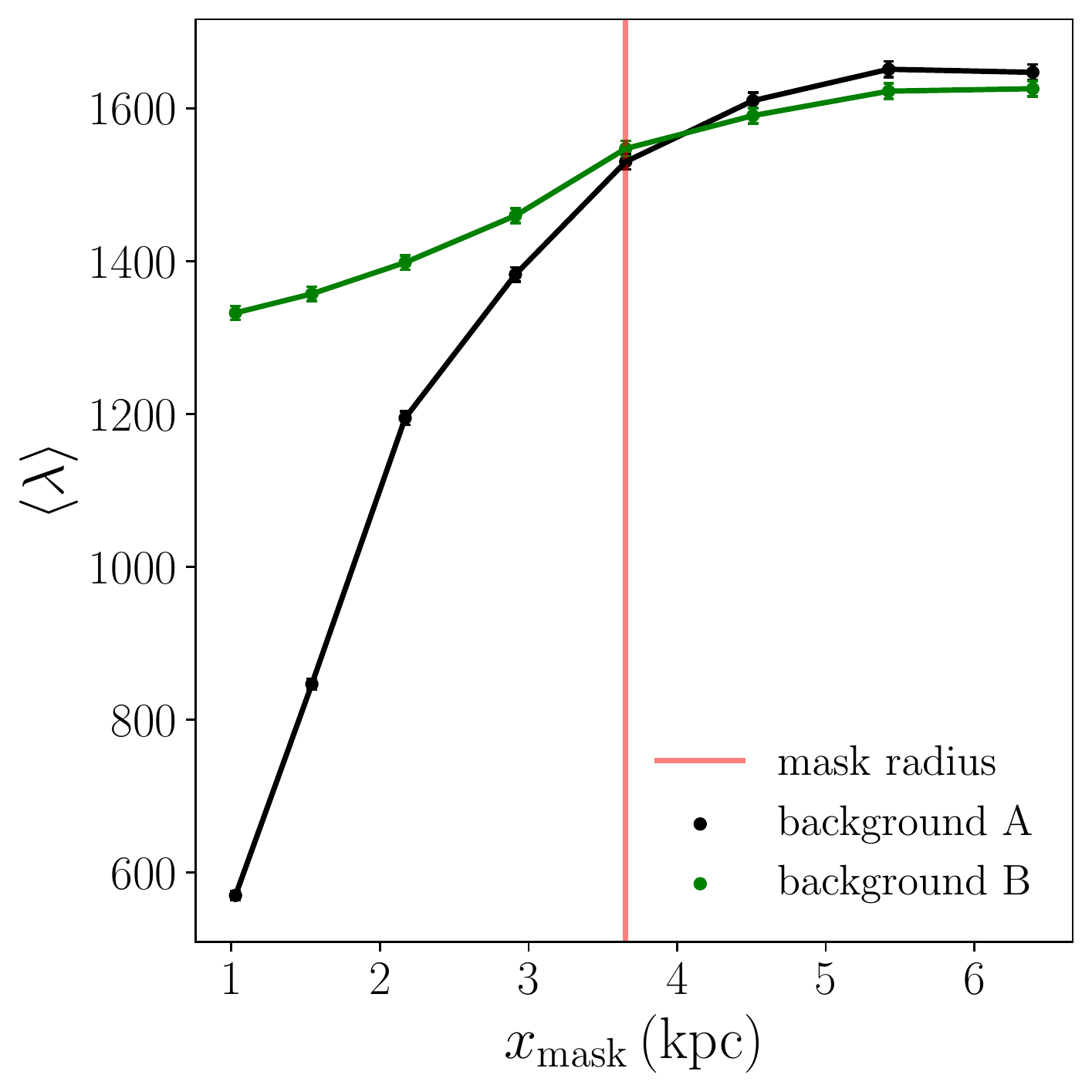}
    \caption{Mean test statistics from background-only pseudo-data for a series of candidate fits of the morphological parameters of the background model. Each fit comes from minimizing Eq.~\eqref{eq:chi^2 ellipse} 
    outside of the candidate signal region mask that intersects the semi-major axis of M31 at $x_{\rm mask}$, assuming the presence (green) or absence (black) of dark matter signal in the data. The distributions of test statistics are constructed for a signal from dark matter with $m_\chi = 38.6 \unit{GeV}$ and $\langle \sigma v \rangle = 1.1 \times 10^{-25}\unit{cm^3/s}$ and default diffusion normalization of $D_0 = 1 \times 10^{28}\unit{cm^2/s}$. The size of the signal-region mask that we select is shown with the red vertical line.}
    \label{fig:mean_ts}
\end{figure}

Figure~\ref{fig:mean_ts} shows the mean test statistic using candidate Fits A and B as a function of the distance from the origin that the signal region mask intersects the semi-major axis. For a mask which intersects the semi-major axis at $x=3.65 \unit{kpc}$, the means of the distributions of the test statistic from Fits A and B agree within statistical noise.  Selecting this signal region mask (shown with a red contour in Figure~\ref{fig:basemask}), we set $\boldsymbol{\hat{\mu}}$ to the best-fit morphological parameters of Fit B. The values of the morphological parameters from this fit are shown in  Table~\ref{tab:annulusparameters} (the ``Signal-Region Masked" column of the ``Full Map Analysis" section). To construct our signal plus background and background-only hypotheses, we will use the background model with the morphological parameters fixed to $\boldsymbol{\hat{\mu}}$ and the coefficients $\boldsymbol{w} = (w_1, w_2)$ free to float to their most likely values for each hypothesis.

\subsection{Limits on a Signal Model}\label{sec:CLs overview}

Having fixed our background model morphology, we now describe our statistical approach to setting limits on dark matter annihilation in M31, using the data in the entire search region. In order to maximize the statistical power in the morphology of the signal when setting limits, we use the $CL_s$ method \cite{2002JPhG...28.2693R} with pixel-level radio data and templates. 
 
 As in Section~\ref{sec:ring background}, we parameterize our signals using the cross section $\langle \sigma v\rangle$, and the parameter vector $\boldsymbol{\theta}$ which includes the dark matter mass $m_\chi$ and diffusion coefficient $D_0$. We will use the test statistic $\lambda_{\langle \sigma v\rangle, \boldsymbol{\theta}}$ from Eq.~\eqref{eq:TS main} to distinguish between background-like and signal plus background-like intensity maps. 

Statistical inference for a signal parameterized by $\langle \sigma v \rangle$ and $\boldsymbol{\theta}$ requires the probability distributions of $\lambda_{\langle \sigma v\rangle, \boldsymbol{\theta}}$ under our background-only and signal plus background hypotheses. We construct these probability distributions by generating an ensemble of simulated observations of M31 under each hypothesis and calculating $\lambda_{\langle \sigma v\rangle, \boldsymbol{\theta}}$ for each simulated observation. The simulated observations are generated under the background (signal plus background) hypothesis by superimposing $\hat{\Phi}^b$ ($\hat{\Phi}^{s+b}$) with randomly drawn noise maps.
The probability distribution from which the noise maps are drawn has correlations between nearby pixels, as expected due to the Gaussian beam of the observations (described in Section~\ref{sec:data}). More details on our procedure for producing pseudo-data are described in Appendix~\ref{sec:simulating intensity maps}.

\begin{figure*}
\centering
\subfigure[]{\label{fig:CL_hist_1} \includegraphics[width=0.9\columnwidth]{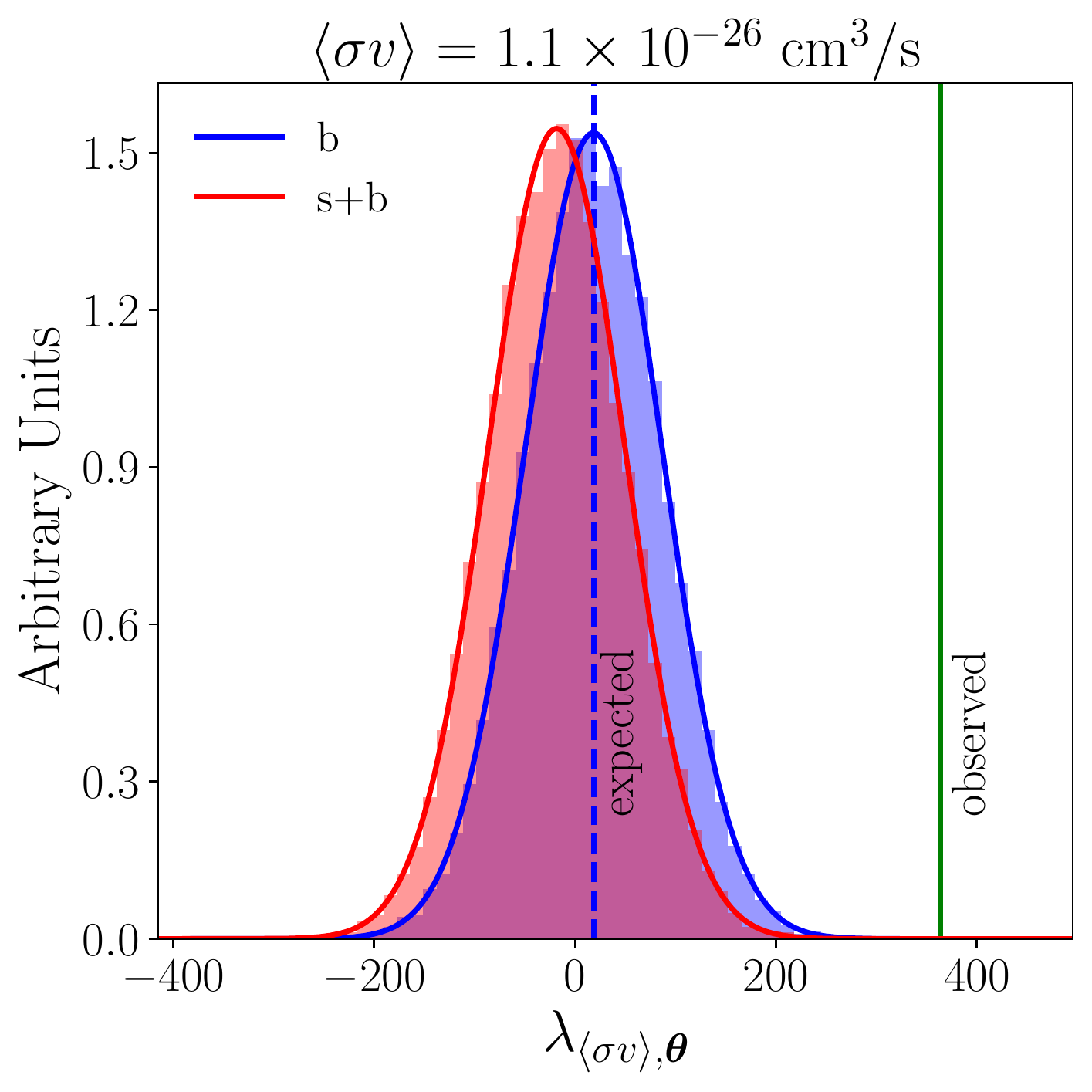}}
\subfigure[]{\label{fig:CL_hist_2} \includegraphics[width=0.9\columnwidth]{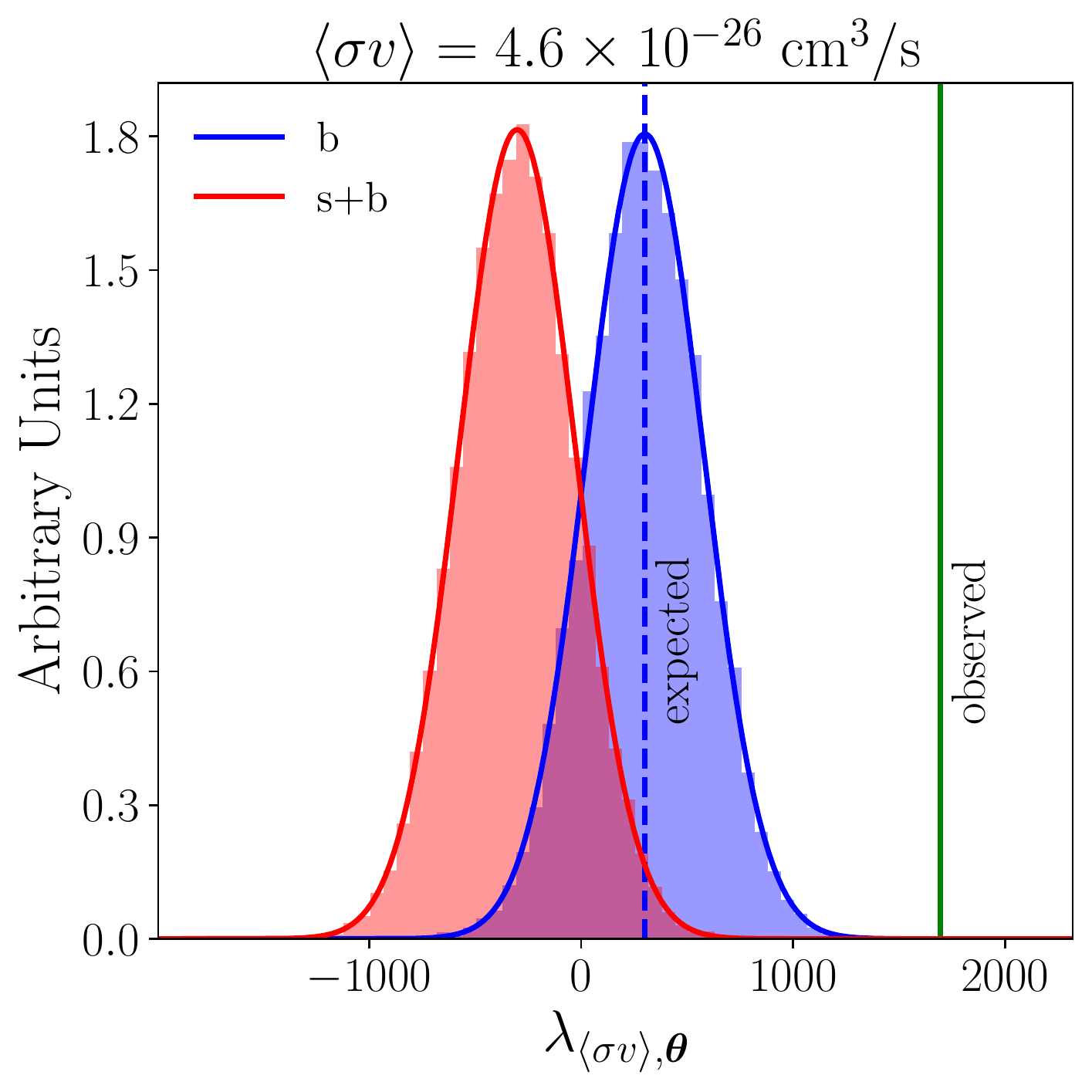}}
\subfigure[]{\label{fig:CL_curves_1} \includegraphics[width=0.9\columnwidth]{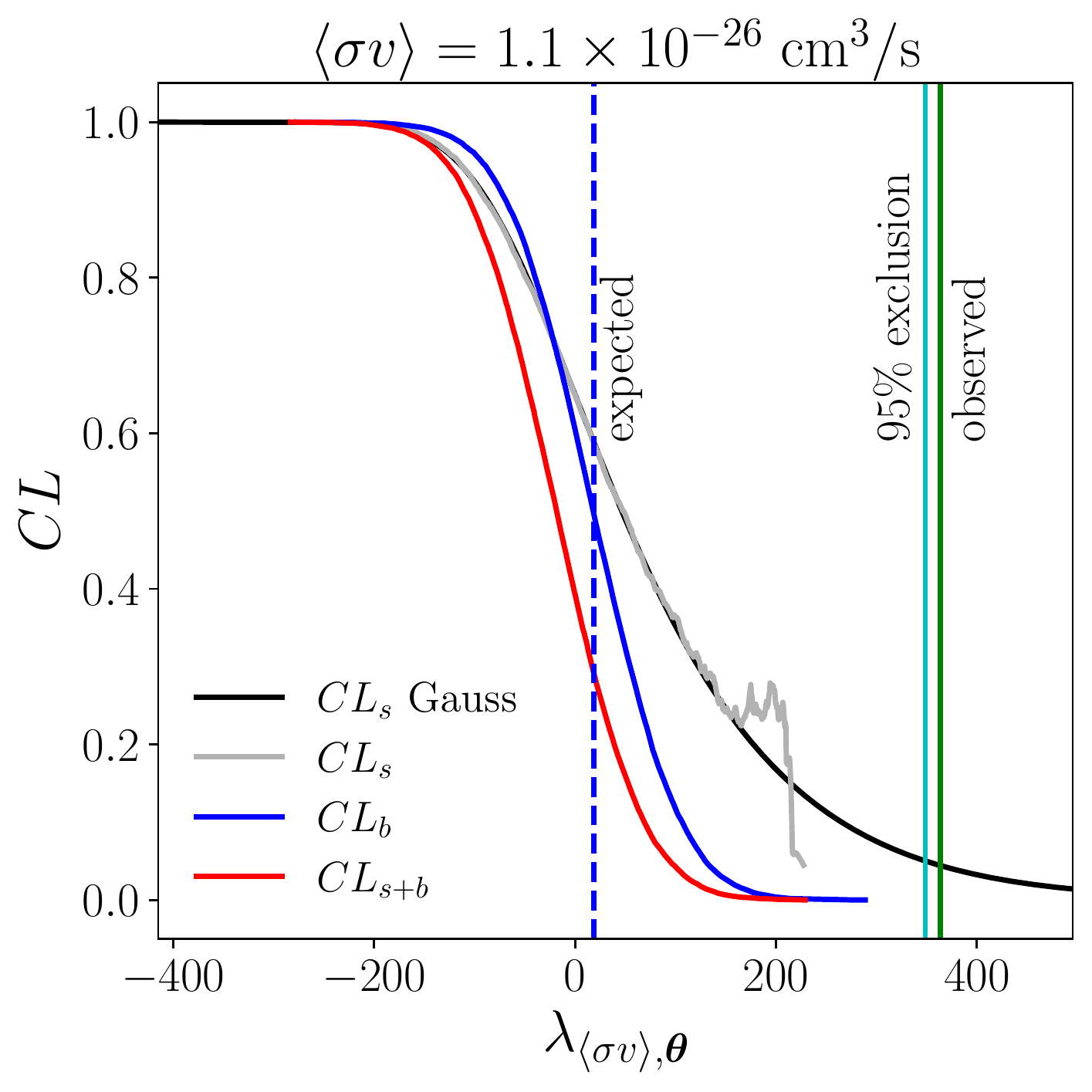}}
\subfigure[]{\label{fig:CL_curves_2} \includegraphics[width=0.9\columnwidth]{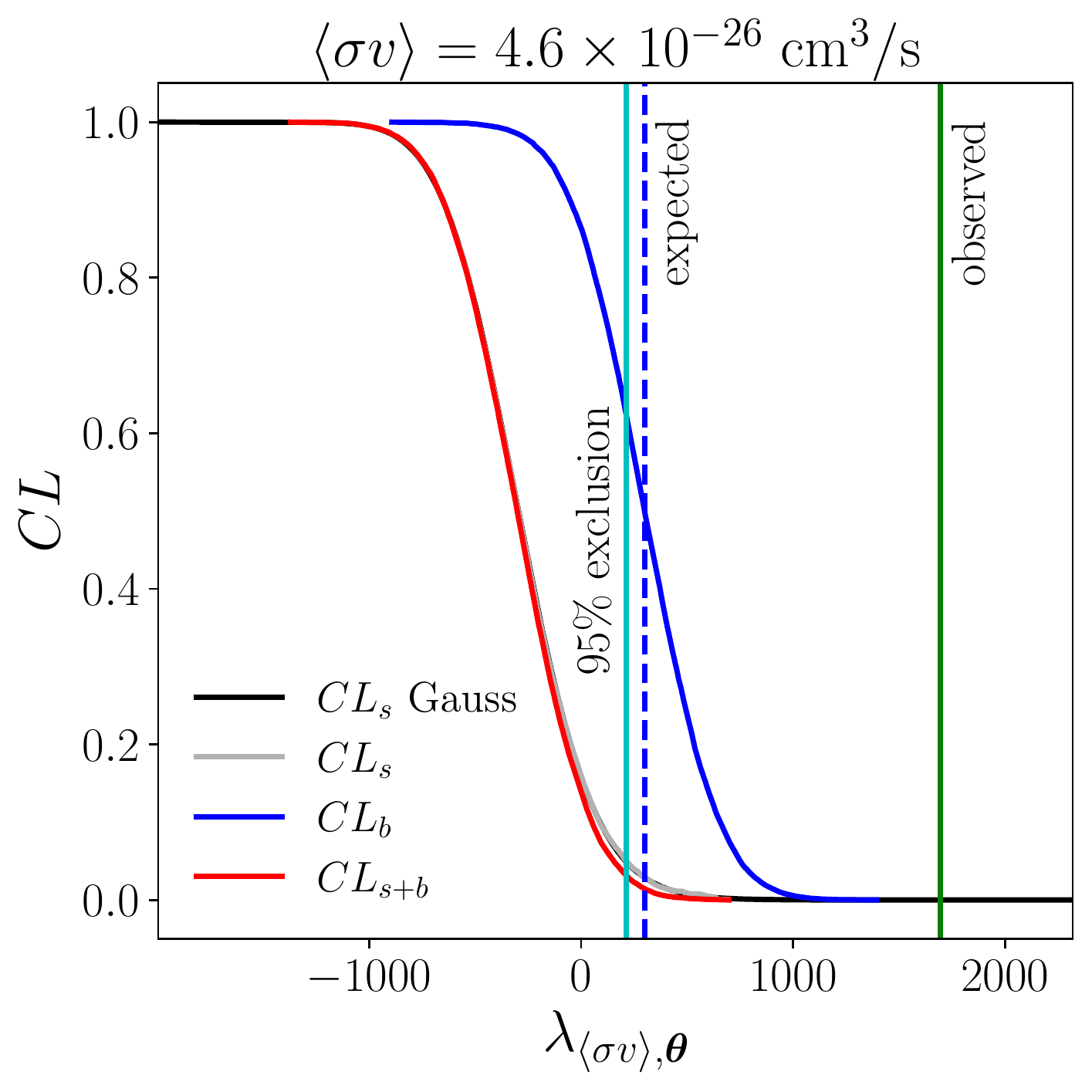}}
\caption{Top row: example histograms of $\lambda_{\langle \sigma v\rangle, \boldsymbol{\theta}}$ for background and signal plus background hypotheses for a signal parameterized by $m_\chi = 38.6 \unit{GeV}$ and $D_0 = 1 \times 10^{28} \unit{cm^2/s}$. Plot (a) has $\langle \sigma v \rangle = 1.1 \times 10^{-26} \unit{cm^3/s}$ and (b) has $\langle \sigma v \rangle = 4.6 \times 10^{-26} \unit{cm^3/s}$. The value of $\lambda_{\langle \sigma v\rangle, \boldsymbol{\theta}}$ for the data is shown with the vertical green lines and the median expected test statistics from the background only hypothesis are shown with the dashed blue vertical lines. Each distribution is constructed from $N = 20000$ independent simulated maps. The smooth curves are Gaussian approximations of the distributions. Bottom row: example $CL$ curves for the signal models shown above. The black curve is an approximation of $CL_s$ derived from the Gaussian approximations of the distributions of the test statistic. The test statistic that is excluded at 95\% confidence according to the Gaussian approximation of $CL_s$ is shown with a vertical cyan line.}
\label{fig:CL_results}
\end{figure*}

In Figure~\ref{fig:CL_results}, we show sample distributions of $\lambda_{\langle \sigma v\rangle, \boldsymbol{\theta}}$ for $m_\chi = 38.6\,\unit{GeV}$, $D_0 = 1 \times 10^{28}\,\unit{cm^2/s}$ and two choices of $\langle \sigma v \rangle$ ($1.1 \times 10^{-26}\unit{cm^3/s}$ and $4.6 \times 10^{-26} \unit{cm^3/s}$). In these examples, the blue and red histograms are the distributions of the test statistic assuming background and signal plus background (respectively) in arbitrary units. The solid curves are the Gaussian approximations of each distribution. 
The vertical green line in each plot is the value of $\lambda_{\langle \sigma v\rangle, \boldsymbol{\theta}}^{\rm(obs)} \equiv \lambda_{\langle \sigma v\rangle, \boldsymbol{\theta}}(\{d_i^{\rm (obs)}\})$, evaluated on the actual M31 radio data. 

For an arbitrary signal hypothesis parameterized by $\langle 
\sigma v \rangle$ and $\boldsymbol{\theta}$, our distributions of $\lambda_{\langle \sigma v\rangle, \boldsymbol{\theta}}$ can be used to approximate the probability distribution of $\lambda_{\langle \sigma v \rangle, \boldsymbol{\theta}}$ assuming background:
\begin{equation}
    p_{\langle \sigma v \rangle, \boldsymbol{\theta}}(\lambda|b) \equiv p(\lambda_{\langle \sigma v \rangle, \boldsymbol{\theta}} = \lambda|b),
\end{equation}
and signal plus background:
\begin{equation}
    p_{\langle \sigma v \rangle, \boldsymbol{\theta}}(\lambda|s+b, \langle \sigma v \rangle, \boldsymbol{\theta}) \equiv p\left(\lambda_{\langle \sigma v \rangle, \boldsymbol{\theta}} = \lambda|s+b, \langle \sigma v \rangle, \boldsymbol{\theta}\right).
\end{equation}
A given observation with test-statistic $\lambda^{\rm(obs)}_{\langle \sigma v \rangle, \boldsymbol{\theta}}$ then has a $CL_{b}$ value given by the probability of seeing data more background-like than observed, assuming the background-only hypothesis is correct:
\begin{equation}
CL_{b}\left(\lambda^{\rm(obs)}_{\langle \sigma v \rangle, \boldsymbol{\theta}}, \langle \sigma v \rangle, \boldsymbol{\theta}\right) = \int\displaylimits_{\lambda^{\rm(obs)}_{\langle \sigma v \rangle, \boldsymbol{\theta}}}^\infty d\lambda \; p_{\langle \sigma v \rangle, \boldsymbol{\theta}}(\lambda|b).
\end{equation}
Similarly, the $CL_{s+b}$ value for the observation is the probability of seeing a more background-like intensity map than that observed, assuming that the signal plus background hypothesis is correct:
\begin{equation} 
CL_{s+b}\left(\lambda^{\rm(obs)}_{\langle \sigma v \rangle, \boldsymbol{\theta}}, \langle \sigma v \rangle, \boldsymbol{\theta}\right) = \int\displaylimits_{\lambda^{\rm(obs)}_{\langle \sigma v \rangle, \boldsymbol{\theta}}}^\infty d\lambda \; p_{\langle \sigma v \rangle, \boldsymbol{\theta}}(\lambda|s+b, \langle \sigma v \rangle, \boldsymbol{\theta}).
\end{equation}
The ratio $CL_s\left(\lambda^{\rm(obs)}_{\langle \sigma v \rangle, \boldsymbol{\theta}}, \langle \sigma v \rangle, \boldsymbol{\theta}\right) \equiv CL_{s+b}/CL_b$ can then be interpreted as the probability of signal parameters greater than $\langle \sigma v\rangle$, given data. A 95\% confidence level exclusion therefore corresponds to a signal for which 
\begin{equation} \label{eq:CLs main}
    CL_s(\lambda^{\rm(obs)}_{\langle \sigma v \rangle, \boldsymbol{\theta}}, \langle \sigma v \rangle, \boldsymbol{\theta}) = \frac{CL_{s+b}(\lambda^{\rm(obs)}_{\langle \sigma v \rangle, \boldsymbol{\theta}}, \langle \sigma v \rangle, \boldsymbol{\theta})}{ CL_b(\lambda^{\rm(obs)}_{\langle \sigma v \rangle, \boldsymbol{\theta}}, \langle \sigma v \rangle, \boldsymbol{\theta})} = 0.05.
\end{equation}
The expected $95\%$ confidence limits correspond to signal parameters for which the median test statistic under the background hypothesis ($CL_{b} = 0.5$) leads to $CL_s = 0.05$. The $1$ and $2\sigma$ errors of this expected limit are calculated using the corresponding percentiles of the background distribution.

Example $CL$ curves are shown in Figures~\ref{fig:CL_curves_1} and \ref{fig:CL_curves_2} (corresponding to the distributions of the test statistics in Figures~\ref{fig:CL_hist_1} and \ref{fig:CL_hist_2}, respectively). The $CL_s$ curve of each plot (shown in grey) is dominated by statistical noise in the simulated intensity maps when $CL_b$ and $CL_{s}$ are small.
As seen in this example, we generically find that $\lambda^{\rm(obs)}_{\langle \sigma v \rangle, \boldsymbol{\theta}}$ is $5-6\sigma$ larger than the mean of the background-only distribution for our search region,\footnote{To obtain $CL$ values not dominated by the finite statistics in our simulated intensity maps would require $\sim 10^9$ maps. Instead, we set limits in this regime by extrapolating the test statistic distributions by fitting them to Gaussians and calculating an approximation of $CL_s$ using these extrapolations (shown in black in Figures~\ref{fig:CL_curves_1} and \ref{fig:CL_curves_2}).} 
implying that $CL_b$ of the observed test statistic is $\sim 2\times 10^{-8}$.

The fact that the {\it observed} test statistics are located on the tail of the $CL_b$ distributions is a consequence of observed radio intensities that are much less signal-like than the signal-free background model. We confirmed that this issue exists for all values of $m_\chi$ and $D_0$ studied in this work. This means that while the background model does not describe the observations well, the observed deviations away from the background-only model are not compatible with the morphology of any signal. This will lead to limits that are much stronger than expected. 

\begin{figure*}
\centering
\subfigure[]{\label{fig:b_sb_data_radial} \includegraphics[width=0.65\columnwidth]{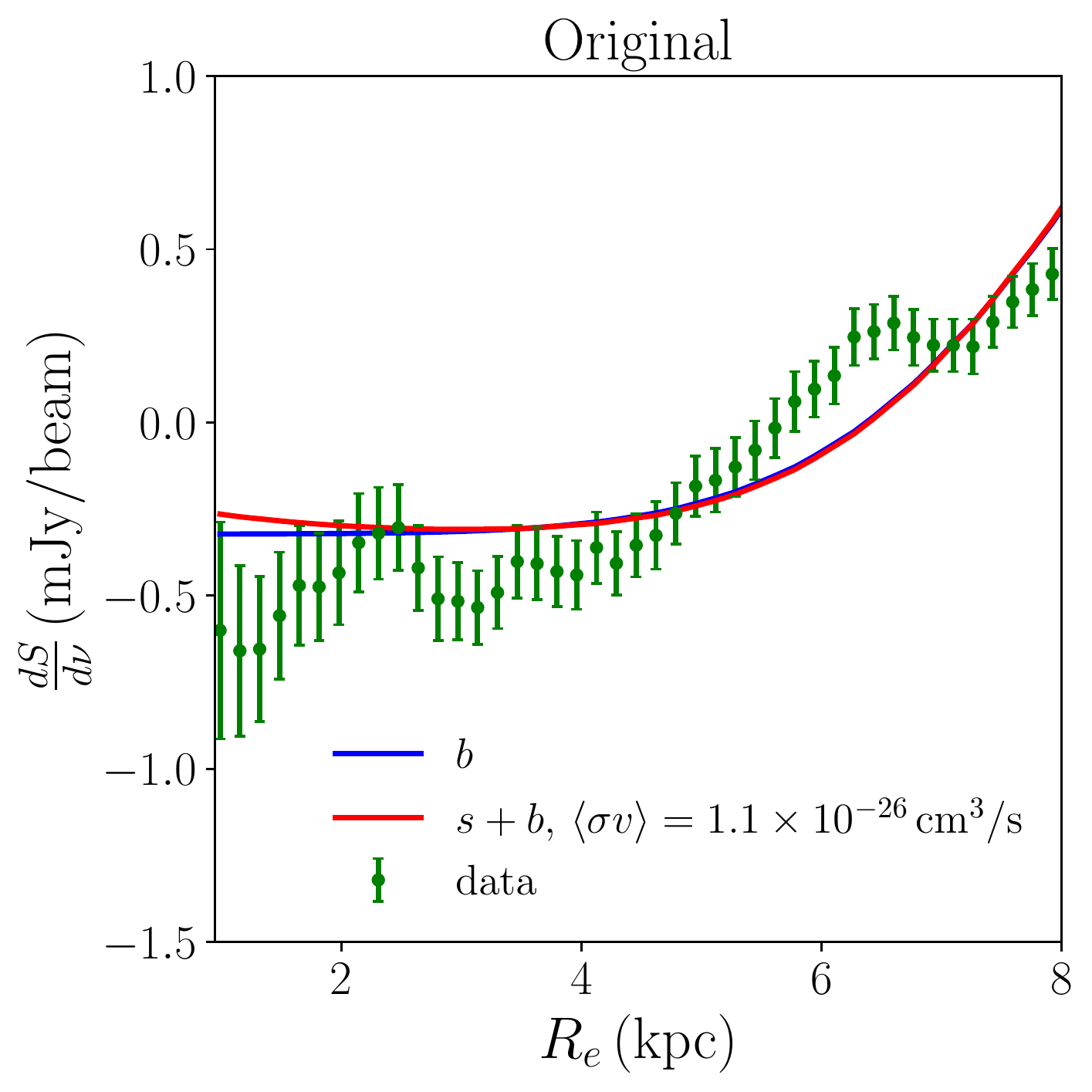}}
\subfigure[]{\label{fig:b_sb_data_radial__right} \includegraphics[width=0.65\columnwidth]{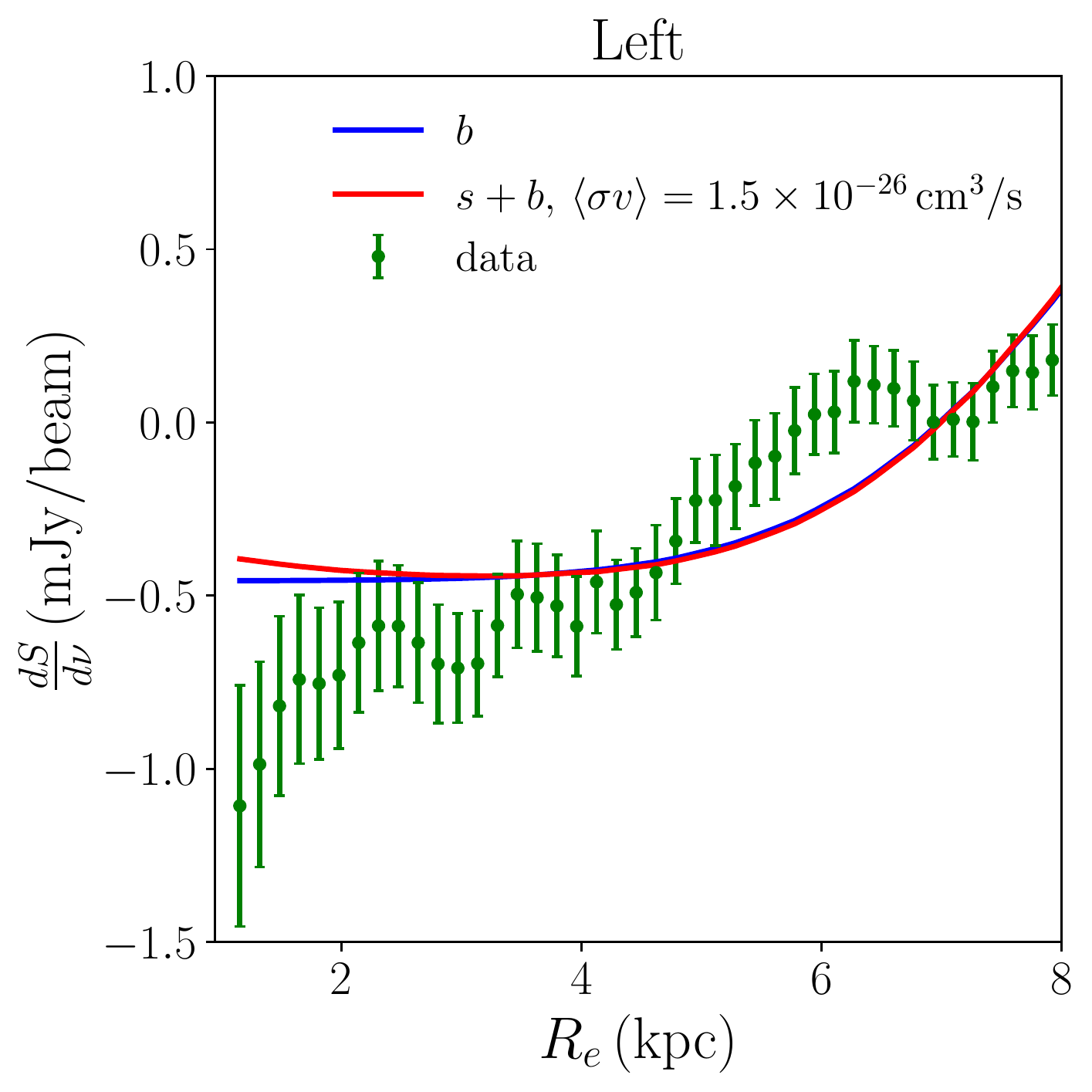}}
\subfigure[]{\label{fig:b_sb_data_radial__left} \includegraphics[width=0.65\columnwidth]{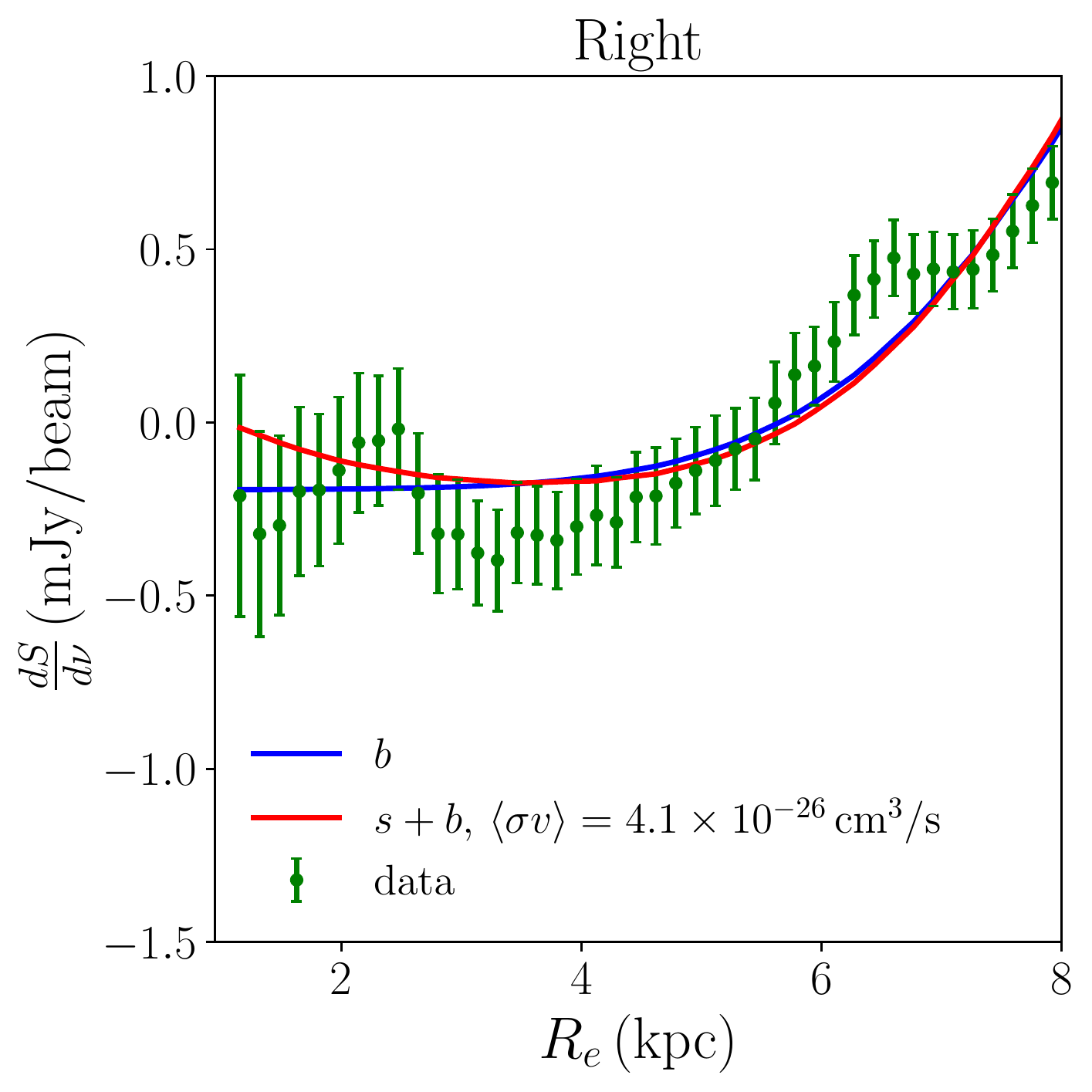}}
\caption{Radio Flux averaged over concentric elliptical annuli as function of $R_e(\boldsymbol{x}, \mu_1)$ (with $\mu_1$ given by the global fit of the Full Map analysis) along with the best fit background model and the excluded signal plus background model for (a) the original search region, (b) the left half search region, and (c) the right half search region. All signal plus background models shown have $m_\chi=38.6 \unit{GeV}$ and $D_0 = 1 \times 10^{28} \unit{cm^2/s}$. For the excluded signal plus background model, we take the lowest value of $\langle \sigma v \rangle$ that leads to $95 \%$ exclusion for the values of mass and diffusion normalization plotted.}
\label{fig:radial_distributions}
\end{figure*}

This issue requires further investigation. In Figure~\ref{fig:b_sb_data_radial}, we show our best fit background model and signal plus background model with $\langle \sigma v \rangle = 1.1 \times 10^{-26} \unit{cm^3/s}$ for $m_\chi = 38.6 \unit{GeV}$ and $D_0 = 1 \times 10^{28}\unit{cm^2/s}$ along with the data in the search region as a function of elliptical distance from the center of the map (note that this choice of signal parameters is excluded at $95\%$ confidence).\footnote{To produce this plot, we compute the average flux per beam in pixels in the search region in concentric elliptical annuli with the same eccentricity as the  globally fit ring model.} As can be seen, the residuals of the data with respect to the background-only model are negative for $R_e\lesssim 4 \unit{kpc}$, but the signal plus background model predicts a larger flux than the background-only model in this region. 

In Figures~\ref{fig:b_sb_data_radial__right} and \ref{fig:b_sb_data_radial__left} we show elliptically averaged radio emission as well as best fit models in the left search region (masking the $x>0$ data) and right search region (masking $x<0$), respectively. As can be seen, the large negative excursion in the left search region is the source of the negative residuals we identified in Fig.~\ref{fig:b_sb_data_radial} and can be traced specifically to the large negative observed flux located around $(x,y)\sim(-2,1)\,\unit{kpc}$. This negative flux is clearly visible in Figures~\ref{fig:intensity_map} and \ref{fig:data_basemask}. The low flux measurement (well in excess of a $2\sigma$ deviation given the expected measurement errors) may be due to the over-subtraction of a point source external to M31.

Critically, given our understanding of the dark matter distribution within M31, dark matter emission cannot create such a region of low emission close to the center of M31, even if the overall baseline of zero radio flux was mismeasured. Thus, considering only the part of the data without this region of anomalously low emission will not set overly-optimistic limits on dark matter annihilation. Indeed, we will see in the next section that it sets more conservative and weaker bounds.

\section{Constraints on Annihilating Dark Matter in M31} \label{sec:limits}

\begin{figure}
\includegraphics[width=0.9\columnwidth]{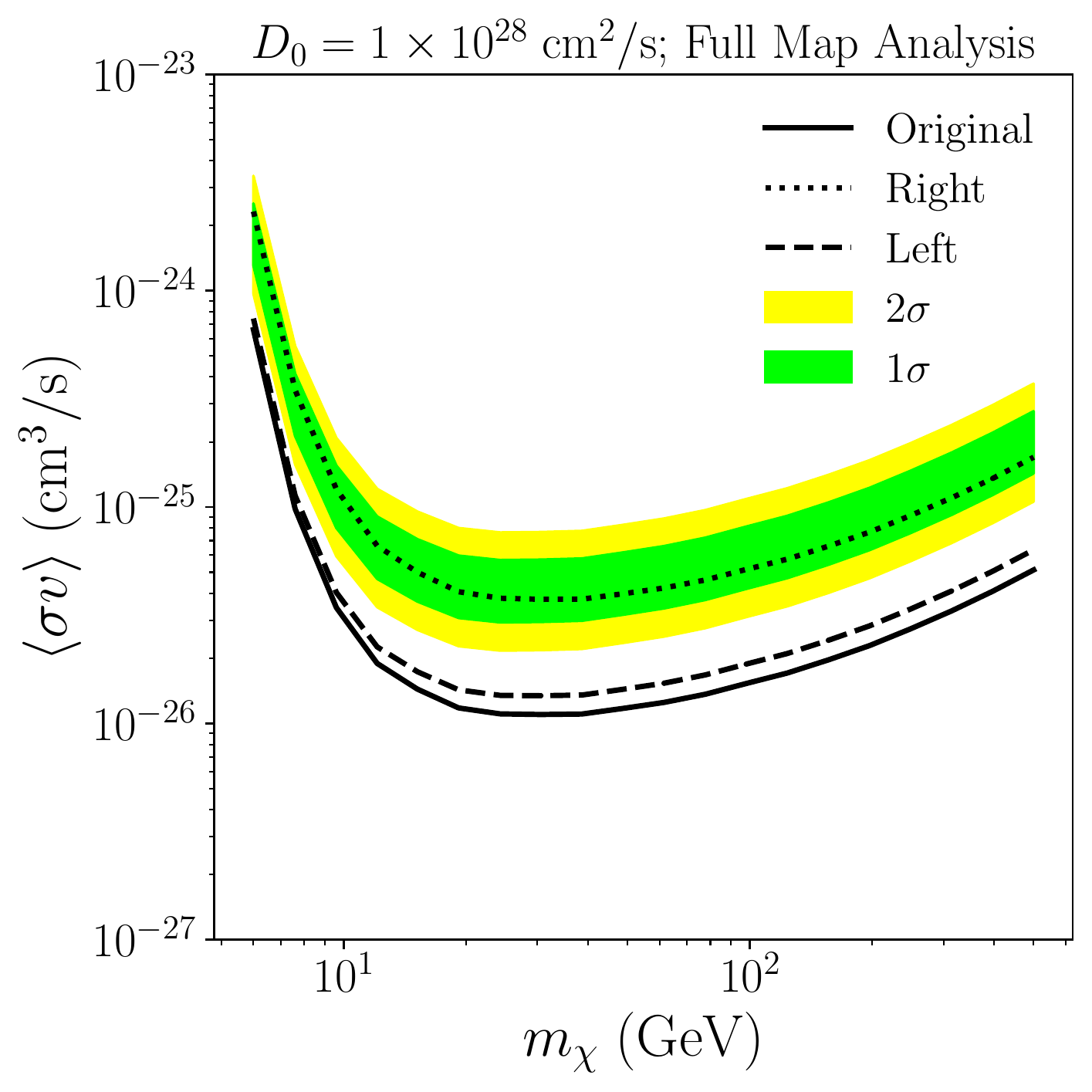}
\caption{95\% confidence limits on dark matter annihilation assuming $D_0 = 1 \times 10^{28}\unit{cm^2/s}$ from our Full Map analysis. The $1\sigma$ and $2\sigma$ expected limits from the original search region are shown in green and yellow, with the observed limits for this search region shown with a solid line (labeled ``original''). The dotted and dashed lines are the actual limits from the data in the right ($x>0$) and left ($x<0$) half of the search region, respectively.}
\label{fig:limits_weighted_dl}
\end{figure}

 In light of the findings from last section, we first calculate limits for the original search region (within the black contour in Fig.~\ref{fig:data_basemask}) and compare them to the limits from the right and left search regions. We set $95\%\,CL_s$ exclusion limits for dark matter with mass in the range $[6-500]\,\unit{GeV}$, annihilating to $b \bar{b}$. In Figure~\ref{fig:limits_weighted_dl} we show the limits on $\langle \sigma v\rangle$ as a function of $m_\chi$, assuming the default diffusion parameter normalization $D_0 = 1 \times 10^{28}\unit{cm^2/s}$. This includes the observed limits for the original, right and left search regions with black curves of various styles, alongside the $1$ and $2\sigma$ variation around the expected limits for the original search region. As expected, the observed limit for the original search region is far stronger than the $2\sigma$ variation assuming the background hypothesis. The limits from the left search region are nearly as strong as those from the original search region while the limits from the right search region are approximately as strong as expected, confirming that the stronger-than-expected limits come entirely from the left-hand side of the M31 emission, where the region of negative flux is located.

\begin{figure}
    \centering
    \includegraphics[width=\columnwidth]{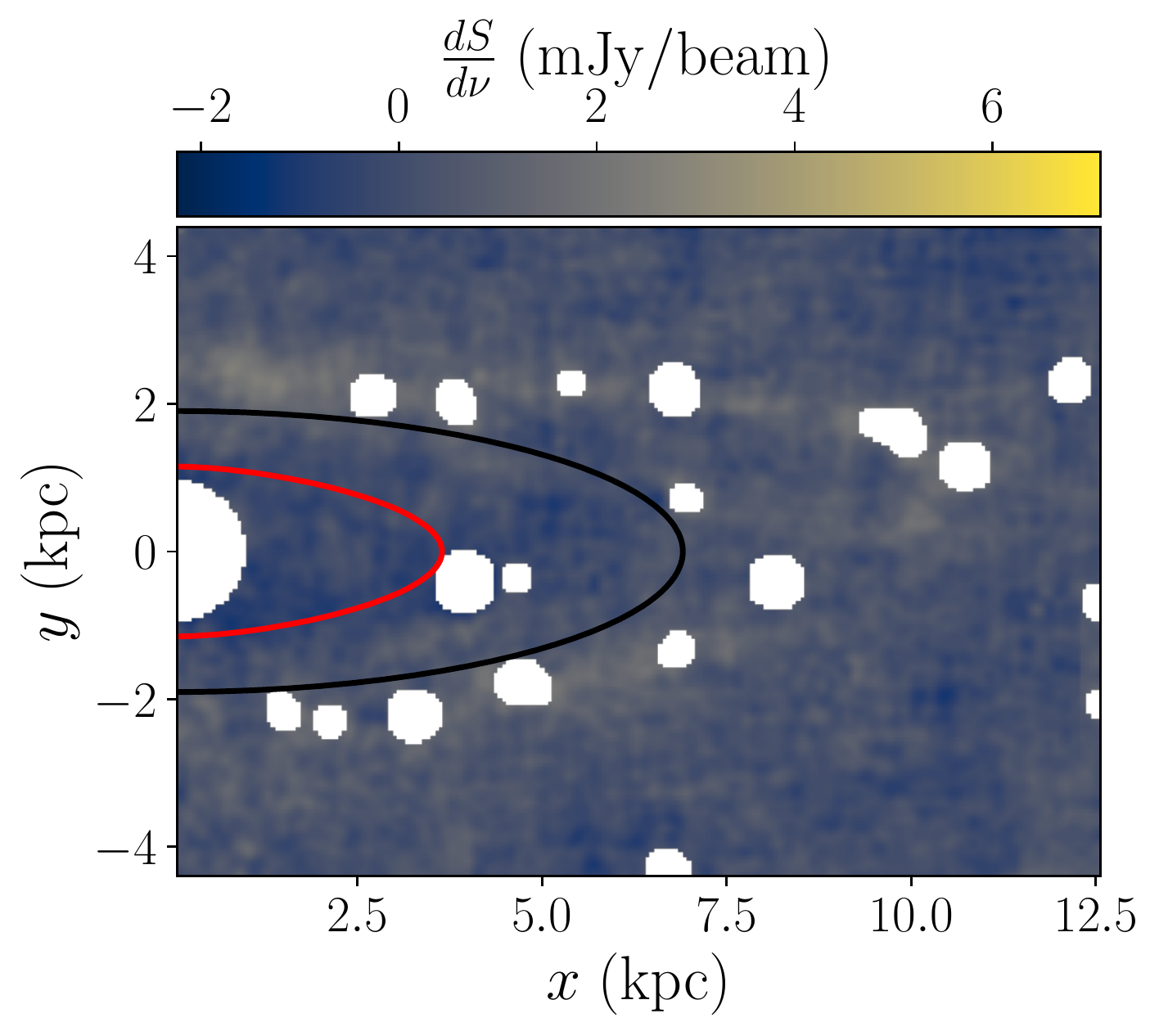}
    \caption{As Figure~\ref{fig:data_basemask} but for the right-only analysis. The search region contour is recalculated with the left side of the image masked, and thus differs slightly from the search region of the full map analysis.}
    \label{fig:data_basemask_left}
\end{figure}

\begin{figure}
    \centering
    \includegraphics[width=0.9\columnwidth]{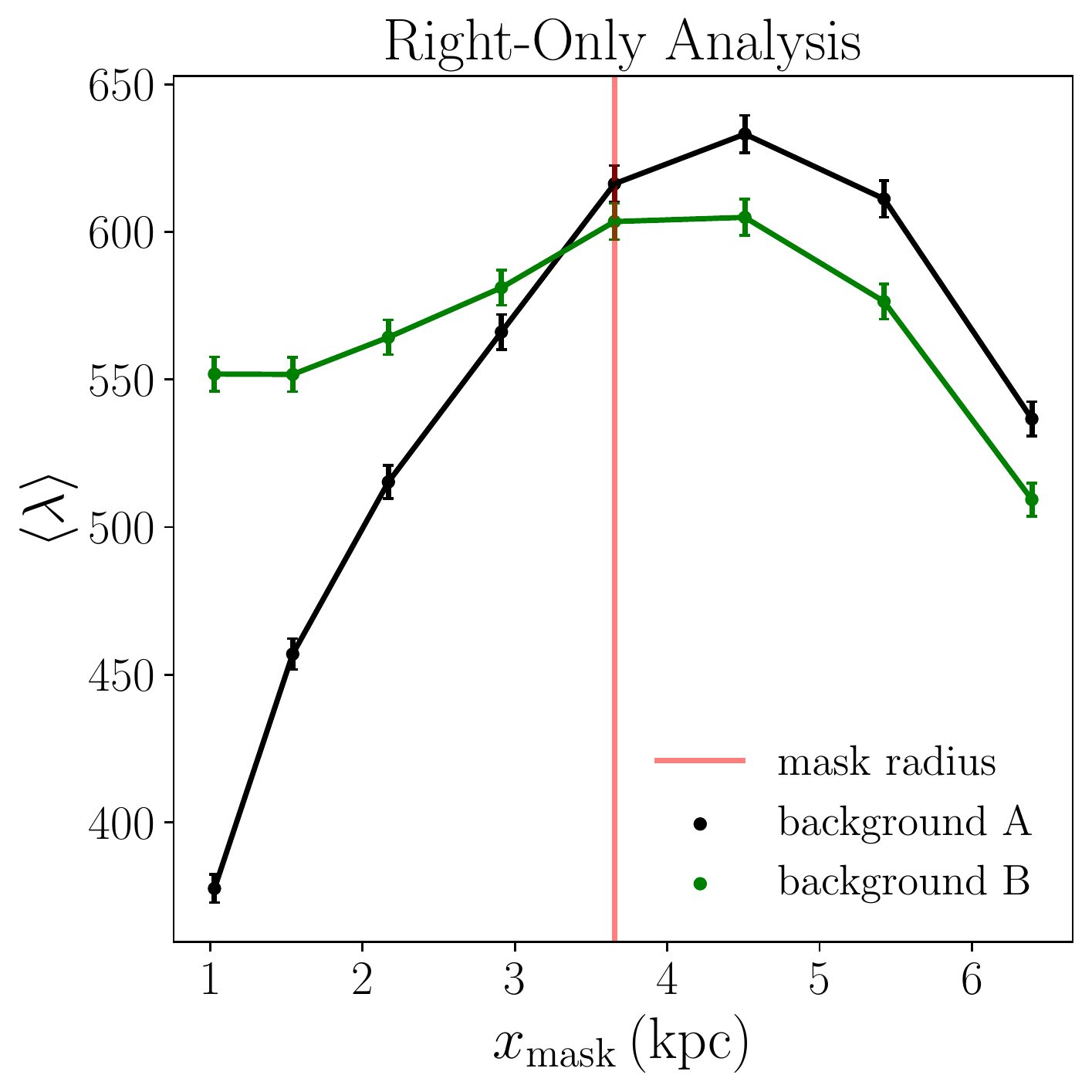}
    \caption{As Figure~\ref{fig:mean_ts}, but for our analysis with the left side of the data masked.}
    \label{fig:mean_ts_left}
\end{figure}

To get the most accurate limits using only the right side of the map, we recalculate the search region and re-fix the background morphological parameters with the left side of the data masked, using the steps described in Sections~\ref{sec:masking} and \ref{sec:ring background}. The resulting morphological parameter values used to recalculate the search region are shown in the ``Right-Only Analysis" section of Table~\ref{tab:annulusparameters} in the second column (labeled ``Global Fit"). The new search region is bounded by the black contour in Figure~\ref{fig:data_basemask_left}. We show the resulting test statistics as a function of candidate signal region mask size for background Fits A and B in Figure~\ref{fig:mean_ts_left}. Based on this, we select the signal region mask for the right-only analysis which intersects the semi-major axis at $x=3.65 \unit{kpc}$, where the mean test statistics from Fits A and B agree within statistical error. This turns out to be the same signal region mask that we found for our full map analysis. This signal region mask is bounded by the red contour in Figure~\ref{fig:data_basemask_left}. We fix the morphological parameters of the background model $\boldsymbol{\mu}$ to the best fit values from background model B with this signal region mask. The resulting values for the morphological parameters are shown in the third column (labeled ``Signal-Region Masked") of the second half (labeled ``Right-Only Analysis") of Table~\ref{tab:annulusparameters}.

\begin{figure*}
\centering
\subfigure[]{\label{fig:CL_hist_left_left_2} \includegraphics[width=0.9\columnwidth]{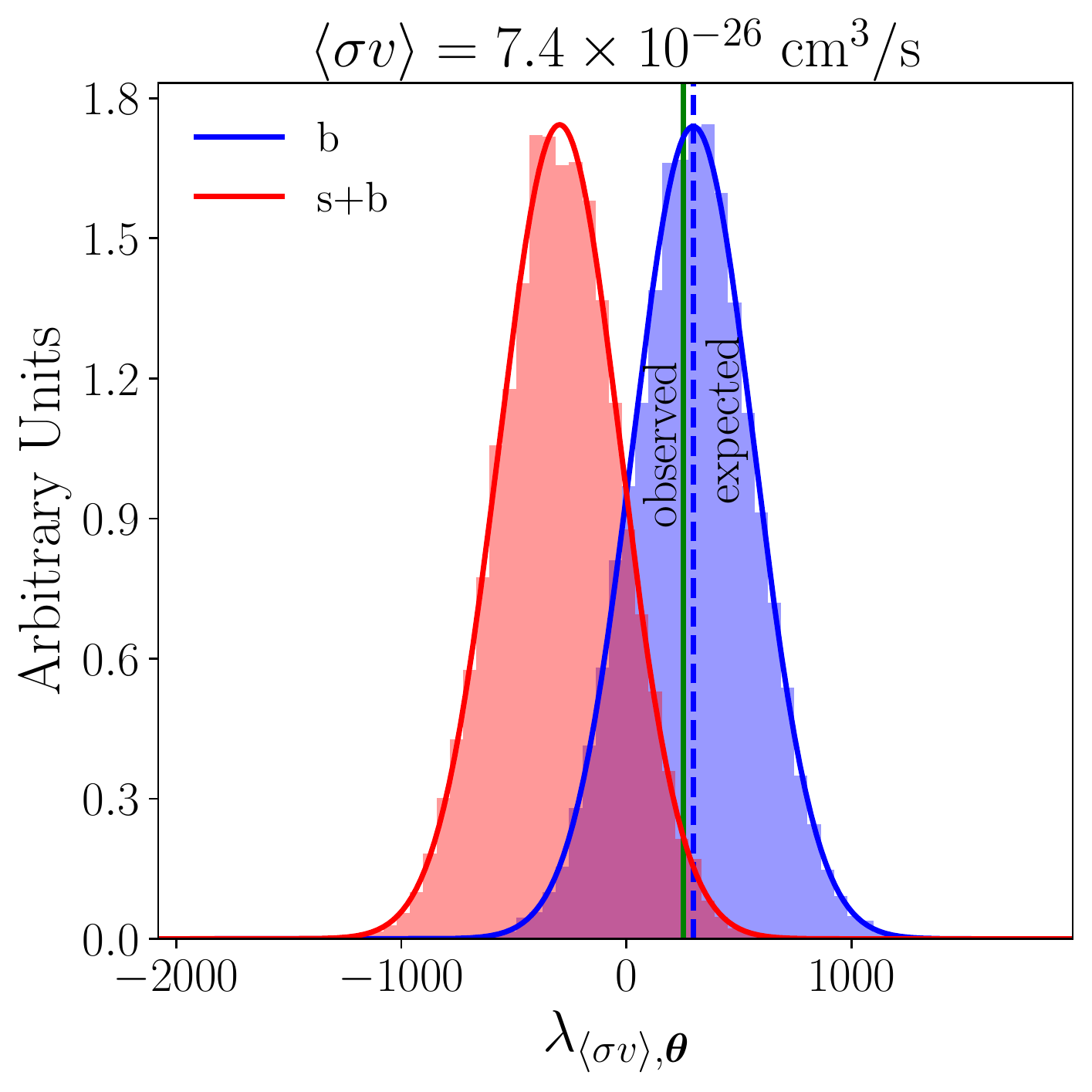}}
\subfigure[]{\label{fig:CL_curves_left_left_2} \includegraphics[width=0.9\columnwidth]{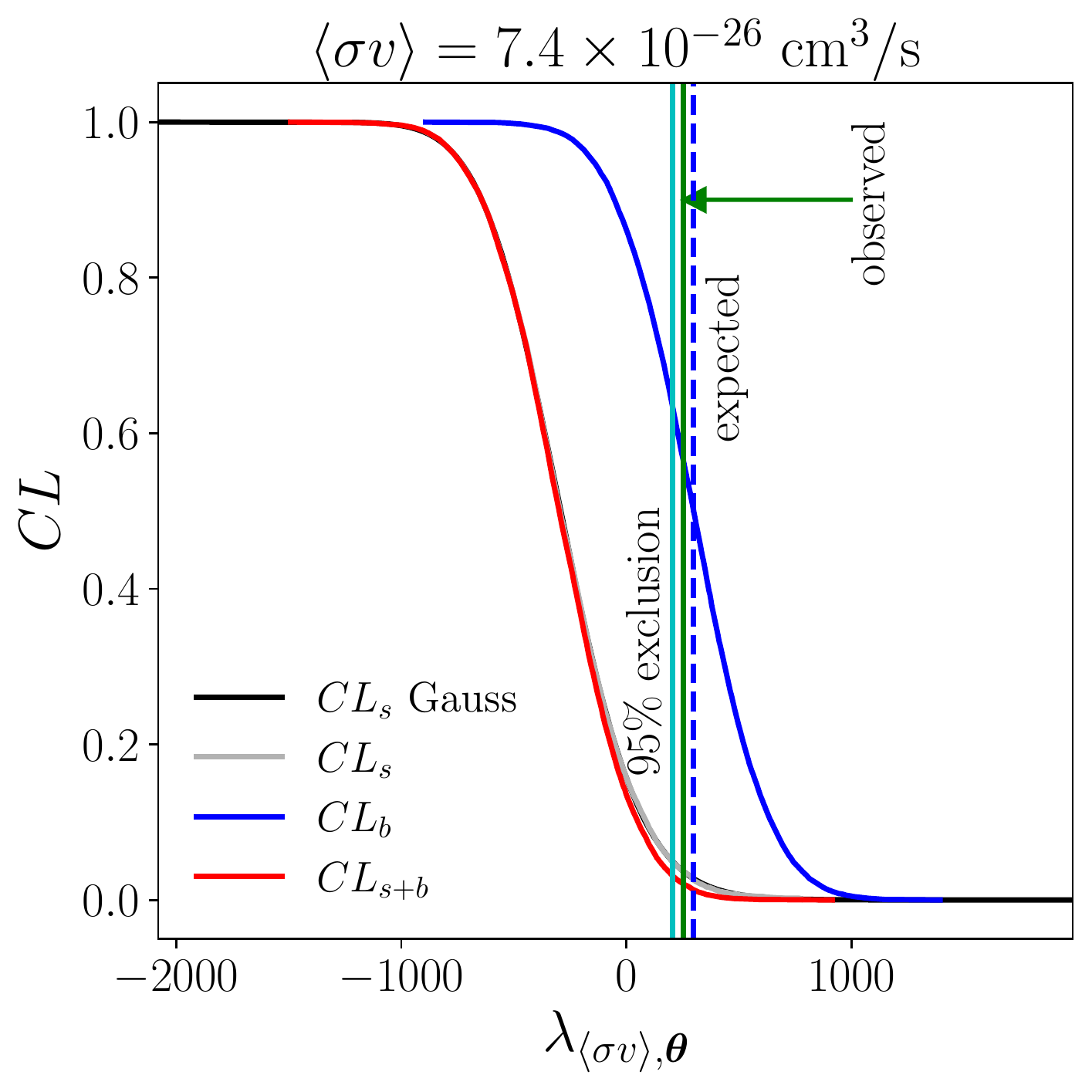}}
\caption{Same as Figure~\ref{fig:CL_results} but for the right-only analysis. The cross-section shown here is close to the expected and actual $95\%$ confidence limit. The expected limit is almost the same as the actual limit since the test statistic from the data is very close to the $50^{\rm th}$ percentile test statistic from background pseudo-data.}
\label{fig:sample_TS_dists_left}
\end{figure*}

Using this background model for the right-only analysis, we show in Figures~\ref{fig:CL_hist_left_left_2} and \ref{fig:CL_curves_left_left_2} examples of the test statistic distributions and $CL$ curves for pseudo-data in the right search region and show sample distributions, for $m_\chi = 38.6 \unit{GeV}$, $D_0 = 1 \times 10^{28} \unit{cm^2/s}$, and $\langle \sigma v \rangle = 7.4 \times 10^{-26} \unit{cm^3/s}$. This is the cross section that is excluded at approximately 95\% confidence for these values of $m_\chi$ and $D_0$.

\begin{figure*}
\centering
\subfigure[]{\label{fig:limits_varying_sphavg} \includegraphics[width=\columnwidth]{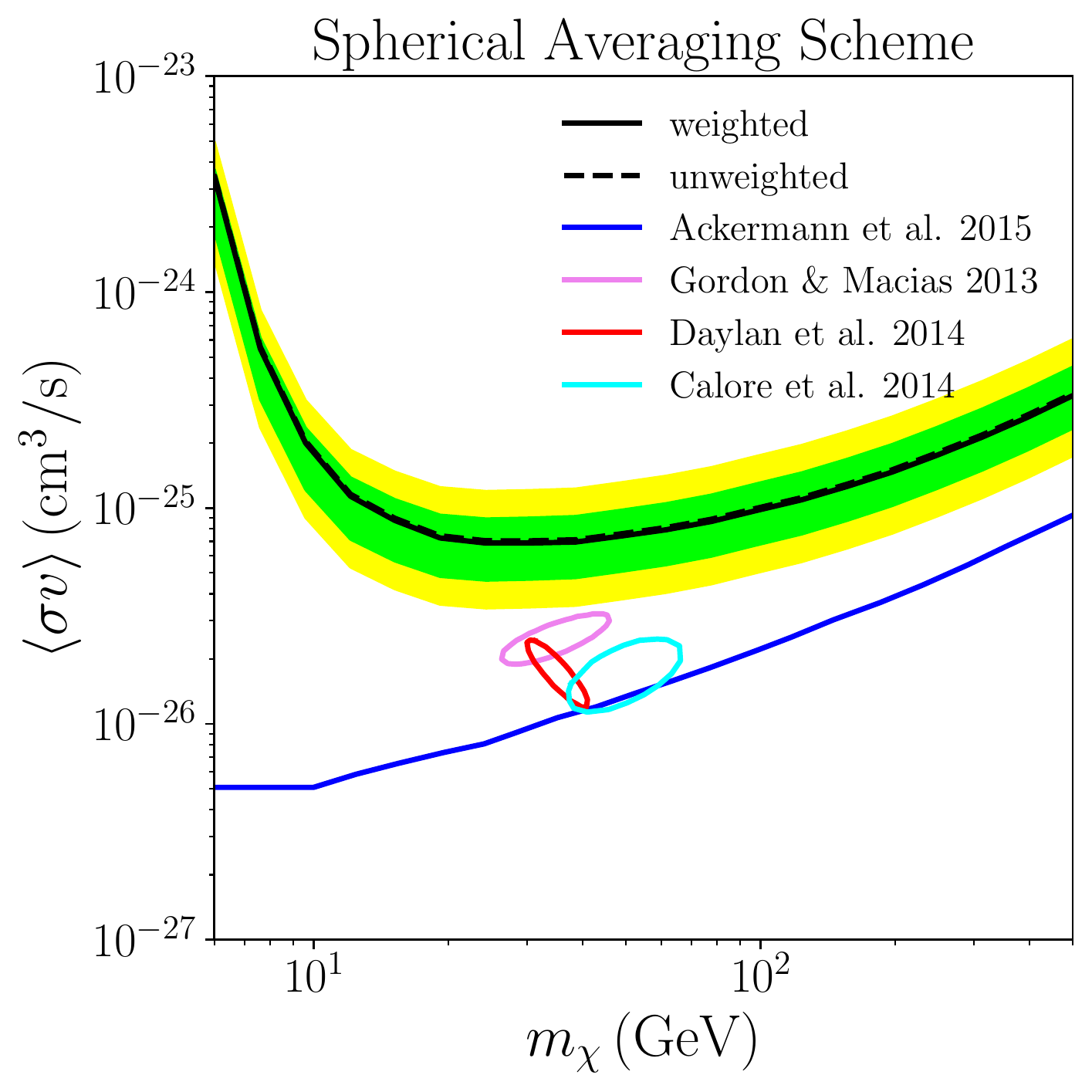}}
\subfigure[]{\label{fig:limits_weighted_dl_left_left_varying_D0} \includegraphics[width=\columnwidth]{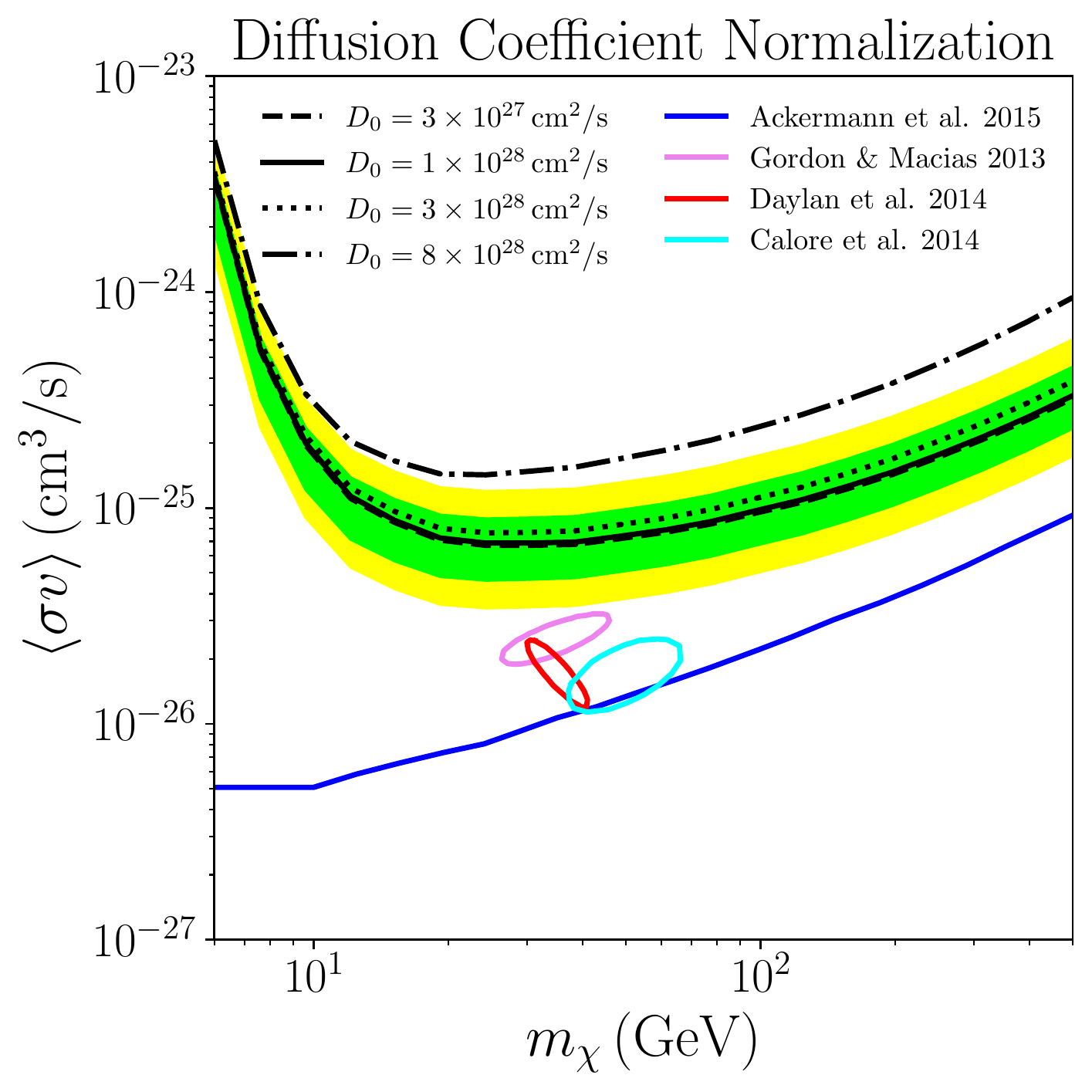}}
\caption{Expected and actual 95\% confidence limits from the right-only analysis, using the data from the search region shown in Figure~\ref{fig:data_basemask_left}. 
The two panels show the variation of the observed limits due to (a) changes in the averaging procedure (introduced in Section~\ref{sec:solving diff loss}) for our default diffusion coefficient normalization and (b) changes in the diffusion coefficient normalization, $D_0$ for our weighed averaging scheme. Both panels have the expected limits obtained using the default value of $D_0$ ($1\times 10^{28}\unit{cm^2/s}$) and the weighted averaging scheme. The contours show best fits to the GCE \cite{2016PDU....12....1D, 2015JCAP...03..038C, 2013PhRvD..88h3521G} and the solid blue lines show limits from dwarfs using {\it Fermi} Pass 8 data \cite{2015PhRvL.115w1301A}.}
\label{fig:varying assumptions}
\end{figure*}

In Figure~\ref{fig:varying assumptions}, we show our limits on $\langle \sigma v\rangle$ as a function of $m_\chi$, using the right-only analysis. These results constitute our most conservative and robust limits on dark matter annihilation using the radio observations of M31. In both panels, the green and yellow bands quantify the $1$ and $2\sigma$ statistical error of our expected limits for our default value of $D_0$ and for the weighted averaging scheme, introduced in Section~\ref{sec:solving diff loss} (and otherwise default parameters). 
Each panel quantifies the effects of different systematics on our results. In Figure~\ref{fig:limits_varying_sphavg} we show the observed $95\%$ confidence exclusion limit from each spherically averaging procedure for our default value of $D_0$. As can be seen, the limits do not depend on the averaging scheme used.
In Figure~\ref{fig:limits_weighted_dl_left_left_varying_D0}, we show the observed limits as $D_0$ is varied. The limits are relatively insensitive to variations of the diffusion coefficient in the range $3\times 10^{27} \unit{cm^2/s} \leq D_0\leq 3 \times 10^{28} \unit{cm^2/s}$. The limits become about a factor of three weaker when $D_0$ changes from $3 \times 10^{28}\unit{cm^2/s}$ to $8 \times 10^{28}\unit{cm^2/s}$ for $m_\chi\gtrsim 10 \unit{GeV}$. We also show best fit contours to the GCE from previous analyses \cite{2016PDU....12....1D, 2015JCAP...03..038C, 2013PhRvD..88h3521G}, as well as limits from Milky Way dwarfs using {\it Fermi} Pass 8 data \cite{2015PhRvL.115w1301A}. Our limits do not exclude parameters that fit the GCE. Although our limits are weaker than the dwarf limits, they are robust to astrophysical uncertainties.

\section{Conclusion} \label{sec:C}

In this work, we have set robust and conservative limits on dark matter annihilation to $b\bar{b}$ using the $8.35\,\unit{GHz}$ Effelsberg radio map of M31, which is sensitive to the predicted synchrotron emission of the $e^\pm$ produced in the cascade $b$ decays. These limits are based on a numeric solution to the diffusion-loss equation that accommodates non-uniform parameters, and an astrophysical model that uses observations of the gas, dust, starlight, and magnetic fields of M31. Our ISRF model for the starlight is derived directly from \textit{ugriz} luminosity data, which led to notably larger values for $\rho_*$ in the center of M31 compared to previous works.

Unlike previous studies, our numerical solution to the diffusion-loss equation allows for position dependent diffusion and loss coefficients. Our method still requires spherical averaging of the background model. Though we have shown that our final limits are insensitive to the averaging procedure, additional work is needed to develop a numeric solution which is adapted to the axisymmetry of M31.

Our limits are based on the morphology of the observed flux, and our results are independent of the true zero-flux level of the intensity map. The limits are based on the radio flux only interior to the bright ring of radio emission in M31, allowing us to use data-driven models of the background based on signal-poor regions of the observed intensity map. Due to a localized anomaly of low radio flux in the search region, we choose conservatively to select only the half of the dataset without this anomaly with which to set our limits. 

Our limits on dark matter annihilation in M31 do not exclude best-fit models to the GCE (shown as contours in Figures~\ref{fig:limits_varying_sphavg}\& \ref{fig:limits_weighted_dl_left_left_varying_D0}) and are weaker than those found in previous radio studies \cite{2013PhRvD..88b3504E, 2016PhRvD..94b3507C, 2021MNRAS.501.5692C, 2022PhRvD.106b3023E}. 
These weaker limits are due in part to the fact that in our analysis we mask the center of the galaxy, where the signal intensity is maximum -- however, this choice minimizes the sensitivity to unknown astrophysical parameters at the galactic center. The weaker limits are also likely due to the differences in our astrophysical model of M31 compared to previous work. In particular, the core of M31 is much more luminous in starlight than a simple scaling of the comparable region of the Milky Way would suggest. This increased starlight flux results in increased energy losses of $e^\pm$ into X-rays through inverse Compton scattering, reducing the flux of dark matter-induced radio waves. Though beyond the scope of this work, this suggests that an analysis of constraints from X-ray emission in the center of M31 from dark matter annihilation may set interesting limits.

The sensitivity of the limits to the astrophysical conditions within M31 are notable; though in this work we have taken care to construct an accurate model of M31 based on observations, future measurements and astronomical input would likely improve the model and the resulting limits. Similar analysis is likely necessary for constraints on dark matter annihilation via radio waves in other systems beyond M31.   

\section*{Acknowledgements}

This work was supported by DOE grant DOE-SC0010008. We thank Andrew Baker for helpful advice and discussion. We also thank the authors of Ref.~\cite{2020A&A...633A...5B},  for providing the data for our analysis.

\appendix

\section{Solving the Diffusion Equation through the Method of Backwards Differences}\label{sec:numerical method}

In this Appendix, we describe our numeric method for solving for the spherically averaged electron phase space density that satisfies Eq.~\eqref{eq:sph sym diff loss}. Using forward differences, the large time-steps required to numerically solve the diffusion-loss equation over the relevant timescales of M31 result in unstable solutions. Backward differences, on the other hand, are unconditionally stable \cite{1992nrfa.book.....P}. Since we are only interested in the equilibrium solution and not the details of the approach to equilibrium, we use backward differences with time-steps large enough that the solution converges only after two time steps. 

It is more convenient to work with $u\equiv r \langle f_e\rangle$, which converts to Eq.~\eqref{eq:sph sym diff loss} to Eq.~\eqref{eq:diff loss eq u}.
The discretized form of Eq.~\eqref{eq:diff loss eq u} with backward differences is
\begin{widetext}
\begin{equation}\label{eq:diff loss finite difference}
\begin{split}
\frac{u_{ij}^{n+1} - u_{ij}^n}{\Delta t} =& D(r_i, E_j)\frac{u_{i+1, j}^{n+1}-2u_{ij}^{n+1} + u_{i-1, j}^{n+1}}{\Delta r^2} + \left.\frac{\partial D}{\partial r}\right|_{r_i, E_j} \left[\frac{u_{i+1, j}^{n+1}-u_{i-1, j}^{n+1}}{2\Delta r} - \frac{u_{ij}^{n+1}}{r}\right] \\
&+ \frac{b(r_i, E_{j+1})u_{i, j+1}^{n+1}-b(r_i, E_{j})u_{ij}^{n+1}}{\Delta E_i} + r_iQ_e(r_i),
\end{split}
\end{equation}
\end{widetext}
where $u_{ij}^n = u(r_i, E_j, t_n)$ and $\Delta t$,  $\Delta r$ and $\Delta E_i=E_{i+1}-E_i$ are the grid spacings for each coordinate. We use $n_E=400$ logarithmically spaced steps for $E$ and $n_r=800$ linearly spaced steps for $r$. 

Combining all terms from Eq.~\eqref{eq:diff loss finite difference} evaluated at time-step $t_{n+1}$ gives
\begin{equation}\label{eq:diff loss finite organized}
\left[\delta_{ik}\delta_{jl} -A_{ik}(E_j)\delta_{jl}-\delta_{ik}B_{jl}(r_i)\right]u_{kl}^{n+1} = u_{ij}^{n}+C(r_i, E_j).
\end{equation}
Here, $A$ and $B$ are given by
\begin{widetext}
\begin{equation}\label{eq:A and B}
\begin{split}
A(E_j) &= 
\begin{pmatrix}
\alpha_0(r_1,E_j) & \alpha_1(r_1,E_j) & 0 & 0 & \dots & 0 & 0 & 0 \\
\alpha_{-1}(r_2,E_j) & \alpha_0(r_2,E_j) & \alpha_1(r_2,E_j) & 0 & \dots & 0 & 0 & 0 \\
0 & \alpha_{-1}(r_3,E_j) & \alpha_0(r_3,E_j) & \alpha_1(r_3,E_j) & \dots & 0 & 0 & 0 \\
\vdots & \vdots & \vdots & \vdots & \ddots &\vdots & \vdots & \vdots\\
0 & 0 & 0 & 0 & \dots &\alpha_{-1}(r_{{n_r}-1},E_j)&\alpha_0(r_{{n_r}-1},E_j) & \alpha_1(r_{{n_r}-1},E_j)\\
0 & 0 & 0 & 0 & \dots & 0 & \alpha_{-1}(r_{n_r},E_j)& \alpha_0(r_{n_r},E_j)
\end{pmatrix},
\end{split}
\end{equation}
\begin{equation}
\begin{split}
B(r_i) &=
\begin{pmatrix}
\beta_0(r_i,E_1) & \beta_1(r_i,E_1) & 0 & 0 & \dots & 0 & 0 & 0 \\
\beta_{-1}(r_i,E_2) & \beta_0(r_i,E_2) & \beta_1(r_i,E_2) & 0 & \dots & 0 & 0 & 0 \\
0 & \beta_{-1}(r_i,E_3) & \beta_0(r_i,E_3) & \beta_1(r_i,E_3) & \dots & 0 & 0 & 0 \\
\vdots & \vdots & \vdots & \vdots & \ddots &\vdots & \vdots & \vdots\\
0 & 0 & 0 & 0 & \dots &\beta_{-1}(r_i,E_{n_E-1})&\beta_0(r_i,E_{n_E-1}) & \beta_1(r_i,E_{n_E-1})\\
0 & 0 & 0 & 0 & \dots & 0 & \beta_{-1}(r_i,E_{n_E})& \beta_0(r_i,E_{n_E})
\end{pmatrix},
\end{split}
\end{equation}


\begin{align}\label{eq:alpha and beta}
\alpha_{-1}(r_i, E_j) &=  \left(\frac{D(r_i, E_j)}{\Delta r^2}-\frac{1}{2\Delta r}\left.\frac{\partial D}{\partial r}\right|_{r_i, E_j}\right)\Delta t &2\leq i\leq n_r \quad & 1\leq j\leq n_E \\
\alpha_{0}(r_i, E_j) &=  \left(-\frac{2D(r_i, E_j)}{\Delta r^2}-\frac{1}{r}\left.\frac{\partial D}{\partial r}\right|_{r_i, E_j}\right)\Delta t &1\leq i\leq n_r \quad & 1\leq j\leq n_E\\
\alpha_{1}(r_i, E_j) &=  \left(\frac{D(r_i, E_j)}{\Delta r^2}+\frac{1}{2\Delta r}\left.\frac{\partial D}{\partial r}\right|_{r_i, E_j}\right)\Delta t &1\leq i\leq n_r-1 \quad & 1\leq j\leq n_E \\
\beta_{-1}(r_i, E_j) &=  0 &1\leq i\leq n_r \quad& 2\leq j\leq n_E \\
\beta_{0}(r_i, E_j) &=  -\frac{b(r_i, E_j)}{\Delta E}\Delta t &2\leq i\leq n_r \quad & 1\leq j\leq n_E\\
\beta_{1}(r_i, E_j) &= \frac{b(r_i, E_{j+1})}{\Delta E}\Delta t &1\leq i\leq n_r \quad &1\leq j\leq n_E-1,
\end{align}
\end{widetext}
and the function $C$ is given by
\begin{equation}\label{C}
C(r_i, E_j) = r_iQ_e(r_i, E_j)\Delta t.
\end{equation}
The matrices in Eq.~\eqref{eq:A and B} are constructed using the boundary conditions 
\begin{equation}\label{eq:boundary conditions}
\begin{split}
u(0, E) &= 0\\
u(r_{n_r+1}, E) &=0 \\
u(r, E_{n_E + 1}) & = 0,
\end{split}
\end{equation}
where $r_{n_r+1} = 49.9\,\unit{kpc}$ and $E_{n_E + 1} = m_\chi$. To update $u$ from time-step $n$ to $n+1$, we must solve Eq.~\eqref{eq:diff loss finite organized} for $u_{ij}^{n+1}$ given $u_{ij}^n$, $A$, $B$ and $C$.

It is convenient to flatten the two lower indices in $u^n_{ij}$ into a single lowered index by reshuffling the $i\in[1,n_r]$ and $j \in [1,n_E]$ indices of $r$ and $E$ into a single index $a \in [1,n_r\times n_E]$ as 
\[
a = i+(j-1)\times n_r.
\]
With this reordering, the phase space density can be encoded as a vector at time-step $n$. Using this redefinition, the vector $\boldsymbol{\mathcal{U}}^n$ at time-step $n$ has components
\begin{equation}
    \mathcal{U}^n_a = u_{ij}^n.
\end{equation}
Eq.~\eqref{eq:diff loss finite organized} can then be written as a matrix equation in the $n_r\times n_E$ vector indices
\begin{equation}\label{eq:diff loss vector}
\mathcal{M}\boldsymbol{\mathcal{U}^{n+1} } = \boldsymbol{\mathcal{U}^n} + \boldsymbol{\mathcal{C}}.
\end{equation}
The matrix $\boldsymbol{\mathcal{C}}$ has been redefined from $C_{ij}^n\equiv C(r_i, E_j)$ in a manner identical to $u_{ij}^n$, with $\mathcal{C}_a^n = C_{ij}^n$, $a = i+(j-1)\times n_r$.
The matrix $\mathcal{M}$ is defined as
\begin{equation}\label{eq:M block}
\mathcal{M}= \mathbb{1}-\mathcal{A}-\mathcal{B}
\end{equation}
The matrices $\mathcal{A}$ and $\mathcal{B}$ are $n_E\times n_E$ block matrices with $n_r\times n_r$ blocks:
\begin{equation}\label{eq:block A}
\mathcal{A} \equiv
\begin{pmatrix}
A(E_1) & 0 & 0 & \dots & 0 & 0\\
0 & A(E_2) & 0 & \dots & 0 & 0\\
0 & 0 & A(E_3) & \dots & 0 & 0\\
\vdots & \vdots &\vdots & \ddots & \vdots & \vdots \\
0 & 0 & 0 & \dots & A(E_{n_E-1}) & 0\\
0 & 0 & 0 & \dots & 0 & A(E_{n_E})\\
\end{pmatrix}
\end{equation}
where $A(E_j)$ is defined in Eq.~\eqref{eq:A and B} and
\begin{widetext}
\begin{equation}\label{eq:block B}
\mathcal{B} \equiv
\begin{pmatrix}
B_0(E_1) & B_1(E_1) & 0 & \dots & 0 & 0\\
B_{-1}(E_2) & B_0(E_2) & B_1(E_2) & \dots & 0 & 0\\
0 & B_{-1}(E_3) & B_0(E_3) & \dots & 0 & 0\\
\vdots & \vdots &\vdots & \ddots & \vdots & \vdots \\
0 & 0 & 0 & \dots & B_0(E_{n_E-1}) & B_1(E_{n_E-1}) \\
0 & 0 & 0 & \dots & B_{-1}(E_{n_E}) & B_0(E_{n_E}) \\
\end{pmatrix}.
\end{equation}
\end{widetext}

The block submatrices are
\begin{widetext}
\begin{equation}\label{eq:blocks of B}
B_m(E_j) = 
\begin{pmatrix}
\beta_m(r_1, E_j) & 0 & 0 & \dots & 0 & 0\\
0 & \beta_m(r_2, E_j) & 0 & \dots & 0 & 0\\
0 & 0 & \beta_m(r_3, E_j) & \dots & 0 & 0\\
\vdots & \vdots &\vdots & \ddots & \vdots & \vdots \\
0 & 0 & 0 & \dots & \beta_m(r_{n_r-1}, E_j) & 0\\
0 & 0 & 0 & \dots & 0 & \beta_m(r_{n_r}, E_j)\\
\end{pmatrix}
\quad \quad m=-1, 0, 1.
\end{equation}
\end{widetext}

Starting from the first row, we reduce $\mathcal{M}$ to upper-echelon form. The components of $\boldsymbol{\mathcal{U}}^{n+1}$ can then be solved for, starting from the final component and recursively solving for the other components in reverse order. For the most general matrix, this procedure would require $\mathcal{O}(n_E\times n_r)^3$ operations, which would be computationally prohibitive. In our case, the matrix $\mathcal{M}$ is tridiagonal with a fringe, which requires only $\mathcal{O}(n_r^2\times n_E)$ operations. 

Initially, we set the time-step to an approximation of the maximum timescale of the problem:
\begin{equation}
    \Delta t = \Delta t_0 = \max{(\tau_D^{\rm max}, \tau_b^{\rm max})},
\end{equation}
where $\tau_D^{\rm max}$ and $\tau^{\rm max}_b$ are the maximum diffusion and loss time-scales:
\begin{equation}
\begin{split}
     \tau_b^{\rm max} \simeq & \max\displaylimits_{ij} {\tau_b(r_i, E_j)}\\
    \tau_D^{\rm max} \simeq & \max\displaylimits_{ij}{\tau_D(r_i, E_j)}
\end{split}
\end{equation}

As we need only a rough estimate of the time scales for our initial time-step, we use simplified equations for $\tau_b$ and $\tau_D$:
\begin{equation}
    \begin{split}
        \tau_b =& \frac{m_\chi}{b}> \frac{\max{E}}{b}, \\
        \tau_D = & \frac{r_s^2}{\bar{D}},
    \end{split}
\end{equation}
where $r_s$ is the scale radius of the dark matter distribution, given in Section~\ref{sec:DM_production}. 

Using $\Delta t=\Delta t_0$, we iteratively solve for $\boldsymbol{\mathcal{U}^{n+1}}$ from $\boldsymbol{\mathcal{U}^{n}}$ until each component of the two vectors is different by less than 1 part in $10^3$. We then reduce the time-step by a factor of 2 and repeat, starting with the final result from the last time-step and iterating until the same convergence criteria is met. We repeat this procedure -- reducing the time-step by a factor of 2, and achieving convergence of the solution -- until there have been at least 5 different values of $\Delta t$ and $\boldsymbol{\mathcal{U}^n}$ converges in one step for 3 values of $\Delta t$ in a row. We find that these convergence criteria are conservative as convergence is achieved after 5 values of $\Delta t$ for all solutions that we examined.

\section{Simulating Intensity Maps}\label{sec:simulating intensity maps}

To generate synthetic data from our background and signal plus background models of M31, correlations between pixels due to the beam size must be correctly modelled. The rms noise is given by $\sigma_{\rm rms} =0.25 \unit{mJy/beam}$ in the central region of the radio map of M31 and $\sigma_{\rm rms} =0.3 \unit{mJy/beam}$ towards the outside of the map \citep{2020A&A...633A...5B} (see Section~\ref{sec:data}). This noise level is independent of the total flux, thus the simulated measurements in pixel $i$ for an intensity model with flux $\Phi_i$ is given by
\begin{equation}\label{eq:simulated data general}
s_i = \Phi_i + r_i
\end{equation}
where $r_i$ is the flux from noise in pixel $i$. These values can be positive or negative. The number of photons collected per beam is large enough that Poisson noise is negligible compared to the rms noise.

In general, the expected observed noise in pixel $i$ can be written as
\begin{equation}\label{eq:observed noise general}
r_i  = \int dx dy K(x-x_i, y-y_i) \tilde{r}(x, y)
\end{equation}
where $K(x-x_i, y-y_i)$ is the shape of the beam centered at pixel $i$ and $\tilde{r}(x,y)$ is the noise before convolution with the beam. We assume that the beam is a Gaussian, given by
\begin{equation}\label{eq:gauss beam}
K(\Delta x, \Delta y) = \frac{1}{2\pi \sigma_b^2} \exp{\left[-\frac{\Delta x^2 + \Delta y^2}{2\sigma_b^2}\right]},
\end{equation}
where
\begin{equation}\label{eq:sigma beam}
\sigma_b = \frac{(HPBW)}{2\sqrt{2\ln{(2)}}}
\end{equation}
 and $HPBW$ is the half-power beam-width projected onto the plane of the sky and is given by $0.34 \unit{kpc}$.

We assume that the noise before convolution is Gaussian distributed and only correlated over length scales much smaller than the size of the beam. Under these conditions the integral in Eq.~\eqref{eq:observed noise general} can be discretized as
\begin{equation}\label{eq:discrete convolution}
r_i  = \delta x \delta y \sum_{\alpha}K(x_\alpha-x_i, y_\alpha-y_i)\tilde{r}(x_\alpha, y_\alpha)
\end{equation}
where we have denoted the discretized coordinates with Greek indices and $\delta x$ and $\delta y$ are the grid-spacing for these coordinates (chosen to have the same value). These spacings are chosen to be much smaller than the beam but larger than the correlation length of $\tilde{r}$ so that
\begin{equation}\label{eq:corr of real r}
\langle \tilde{r}_\alpha \tilde{r}_\beta \rangle = \delta_{\alpha \beta} \tilde{\sigma}_{\alpha}^2
\end{equation}
where $\tilde{r}_\alpha \equiv \tilde{r}(x_\alpha, y_\alpha)$ and $\tilde{\sigma}_\alpha$ is related to $\sigma_{\mathrm{rms}, i}$ through 
\begin{equation}\label{eq:sigma rms vs sigma alpha}
\sigma_{\mathrm{rms}, i}^2=  \langle r_i^2 \rangle = \delta x^2 \delta y^2 \sum_\alpha \tilde{\sigma}_\alpha^2 K(x_\alpha-x_i, y_\alpha -y_i)^2.
\end{equation}
To solve for $\tilde{\sigma}_\alpha^2$, we make the approximation that $\tilde{\sigma}_\alpha$ is constant over the relevant regions of $K$ leading to 
\begin{equation}\label{eq:integ of beam}
\tilde{\sigma}^2_\alpha = \frac{4 \pi \sigma_b^2}{\delta x \delta y}\sigma_{\mathrm{rms},i}^2.
\end{equation}
for $\boldsymbol{x_\alpha}$ near pixel $i$. To generate noise for our synthetic data, we randomly sample each $\tilde{r}_\alpha$ from a Gaussian with a standard deviation given by $\tilde{\sigma}_\alpha$ and substitute the result into Eq.~\eqref{eq:discrete convolution}. For $\tilde{\sigma}_\alpha$, we use Eq.~\eqref{eq:integ of beam} where $i$ is the pixel closest to the point $\boldsymbol{x_\alpha}$. To avoid edge effects, we allow $x_\alpha$ and $y_\alpha$ to vary beyond the boundaries of the field of view by $5\sigma_b$.

To make an ensemble of pseudo-data assuming a particular hypothesis, we generate an ensemble of random noise maps and add them to a map of the intensity predicted by the hypothesis. For each combination of signal parameters that we test, we construct an ensemble of background-only maps from a set of $2\times 10^4$ random noise maps and we make an equal sized ensemble of signal plus background maps from an independent set of $2\times 10^4$ random noise maps. For each combination of signal parameters, we use the same set of random noise maps to construct our signal plus background pseudo-data, as we do not need to compare the ensembles of pseudo-data from one signal hypothesis to another.

\bibliographystyle{JHEP}
\bibliography{M31_DMRadio_bib}     

\providecommand{\href}[2]{#2}\begingroup\raggedright\begin{thebibliography}{10}

\bibitem{2022PhRvD.106j3526B}
D.J.~Bartlett, A.~Kostić, H.~Desmond, J.~Jasche and G.~Lavaux,
  \emph{Constraints on dark matter annihilation and decay from the large-scale
  structure of the nearby {Universe}},
  \href{https://doi.org/10.1103/PhysRevD.106.103526}{\emph{Physical Review D}
  {\bfseries 106} (2022) 103526}.

\bibitem{2017ApJ...836..208A}
F.-L.~Collaboration, \emph{Observations of {M31} and {M33} with the {Fermi}
  {Large} {Area} {Telescope}: a galactic center excess in {Andromeda}?},
  \href{https://doi.org/10.3847/1538-4357/aa5c3d}{\emph{The Astrophysical
  Journal} {\bfseries 836} (2017) 208}.

\bibitem{2021PhRvD.103b3027K}
C.M.~Karwin, S.~Murgia, I.V.~Moskalenko, S.P.~Fillingham, A.-K.~Burns and
  M.~Fieg, \emph{Dark matter interpretation of the {Fermi}-{LAT} observations
  toward the outer halo of {M31}},
  \href{https://doi.org/10.1103/PhysRevD.103.023027}{\emph{Physical Review D}
  {\bfseries 103} (2021) 023027}.

\bibitem{2015PhRvD..91j2001B}
M.R.~Buckley, E.~Charles, J.M.~Gaskins, A.M.~Brooks, A.~Drlica-Wagner,
  P.~Martin et~al., \emph{Search for {Gamma}-ray {Emission} from {Dark}
  {Matter} {Annihilation} in the {Large} {Magellanic} {Cloud} with the {Fermi}
  {Large} {Area} {Telescope}},
  \href{https://doi.org/10.1103/PhysRevD.91.102001}{\emph{Physical Review D}
  {\bfseries 91} (2015) 102001}.

\bibitem{2016PhRvD..93f2004C}
R.~Caputo, M.R.~Buckley, P.~Martin, E.~Charles, A.M.~Brooks, A.~Drlica-Wagner
  et~al., \emph{Search for {Gamma}-ray {Emission} from {Dark} {Matter}
  {Annihilation} in the {Small} {Magellanic} {Cloud} with the {Fermi} {Large}
  {Area} {Telescope}},
  \href{https://doi.org/10.1103/PhysRevD.93.062004}{\emph{Physical Review D}
  {\bfseries 93} (2016) 062004}.

\bibitem{2011PhRvL.107x1303G}
A.~Geringer-Sameth and S.M.~Koushiappas, \emph{Exclusion of {Canonical}
  {Weakly} {Interacting} {Massive} {Particles} by {Joint} {Analysis} of {Milky}
  {Way} {Dwarf} {Galaxies} with {Data} from the {Fermi} {Gamma}-{Ray} {Space}
  {Telescope}},
  \href{https://doi.org/10.1103/PhysRevLett.107.241303}{\emph{Physical Review
  Letters} {\bfseries 107} (2011) 241303}.

\bibitem{2011PhRvL.107x1302A}
M.~Ackermann, M.~Ajello, A.~Albert, W.B.~Atwood, L.~Baldini, J.~Ballet et~al.,
  \emph{Constraining {Dark} {Matter} {Models} from a {Combined} {Analysis} of
  {Milky} {Way} {Satellites} with the {Fermi} {Large} {Area} {Telescope}},
  \href{https://doi.org/10.1103/PhysRevLett.107.241302}{\emph{Physical Review
  Letters} {\bfseries 107} (2011) 241302}.

\bibitem{2014PhRvD..89d2001A}
M.~Ackermann, A.~Albert, B.~Anderson, L.~Baldini, J.~Ballet, G.~Barbiellini
  et~al., \emph{Dark matter constraints from observations of 25 {MilkyÂ} {Way}
  satellite galaxies with the {Fermi} {Large} {Area} {Telescope}},
  \href{https://doi.org/10.1103/PhysRevD.89.042001}{\emph{Physical Review D}
  {\bfseries 89} (2014) 042001}.

\bibitem{2015PhRvD..91h3535G}
A.~Geringer-Sameth, S.M.~Koushiappas and M.G.~Walker, \emph{Comprehensive
  search for dark matter annihilation in dwarf galaxies},
  \href{https://doi.org/10.1103/PhysRevD.91.083535}{\emph{Physical Review D}
  {\bfseries 91} (2015) 083535}.

\bibitem{2015ApJ...809L...4D}
A.~Drlica-Wagner, A.~Albert, K.~Bechtol, M.~Wood, L.~Strigari,
  M.~Sánchez-Conde et~al., \emph{Search for {Gamma}-{Ray} {Emission} from
  {DES} {Dwarf} {Spheroidal} {Galaxy} {Candidates} with {Fermi}-{LAT} {Data}},
  \href{https://doi.org/10.1088/2041-8205/809/1/L4}{\emph{The Astrophysical
  Journal} {\bfseries 809} (2015) L4}.

\bibitem{2015PhRvL.115h1101G}
A.~Geringer-Sameth, M.G.~Walker, S.M.~Koushiappas, S.E.~Koposov, V.~Belokurov,
  G.~Torrealba et~al., \emph{Indication of {Gamma}-{Ray} {Emission} from the
  {Newly} {Discovered} {Dwarf} {Galaxy} {Reticulum} {II}},
  \href{https://doi.org/10.1103/PhysRevLett.115.081101}{\emph{Physical Review
  Letters} {\bfseries 115} (2015) 081101}.

\bibitem{2015PhRvL.115w1301A}
M.~Ackermann, A.~Albert, B.~Anderson, W.B.~Atwood, L.~Baldini, G.~Barbiellini
  et~al., \emph{Searching for {Dark} {Matter} {Annihilation} from {Milky} {Way}
  {Dwarf} {Spheroidal} {Galaxies} with {Six} {Years} of {Fermi} {Large} {Area}
  {Telescope} {Data}},
  \href{https://doi.org/10.1103/PhysRevLett.115.231301}{\emph{Physical Review
  Letters} {\bfseries 115} (2015) 231301}.

\bibitem{2021PhRvD.104h3026G}
V.~Gammaldi, J.~Pérez-Romero, J.~Coronado-Blázquez, M.~Di~Mauro,
  E.V.~Karukes, M.A.~Sánchez-Conde et~al., \emph{Dark matter search in dwarf
  irregular galaxies with the {Fermi} {Large} {Area} {Telescope}},
  \href{https://doi.org/10.1103/PhysRevD.104.083026}{\emph{Physical Review D}
  {\bfseries 104} (2021) 083026}.

\bibitem{2009ApJ...697.1071A}
F.~Collaboration and W.B.~Atwood, \emph{The {Large} {Area} {Telescope} on the
  {Fermi} {Gamma}-ray {Space} {Telescope} {Mission}},
  \href{https://doi.org/10.1088/0004-637X/697/2/1071}{\emph{The Astrophysical
  Journal} {\bfseries 697} (2009) 1071}.

\bibitem{2011PhLB..697..412H}
D.~Hooper and L.~Goodenough, \emph{Dark matter annihilation in the {Galactic}
  {Center} as seen by the {Fermi} {Gamma} {Ray} {Space} {Telescope}},
  \href{https://doi.org/10.1016/j.physletb.2011.02.029}{\emph{Physics Letters
  B} {\bfseries 697} (2011) 412}.

\bibitem{2016PDU....12....1D}
T.~Daylan, D.P.~Finkbeiner, D.~Hooper, T.~Linden, S.K.N.~Portillo, N.L.~Rodd
  et~al., \emph{The {Characterization} of the {Gamma}-{Ray} {Signal} from the
  {Central} {Milky} {Way}: {A} {Compelling} {Case} for {Annihilating} {Dark}
  {Matter}}, \href{https://doi.org/10.1016/j.dark.2015.12.005}{\emph{Physics of
  the Dark Universe} {\bfseries 12} (2016) 1}.

\bibitem{2015JCAP...03..038C}
F.~Calore, I.~Cholis and C.~Weniger, \emph{Background model systematics for the
  {Fermi} {GeV} excess},
  \href{https://doi.org/10.1088/1475-7516/2015/03/038}{\emph{Journal of
  Cosmology and Astroparticle Physics} {\bfseries 2015} (2015) 038}.

\bibitem{2014PhRvD..90b3526A}
K.N.~Abazajian, N.~Canac, S.~Horiuchi and M.~Kaplinghat, \emph{Astrophysical
  and {Dark} {Matter} {Interpretations} of {Extended} {Gamma}-{Ray} {Emission}
  from the {Galactic} {Center}},
  \href{https://doi.org/10.1103/PhysRevD.90.023526}{\emph{Physical Review D}
  {\bfseries 90} (2014) 023526}.

\bibitem{2013PhRvD..88h3521G}
C.~Gordon and O.~Macias, \emph{Dark {Matter} and {Pulsar} {Model} {Constraints}
  from {Galactic} {Center} {Fermi}-{LAT} {Gamma} {Ray} {Observations}},
  \href{https://doi.org/10.1103/PhysRevD.88.083521}{\emph{Physical Review D}
  {\bfseries 88} (2013) 083521}.

\bibitem{2018PhRvD..97j3021M}
A.~McDaniel, T.~Jeltema and S.~Profumo, \emph{A {Multi}-{Wavelength} {Analysis}
  of {Annihilating} {Dark} {Matter} as the {Origin} of the {Gamma}-{Ray}
  {Emission} from {M31}},
  \href{https://doi.org/10.1103/PhysRevD.97.103021}{\emph{Physical Review D}
  {\bfseries 97} (2018) 103021}.

\bibitem{2011JCAP...03..010A}
K.N.~Abazajian, \emph{The {Consistency} of {Fermi}-{LAT} {Observations} of the
  {Galactic} {Center} with a {Millisecond} {Pulsar} {Population} in the
  {Central} {Stellar} {Cluster}},
  \href{https://doi.org/10.1088/1475-7516/2011/03/010}{\emph{Journal of
  Cosmology and Astroparticle Physics} {\bfseries 2011} (2011) 010}.

\bibitem{2013MNRAS.436.2461M}
N.~Mirabal, \emph{Dark matter vs. {Pulsars}: {Catching} the impostor},
  \href{https://doi.org/10.1093/mnras/stt1740}{\emph{Monthly Notices of the
  Royal Astronomical Society} {\bfseries 436} (2013) 2461}.

\bibitem{2014JHEAp...3....1Y}
Q.~Yuan and B.~Zhang, \emph{Millisecond pulsar interpretation of the {Galactic}
  center gamma-ray excess},
  \href{https://doi.org/10.1016/j.jheap.2014.06.001}{\emph{Journal of High
  Energy Astrophysics} {\bfseries 3-4} (2014) 1}.

\bibitem{2019JCAP...09..042M}
O.~Macias, S.~Horiuchi, M.~Kaplinghat, C.~Gordon, R.M.~Crocker and D.M.~Nataf,
  \emph{Strong {Evidence} that the {Galactic} {Bulge} is {Shining} in {Gamma}
  {Rays}}, \href{https://doi.org/10.1088/1475-7516/2019/09/042}{\emph{Journal
  of Cosmology and Astroparticle Physics} {\bfseries 2019} (2019) 042}.

\bibitem{2018NatAs...2..819B}
R.~Bartels, E.~Storm, C.~Weniger and F.~Calore, \emph{The {Fermi}-{LAT} {GeV}
  {Excess} {Traces} {Stellar} {Mass} in the {Galactic} {Bulge}},
  \href{https://doi.org/10.1038/s41550-018-0531-z}{\emph{Nature Astronomy}
  {\bfseries 2} (2018) 819}.

\bibitem{2018NatAs...2..387M}
O.~Macias, C.~Gordon, R.M.~Crocker, B.~Coleman, D.~Paterson, S.~Horiuchi
  et~al., \emph{Galactic {Bulge} {Preferred} {Over} {Dark} {Matter} for the
  {Galactic} {Center} {Gamma}-{Ray} {Excess}},
  \href{https://doi.org/10.1038/s41550-018-0414-3}{\emph{Nature Astronomy}
  {\bfseries 2} (2018) 387}.

\bibitem{2015JCAP...05..056L}
S.K.~Lee, M.~Lisanti and B.R.~Safdi, \emph{Distinguishing {Dark} {Matter} from
  {Unresolved} {Point} {Sources} in the {Inner} {Galaxy} with {Photon}
  {Statistics}},
  \href{https://doi.org/10.1088/1475-7516/2015/05/056}{\emph{Journal of
  Cosmology and Astroparticle Physics} {\bfseries 2015} (2015) 056}.

\bibitem{2013PhRvD..88h3009H}
D.~Hooper, I.~Cholis, T.~Linden, J.~Siegal-Gaskins and T.~Slatyer,
  \emph{Millisecond {Pulsars} {Cannot} {Account} for the {Inner} {Galaxy}'s
  {GeV} {Excess}},
  \href{https://doi.org/10.1103/PhysRevD.88.083009}{\emph{Physical Review D}
  {\bfseries 88} (2013) 083009}.

\bibitem{2016PhRvL.117k1101C}
E.~Carlson, T.~Linden and S.~Profumo, \emph{Putting {Things} {Back} {Where}
  {They} {Belong}: {Tracing} {Cosmic}-{Ray} {Injection} with {H2}},
  \href{https://doi.org/10.1103/PhysRevLett.117.111101}{\emph{Physical Review
  Letters} {\bfseries 117} (2016) 111101}.

\bibitem{2017ApJ...840...43A}
T.F.-L.~Collaboration, \emph{The {Fermi} {Galactic} {Center} {GeV} {Excess} and
  {Implications} for {Dark} {Matter}},
  \href{https://doi.org/10.3847/1538-4357/aa6cab}{\emph{The Astrophysical
  Journal} {\bfseries 840} (2017) 43}.

\bibitem{2021PhRvD.103f3029D}
M.~Di~Mauro, \emph{The characteristics of the {Galactic} center excess measured
  with 11 years of {Fermi}-{LAT} data},
  \href{https://doi.org/10.1103/PhysRevD.103.063029}{\emph{Physical Review D}
  {\bfseries 103} (2021) 063029}.

\bibitem{2019arXiv190408430L}
R.K.~Leane and T.R.~Slatyer, \emph{Dark {Matter} {Strikes} {Back} at the
  {Galactic} {Center}},
  \href{https://doi.org/10.1103/PhysRevLett.123.241101}{\emph{Physical Review
  Letters} {\bfseries 123} (2019) 241101}.

\bibitem{2020PhRvD.101b3014C}
L.J.~Chang, S.~Mishra-Sharma, M.~Lisanti, M.~Buschmann, N.L.~Rodd and
  B.R.~Safdi, \emph{Characterizing the {Nature} of the {Unresolved} {Point}
  {Sources} in the {Galactic} {Center}},
  \href{https://doi.org/10.1103/PhysRevD.101.023014}{\emph{Physical Review D}
  {\bfseries 101} (2020) 023014}.

\bibitem{2021arXiv211006931M}
S.~Mishra-Sharma and K.~Cranmer, \emph{A neural simulation-based inference
  approach for characterizing the {Galactic} {Center}
  \${\textbackslash}gamma\$-ray excess}, {\emph{arXiv:2110.06931 [astro-ph,
  physics:hep-ph]} (2021) }.

\bibitem{2020A&A...633A...5B}
R.~Beck, E.M.~Berkhuijsen, R.~Gießübel and D.D.~Mulcahy, \emph{Magnetic
  fields and cosmic rays in {M} 31. {I}. {Spectral} indices, scale lengths,
  {Faraday} rotation, and magnetic field pattern},
  \href{https://doi.org/10.1051/0004-6361/201936481}{\emph{Astronomy \&
  Astrophysics} {\bfseries 633} (2020) A5}.

\bibitem{2013PhRvD..88b3504E}
A.E.~Egorov and E.~Pierpaoli, \emph{Constraints on dark matter annihilation by
  radio observations of {M31}},
  \href{https://doi.org/10.1103/PhysRevD.88.023504}{\emph{Physical Review D}
  {\bfseries 88} (2013) 023504}.

\bibitem{2016PhRvD..94b3507C}
M.H.~Chan, \emph{Revisiting the constraints on annihilating dark matter by
  radio observational data of {M31}},
  \href{https://doi.org/10.1103/PhysRevD.94.023507}{\emph{Physical Review D}
  {\bfseries 94} (2016) 023507}.

\bibitem{2021MNRAS.501.5692C}
M.H.~Chan, C.F.~Yeung, L.~Cui and C.S.~Leung, \emph{Analysing the radio flux
  density profile of the {M31} galaxy: a possible dark matter interpretation},
  \href{https://doi.org/10.1093/mnras/staa4004}{\emph{Monthly Notices of the
  Royal Astronomical Society} (2020) staa4004}.

\bibitem{2022PhRvD.106b3023E}
A.E.~Egorov, \emph{Updated constraints on dark matter ({WIMP}) annihilation by
  radio observations of {M31}},  May, 2022.

\bibitem{2022arXiv221202999B}
K.~Boshkayev, T.~Konysbayev, Y.~Kurmanov, O.~Luongo, M.~Muccino, H.~Quevedo
  et~al., {\emph{Numerical analyses of {M31} dark matter profiles},  Tech. Rep.
  } (Dec., 2022), \href{https://doi.org/10.48550/arXiv.2212.02999}{DOI}.

\bibitem{2014A&A...571A..61G}
R.~Gießübel and R.~Beck, \emph{The magnetic field structure of the central
  region in {M31}},
  \href{https://doi.org/10.1051/0004-6361/201323211}{\emph{Astronomy \&
  Astrophysics} {\bfseries 571} (2014) A61}.

\bibitem{2011ApJ...739...20C}
S.~Courteau, L.M.~Widrow, M.~McDonald, P.~Guhathakurta, K.M.~Gilbert, Y.~Zhu
  et~al., \emph{{THE} {LUMINOSITY} {PROFILE} {AND} {STRUCTURAL} {PARAMETERS}
  {OF} {THE} {ANDROMEDA} {GALAXY}},
  \href{https://doi.org/10.1088/0004-637X/739/1/20}{\emph{The Astrophysical
  Journal} {\bfseries 739} (2011) 20}.

\bibitem{2012A&A...546A...4T}
A.~Tamm, E.~Tempel, P.~Tenjes, O.~Tihhonova and T.~Tuvikene, \emph{Stellar mass
  map and dark matter distribution in {M31}},
  \href{https://doi.org/10.1051/0004-6361/201220065}{\emph{Astronomy \&
  Astrophysics} {\bfseries 546} (2012) A4}.

\bibitem{2015ICRC...34..492M}
I.V.~Moskalenko, G.~Jóhannesson, E.~Orlando, T.A.~Porter, A.W.~Strong and
  A.E.~Vladimirov, \emph{{GALPROP} {Code} for {Galactic} {Cosmic} {Ray}
  {Propagation} and {Associated} {Photon} {Emissions}}, .

\bibitem{2017JCAP...09..027M}
A.~McDaniel, T.~Jeltema, S.~Profumo and E.~Storm, \emph{Multiwavelength
  {Analysis} of {Dark} {Matter} {Annihilation} and {RX}-{DMFIT}},
  \href{https://doi.org/10.1088/1475-7516/2017/09/027}{\emph{Journal of
  Cosmology and Astroparticle Physics} {\bfseries 2017} (2017) 027}.

\bibitem{2005MNRAS.356..979M}
A.W.~McConnachie, M.J.~Irwin, A.M.N.~Ferguson, R.A.~Ibata, G.F.~Lewis and
  N.~Tanvir, \emph{Distances and metallicities for 17 {Local} {Group}
  galaxies},
  \href{https://doi.org/10.1111/j.1365-2966.2004.08514.x}{\emph{Monthly Notices
  of the Royal Astronomical Society} {\bfseries 356} (2005) 979}.

\bibitem{2001ChPhL..18.1420M}
J.~Ma, \emph{Structure and {Inclination} {Angle} of the {Spiral} {Galaxy}
  {M31}}, \href{https://doi.org/10.1088/0256-307X/18/10/339}{\emph{Chinese
  Physics Letters} {\bfseries 18} (2001) 1420}.

\bibitem{2022arXiv220311601B}
C.~Bierlich, S.~Chakraborty, N.~Desai, L.~Gellersen, I.~Helenius, P.~Ilten
  et~al., {\emph{A comprehensive guide to the physics and usage of {PYTHIA}
  8.3},  Tech. Rep. } (Mar., 2022),
  \href{https://doi.org/10.48550/arXiv.2203.11601}{DOI}.

\bibitem{1996ApJ...462..563N}
J.F.~Navarro, C.S.~Frenk and S.D.M.~White, \emph{The {Structure} of {Cold}
  {Dark} {Matter} {Halos}}, {\emph{The Astrophysical Journal} {\bfseries 462}
  (1996) 563}.

\bibitem{1997ApJ...490..493N}
J.F.~Navarro, C.S.~Frenk and S.D.M.~White, \emph{A {Universal} {Density}
  {Profile} from {Hierarchical} {Clustering}},
  \href{https://doi.org/10.1086/304888}{\emph{The Astrophysical Journal}
  {\bfseries 490} (1997) 493}.

\bibitem{1996MNRAS.278..488Z}
H.~Zhao, \emph{Analytical models for galactic nuclei},
  \href{https://doi.org/10.1093/mnras/278.2.488}{\emph{Monthly Notices of the
  Royal Astronomical Society} {\bfseries 278} (1996) 488}.

\bibitem{1963AJ.....68..435B}
W.~Baade and H.H.~Swope, \emph{Variable star field 96' south preceeding the
  nucleous of the {Andromeda} galaxy.},
  \href{https://doi.org/10.1086/108996}{\emph{The Astronomical Journal}
  {\bfseries 68} (1963) 435}.

\bibitem{2010A&A...511A..89C}
E.~Corbelli, S.~Lorenzoni, R.~Walterbos, R.~Braun and D.~Thilker, \emph{A
  wide-field {H} {I} mosaic of {Messier} 31. {II}. {The} disk warp, rotation,
  and the dark matter halo},
  \href{https://doi.org/10.1051/0004-6361/200913297}{\emph{Astronomy and
  Astrophysics} {\bfseries 511} (2010) A89}.

\bibitem{2021ApJ...920...84L}
S.~Li, A.G.~Riess, M.P.~Busch, S.~Casertano, L.M.~Macri and W.~Yuan, \emph{A
  {Sub}-2\% {Distance} to {M31} from {Photometrically} {Homogeneous}
  {Near}-infrared {Cepheid} {Period}-{Luminosity} {Relations} {Measured} with
  the {Hubble} {Space} {Telescope}},
  \href{https://doi.org/10.3847/1538-4357/ac1597}{\emph{The Astrophysical
  Journal} {\bfseries 920} (2021) 84}.

\bibitem{1998APh.....9..227C}
S.~Colafrancesco and P.~Blasi, \emph{Clusters of {Galaxies} and the {Diffuse}
  {Gamma} {Ray} {Background}},
  \href{https://doi.org/10.1016/S0927-6505(98)00018-8}{\emph{Astroparticle
  Physics} {\bfseries 9} (1998) 227}.

\bibitem{1998ApJ...509..212S}
A.W.~Strong and I.V.~Moskalenko, \emph{Propagation of cosmic-ray nucleons in
  the {Galaxy}}, \href{https://doi.org/10.1086/306470}{\emph{The Astrophysical
  Journal} {\bfseries 509} (1998) 212}.

\bibitem{2007ARNPS..57..285S}
A.W.~Strong, I.V.~Moskalenko and V.S.~Ptuskin, \emph{Cosmic-{Ray} {Propagation}
  and {Interactions} in the {Galaxy}},
  \href{https://doi.org/10.1146/annurev.nucl.57.090506.123011}{\emph{Annual
  Review of Nuclear and Particle Science} {\bfseries 57} (2007) 285}.

\bibitem{2012ApJ...752...68V}
A.E.~Vladimirov, G.~Jóhannesson, I.V.~Moskalenko and T.A.~Porter,
  \emph{Testing the {Origin} of {High}-energy {Cosmic} {Rays}},
  \href{https://doi.org/10.1088/0004-637X/752/1/68}{\emph{The Astrophysical
  Journal} {\bfseries 752} (2012) 68}.

\bibitem{1995ApJ...441..209H}
U.~Heinbach and M.~Simon, \emph{Propagation of {Galactic} {Cosmic} {Rays} under
  {Diffusive} {Reacceleration}},
  \href{https://doi.org/10.1086/175350}{\emph{The Astrophysical Journal}
  {\bfseries 441} (1995) 209}.

\bibitem{1941DoSSR..30..301K}
A.N.~Kolmogorov, V.~Levin, J.C.R.~Hunt, O.M.~Phillips and D.~Williams,
  \emph{The local structure of turbulence in incompressible viscous fluid for
  very large {Reynolds} numbers},
  \href{https://doi.org/10.1098/rspa.1991.0075}{\emph{Proceedings of the Royal
  Society of London. Series A: Mathematical and Physical Sciences} {\bfseries
  434} (1941) 9}.

\bibitem{1998IAUS..184..351H}
P.~Hoernes, R.~Beck and E.M.~Berkhuijsen, \emph{The central regions of the
  galaxy and galaxies: proceedings of the 184th {Symposium} of the
  {International} {Astronomical} {Union}, held in {Tokyo}, {Japan}, {August}
  18-22}, Kluwer Academic, Boston, Mass (1998).

\bibitem{2004A&A...414...53F}
A.~Fletcher, E.M.~Berkhuijsen, R.~Beck and A.~Shukurov, \emph{The magnetic
  field of {M31} from multi-wavelength radio polarization observations},
  \href{https://doi.org/10.1051/0004-6361:20034133}{\emph{Astronomy \&
  Astrophysics} {\bfseries 414} (2004) 53}.

\bibitem{2019A&A...622A..16O}
S.P.~O'Sullivan, J.~Machalski, C.L.~Van~Eck, G.~Heald, M.~Brüggen,
  J.P.U.~Fynbo et~al., \emph{The intergalactic magnetic field probed by a giant
  radio galaxy},
  \href{https://doi.org/10.1051/0004-6361/201833832}{\emph{Astronomy and
  Astrophysics} {\bfseries 622} (2019) A16}.

\bibitem{2012SSRv..166..133H}
M.~Haverkorn and V.~Heesen, \emph{Magnetic {Fields} in {Galactic} {Haloes}},
  \href{https://doi.org/10.1007/s11214-011-9757-0}{\emph{Space Science Reviews}
  {\bfseries 166} (2012) 133}.

\bibitem{1991ApJ...372...54B}
R.~Braun, \emph{The {Distribution} and {Kinematics} of {Neutral} {Gas} in
  {M31}}, \href{https://doi.org/10.1086/169954}{\emph{The Astrophysical
  Journal} {\bfseries 372} (1991) 54}.

\bibitem{2009ApJ...707..916F}
D.J.~Fixsen, \emph{The {Temperature} of the {Cosmic} {Microwave} {Background}},
  \href{https://doi.org/10.1088/0004-637X/707/2/916}{\emph{The Astrophysical
  Journal} {\bfseries 707} (2009) 916}.

\bibitem{2011A&A...526A.155T}
E.~Tempel, T.~Tuvikene, A.~Tamm and P.~Tenjes, \emph{{SDSS} surface photometry
  of {M} 31 with absorption corrections},
  \href{https://doi.org/10.1051/0004-6361/201016067}{\emph{Astronomy and
  Astrophysics} {\bfseries 526} (2011) A155}.

\bibitem{2012MNRAS.426..892G}
B.~Groves, O.~Krause, K.~Sandstrom, A.~Schmiedeke, A.~Leroy, H.~Linz et~al.,
  \emph{The heating of dust by old stellar populations in the bulge of {M31}},
  \href{https://doi.org/10.1111/j.1365-2966.2012.21696.x}{\emph{Monthly Notices
  of the Royal Astronomical Society} {\bfseries 426} (2012) 892}.

\bibitem{2012IAUS..284..112G}
B.~Groves and O.~Krause, \emph{Hot \& cold dust in {M31}: the resolved {SED} of
  {Andromeda}}, .

\bibitem{2000ApJ...537..763S}
A.W.~Strong, I.V.~Moskalenko and O.~Reimer, \emph{Diffuse continuum gamma rays
  from the {Galaxy}}, \href{https://doi.org/10.1086/309038}{\emph{The
  Astrophysical Journal} {\bfseries 537} (2000) 763}.

\bibitem{2001AJ....122..908G}
G.C.~Gómez, R.A.~Benjamin and D.P.~Cox, \emph{A {Reexamination} of the
  {Distribution} of {Galactic} {Free} {Electrons}},
  \href{https://doi.org/10.1086/321180}{\emph{The Astronomical Journal}
  {\bfseries 122} (2001) 908}.

\bibitem{1994ApJ...431..156W}
R.A.M.~Walterbos and R.~Braun, \emph{Diffuse {Ionized} {Gas} in the {Spiral}
  {Galaxy} {M31}}, \href{https://doi.org/10.1086/174475}{\emph{The
  Astrophysical Journal} {\bfseries 431} (1994) 156}.

\bibitem{2009A&A...505..497Y}
J.~Yin, J.L.~Hou, N.~Prantzos, S.~Boissier, R.X.~Chang, S.Y.~Shen et~al.,
  \emph{Milky {Way} versus {Andromeda}: a tale of two disks},
  \href{https://doi.org/10.1051/0004-6361/200912316}{\emph{Astronomy and
  Astrophysics} {\bfseries 505} (2009) 497}.

\bibitem{2017A&A...607A.106M}
A.~Marasco, F.~Fraternali, J.M.~van~der Hulst and T.~Oosterloo,
  \emph{Distribution and kinematics of atomic and molecular gas inside the
  solar circle},
  \href{https://doi.org/10.1051/0004-6361/201731054}{\emph{Astronomy and
  Astrophysics} {\bfseries 607} (2017) A106}.

\bibitem{2019A&A...630A.125V}
F.~Vincenzo, A.~Miglio, C.~Kobayashi, J.T.~Mackereth and J.~Montalban, \emph{He
  abundances in disc galaxies. {I}. {Predictions} from cosmological
  chemodynamical simulations},
  \href{https://doi.org/10.1051/0004-6361/201935886}{\emph{Astronomy and
  Astrophysics} {\bfseries 630} (2019) A125}.

\bibitem{2011hea..book.....L}
M.S.~Longair, \emph{High {Energy} {Astrophysics}}, {\emph{High Energy
  Astrophysics, by Malcolm S. Longair, Cambridge, UK: Cambridge University
  Press, 2011} (2011) }.

\bibitem{2017ApJ...837..140S}
P.~Subedi, W.~Sonsrettee, P.~Blasi, D.~Ruffolo, W.~Matthaeus, D.~Montgomery
  et~al., \emph{Charged particle diffusion in isotropic random magnetic
  fields}, \href{https://doi.org/10.3847/1538-4357/aa603a}{\emph{The
  Astrophysical Journal} {\bfseries 837} (2017) 140}.

\bibitem{2015MNRAS.448.3747R}
M.~Regis, L.~Richter, S.~Colafrancesco, S.~Profumo, W.J.G.~de~Blok and
  M.~Massardi, \emph{Local {Group} {dSph} radio survey with {ATCA} ({II}):
  {Non}-thermal diffuse emission},
  \href{https://doi.org/10.1093/mnras/stv127}{\emph{Monthly Notices of the
  Royal Astronomical Society} {\bfseries 448} (2015) 3747}.

\bibitem{2013MNRAS.435.1598B}
E.M.~Berkhuijsen, R.~Beck and F.S.~Tabatabaei, \emph{How cosmic ray electron
  propagation affects radio-far-infrared correlations in {M} 31 and {M} 33},
  \href{https://doi.org/10.1093/mnras/stt1400}{\emph{Monthly Notices of the
  Royal Astronomical Society} {\bfseries 435} (2013) 1598}.

\bibitem{2001ApJ...563..768Y}
N.E.~Yanasak, M.E.~Wiedenbeck, R.A.~Mewaldt, A.J.~Davis, A.C.~Cummings,
  J.S.~George et~al., \emph{Measurement of the {Secondary} {Radionuclides}
  {10Be}, {26Al}, {36Cl}, {54Mn}, and {14C} and {Implications} for the
  {Galactic} {Cosmic}-{Ray} {Age}},
  \href{https://doi.org/10.1086/323842}{\emph{The Astrophysical Journal}
  {\bfseries 563} (2001) 768}.

\bibitem{2009ApJ...698.1666G}
J.S.~George, K.A.~Lave, M.E.~Wiedenbeck, W.R.~Binns, A.C.~Cummings, A.J.~Davis
  et~al., \emph{Elemental {Composition} and {Energy} {Spectra} of {Galactic}
  {Cosmic} {Rays} {During} {Solar} {Cycle} 23},
  \href{https://doi.org/10.1088/0004-637X/698/2/1666}{\emph{The Astrophysical
  Journal} {\bfseries 698} (2009) 1666}.

\bibitem{2011ApJ...729..106T}
R.~Trotta, G.~Johannesson, I.V.~Moskalenko, T.A.~Porter, R.R.~de~Austri and
  A.W.~Strong, \emph{Constraints on cosmic-ray propagation models from a global
  {Bayesian} analysis},
  \href{https://doi.org/10.1088/0004-637X/729/2/106}{\emph{The Astrophysical
  Journal} {\bfseries 729} (2011) 106}.

\bibitem{2021PhRvD.104h3005G}
Y.~Génolini, M.~Boudaud, M.~Cirelli, L.~Derome, J.~Lavalle, D.~Maurin et~al.,
  \emph{New minimal, median, and maximal propagation models for dark matter
  searches with {Galactic} cosmic rays},
  \href{https://doi.org/10.1103/PhysRevD.104.083005}{\emph{Physical Review D}
  {\bfseries 104} (2021) 083005}.

\bibitem{1970RvMP...42..237B}
G.R.~BLUMENTHAL and R.J.~GOULD, \emph{Bremsstrahlung, {Synchrotron}
  {Radiation}, and {Compton} {Scattering} of {High}-{Energy} {Electrons}
  {Traversing} {Dilute} {Gases}},
  \href{https://doi.org/10.1103/RevModPhys.42.237}{\emph{Reviews of Modern
  Physics} {\bfseries 42} (1970) 237}.

\bibitem{1972Phy....60..145G}
R.J.~Gould, \emph{Energy loss of fast electrons and positrons in a plasma},
  \href{https://doi.org/10.1016/0031-8914(72)90227-3}{\emph{Physica} {\bfseries
  60} (1972) 145}.

\bibitem{1988SoPh..115..313S}
J.~Steinacker, W.~Dröge and R.~Schlickeiser, \emph{Particle {Acceleration} in
  {Impulsive} {Solar} {Flares} - {Part} {One}},
  \href{https://doi.org/10.1007/BF00148731}{\emph{Solar Physics} {\bfseries
  115} (1988) 313}.

\bibitem{1992nrfa.book.....P}
W.H.~Press, ed., \emph{{FORTRAN} numerical recipes}, Cambridge University
  Press, Cambridge [England] ; New York, 2nd ed~ed. (1996).

\bibitem{1998clel.book.....J}
J.D.~Jackson, \emph{Classical {Electrodynamics}}, Wiley (Oct., 1975).

\bibitem{2010A&A...517A..77T}
F.S.~Tabatabaei and E.M.~Berkhuijsen, \emph{Relating dust, gas, and the rate of
  star formation in {M} 31},
  \href{https://doi.org/10.1051/0004-6361/200913593}{\emph{Astronomy and
  Astrophysics} {\bfseries 517} (2010) A77}.

\bibitem{2016MNRAS.456.4128R}
S.~Rahmani, S.~Lianou and P.~Barmby, \emph{Star formation laws in the
  {Andromeda} galaxy: gas, stars, metals and the surface density of star
  formation}, \href{https://doi.org/10.1093/mnras/stv2951}{\emph{Monthly
  Notices of the Royal Astronomical Society} {\bfseries 456} (2016) 4128}.

\bibitem{2002JPhG...28.2693R}
A.L.~Read, \emph{Presentation of search results:
  the\${\textbackslash}less\$i\${\textbackslash}greater\${CL}\${\textbackslash}less\$sub\${\textbackslash}greater\$s\${\textbackslash}less\$/sub\${\textbackslash}greater\$\${\textbackslash}less\$/i\${\textbackslash}greater\$technique},
  \href{https://doi.org/10.1088/0954-3899/28/10/313}{\emph{Journal of Physics
  G: Nuclear and Particle Physics} {\bfseries 28} (2002) 2693}.

\end{thebibliography}\endgroup

\end{document}